\documentclass[aps,prd,twocolumn,floatfix,superscriptaddress]{revtex4-2}
\usepackage{tabularx}
\usepackage{hyperref}
\hypersetup{hypertex=true,
	colorlinks=true,
	linkcolor=blue,
	anchorcolor=blue,
	citecolor=blue}
\usepackage{graphicx}
\usepackage{amsmath}
\usepackage{amssymb}
\usepackage{comment}
\usepackage{xcolor}
\usepackage{xspace}
\usepackage{multirow}
\usepackage{natbib}
\usepackage{subfigure}	
\usepackage[english]{babel}
\usepackage{url}
\usepackage{makecell}
\usepackage{ulem} 

\begin{document}
	\title{Fast resolving Galactic binaries in LISA data and its ability to study the Milky Way}
	\author{Pin Gao}
	\affiliation{Department of Astronomy, Beijing Normal University, Beijing 100875, China}
	\author{Xi-Long Fan}\email{ xilong.fan@whu.edu.cn}
	\affiliation{School of Physics and Technology, Wuhan University, Wuhan 430072, China}
	\author{Zhou-Jian Cao}
	\affiliation{Department of Astronomy, Beijing Normal University, Beijing 100875, China}
	\author{Xue-Hao Zhang}
	\affiliation{Institute of Theoretical Physics and Research Center of Gravitation, Lanzhou 
		University, Lanzhou 730000, China}

	\begin{abstract}
		Resolving individual gravitational waves from tens of millions of double white dwarf (DWD) 
		binaries in the Milky Way is a challenge for future space-based gravitational wave 
		detection programs. 
		By using previous data to define the priors for the next search, we propose 
		an accelerated approach of searching the DWD binaries, and demonstrate it’s effciency 
		based on the GBSIEVER detection pipeline.
		Compared to the traditional GBSIEVER method, our method can obtain $\sim 50\%$ of sources 
		with 2.5\% of the searching time for LDC1-4 data.
		In addition, we find that both methods have a similar ability to detect the Milky 
		Way structure by their confirmed sources. The relative error of distance and chirp mass is 
		about 20\% for DWD binaries whose gravitational wave frequency is higher than 
		$4\times10^{-3}$ Hz, even if they are close to the Galactic centre.
		Finally, we propose a signal-to-noise ratio (SNR) threshold for LISA to confirm the 
		detection of DWD binaries.
		The threshold should be 16 when the gravitational wave frequency is lower than 
		$4\times10^{-3}$ Hz and 9 when the frequency range is from $4\times10^{-3}$ Hz to 
		$1.5\times10^{-2}$ Hz.
	\end{abstract}
	\maketitle

	\section{Introduction}
	\label{sec:intro}
	The first direct detection of gravitational wave (GW) from a binary black hole merger by the 
	Laser Interferometer Gravitational-Wave Observatory (LIGO) gives us access to a new messenger 
	and opens the era of observational gravitational wave astronomy 
	\cite{PhysRevLett.116.061102}. For lower frequencies, we have to increase the arm length of the 
	gravitational wave detector and avoid the influence of ground 
	noise. Laser Interferometric Space Antenna (LISA) \cite{noauthor_lisa_nodate, baker_laser_2019, 
		gong_scientific_2011},
	Taiji \cite{2020IJMPA..3550075R, hu_taiji_2017, the_taiji_scientific_collaboration_chinas_2021} 
	and Tianqin \cite{Luo_2016, mei_tianqin_2020, luo_first_2020} are proposed in this context. 
	They are space-based gravitational wave detectors and have a sensitive frequency domain from 
	$10^{-4}$ Hz to $10^{-1}$ Hz, where there are rich sources of GW, such as coalescing 
	massive black holes, ultracompact galactic binaries, and extreme mass ratio inspirals 
	\cite{noauthor_lisa_nodate, Robson_2019}.
	
	For space-based gravitational wave detectors, data processing is a
	challenge. There are numerous ($\sim10^7$) Galactic binary gravitational wave 
	sources \cite{Cornish_2017} whose signals span the whole LISA band. 
	Most of them are double white dwarf (DWD) binaries. Though DWD signals are simple: nearly 
	constant amplitude sine function with a small amount of change in frequency, the enormous 
	quantity of them, as well as the instrument's frequency and amplitude modulations, 
	make it difficult to break them down. 
	In the last two decades, a series of LISA data challenges have been organized to prepare the 
	LISA data analysis techniques. 
	The first four challenges are
	called mock LISA data challenges (MLDCs) \cite{arnaud_mock_2006, arnaud_overview_2007, 
		Babak_2008, babak_mock_2010}, followed by a series 
	called LISA data challenge (LDC) \cite{noauthor_ldc-manual-sangriapdf_nodate}. The LDC1 
	contains several sections, among 
	which LDC1-4 is our concern's galactic binary gravitational waves. The data
	contains about 30 million binary systems and mock LISA instrumental 
	noise with a signal sampling period of 15 seconds. The total duration of the data 
	is two years (62914560 seconds). Several data analysis methods have been 
	proposed \cite{mohanty_tomographic_2006, stroeer_inference_2007, crowder_extracting_2007-1, 
		prix_f-statistic_2007, zhang_resolving_2021-1, littenberg_detection_2011, blaut_mock_2010, 
		littenberg_global_2020, strub_bayesian_2022, Lu_2023, zhang_resolving_2022}. Some of them 
	are 
	relatively mature:
	\begin{itemize}
		\item[(1)] The blocked-annealed Metropolis–Hasting (BAM) algorithm combined with simulated 
		annealing and Markov chain Monte Carlo (MCMC) has been optimized to 
		search for tens of thousands of overlapping signals across the LISA band and performed 
		almost flawlessly on all the round 1 Mock LISA Data Challenge data sets. In a blockwise
		frequency search, they report about 20000 sources in MLDC-2.1
		data which had 26 million DWD binaries \cite{crowder_extracting_2007-1}. 
		\item[(2)] An extension of BAM incorporating transdimensional
		MCMC was tested on the training data from MLDC-4
		and reported approximately 9000 sources in the [0, 10] mHz band with a
		detection rate of about 90\% \cite{littenberg_detection_2011}.
		\item[(3)] An iterative source subtraction with a maximum likelihood detection method 
		reported about 12000 sources in MLDC-3.1 data, which contains 60 million DWD binaries 
		\cite{blaut_mock_2010}.
		\item[(4)] An interesting new approach that aims to create a time-evolving
		source catalogue rather than using them all at once has been opened recently. In real 
		detection, as LISA data accumulates, this approach could be of great use 
		\cite{littenberg_global_2020}.
		\item[(5)] Particle swarm optimization (PSO) is used in Galactic binary separation by 
		iterative extraction and validation using extended range (GBSIEVER) 
		\cite{zhang_resolving_2021-1}. In the [0.1, 15] mHz frequency domain, the method can find 
		sources from 9291 to 12270, corresponding to the detection rate of 90.26\% and 84.16\%, 
		respectively. 
		\item[(6)] Bayesian approaches are computationally expensive to extract tens of 
		thousands of signals. The authors first find the maximum likelihood estimate and then 
		compute the posterior distribution around the identified maximum likelihood estimate to 
		give a more informative solution \cite{strub_bayesian_2022}.
		\item[(7)] An iterative combinatorial algorithm to search for double white dwarfs in 
		MLDC-3.1 data quickly determine the rough parameters of the target sources in a coarse 
		search process. After that, a fine search process (Particle Swarm Optimization) is used to 
		further determine the parameters of the signals based on the coarse search 
		\cite{Lu_2023}.
	\end{itemize}
	
	Until now, most search methods directly search the whole 2-year mock data and found about 
	$10^4$ candidates. However, waiting for LISA to return all the data before searching is 
	impractical. In fact, once the LISA mission begins taking data, the data analysis could 
	continuously add small chunks of new data to the existing data set, then use previous results 
	to define the priors for the next search. 
	Here we propose an accelerated approach depending on the early search results 
	from the previous LISA data. Similar idea has been used for existing time-evolving LISA 
	Galactic binary GWs extraction algorithm \cite{littenberg_global_2020}.
	We demonstrate our approach based on the GBSIEVER detection pipeline 
	\cite{zhang_resolving_2021-1}, and compare the search results and computation time of the 
	traditional search methods with ours.
	
	The rest of this paper is organized as follows. We briefly describe 
	binary gravitational waves in Section 2. Section 3 introduces our search method
	and develops a way to accelerate the search. Moreover, we show our search results in Section 4. 
	LISA's ability to detect the Milky Way structure is shown in Section 5. 
	According to these search results, we give a suggested signal-to-noise ratio (SNR) threshold 
	value in future actual space-based gravitational wave detection missions in Section 
	6. Finally, our conclusions are presented in Section 7.

	\section{LDC Data description}
	\label{sec:data}
	\subsection{DWD binary gravitational wave}
	There are two double white dwarf (DWD) binaries in LDC1-4:
	\begin{itemize}
		\item[(1)] Detached DWD binaries. They only evolve with the release of orbital 
		energy by GWs. There is no exchange of matter between the two stars \cite{han_binary_2020, 
			postnov_evolution_2014}.
		\item[(2)] Interacting DWD binaries. When the binaries are close enough, two 
		stars exchange matter \cite{han_binary_2020, postnov_evolution_2014}. GWs and masses 
		transfer affect the orbit together \cite{marsh_mass_2004}. We currently have 
		22 known interacting binaries, AM CVn stars, with periods ranging from 5.4 to 65 minutes
		\cite{Sberna_2021, noauthor_lisa_nodate1}.
	\end{itemize}
	DWD binaries gravitational waves in the source frame are defined as:
	\begin{equation}
		h_{+}(t)=\mathcal{A}\left(1+\cos ^{2} \imath\right) \cos \Phi(t), 
	\end{equation}
	\begin{equation}
		h_{\times}(t)=-2 \mathcal{A} \cos \imath \sin \Phi(t), 
	\end{equation}
	\begin{equation}
		\Phi(t)=\phi_{0}+2 \pi f t+\pi \dot{f} t^{2},
	\end{equation}
	where $\mathcal{A}$ represents the GW amplitude, $\imath$ represents 
	the DWD binary orbit to the line of sight from the Solar System
	barycentric (SSB) origin, $\phi_{0}$ represents the initial
	phase, $f$ is the GW frequency, and $\dot{f}$ is the first order derivative of $f$. 
	
	For detached DWD binaries, combining $f$ with $\dot{f}$, we 
	can obtain their chirp masses $\mathcal{M}$: 
	\begin{equation}
		\label{format1}
		\dot{f}=\frac{96}{5} \pi^{8 / 3} \mathcal{M}^{5 / 3} f^{11 / 3}.
	\end{equation}
	With GW amplitudes $\mathcal{A}$, the distances $D$ between binary systems and 
	solar can be calculated:   
	\begin{equation}
		\label{format2}
		\mathcal{A}=\frac{2 \mathcal{M}^{5 / 3}\left(\pi f\right)^{2 / 3}}{D}.
	\end{equation}
	
	\subsection{LISA response}
	LISA rotates around the sun with a cycle of one year, 20 degrees behind the Earth 
	\cite{martens_trajectory_2021}. It has 
	three satellites to form an equilateral triangle. Each side of the triangle is 2.5 million 
	kilometres \cite{noauthor_lisa_nodate}. There are laser beams between each of the two 
	satellites to measure the change in the distance between them. LISA will generate three 
	separate pieces of data on three arms. Because the phase noise of the laser seriously pollutes 
	the detection data, it must combine these data from different arms to reduce the 
	impact of laser phase noise \cite{armstrong_timedelay_1999}. LDC1 uses  
	the first-generation Time-delay interferometry (TDI) Michelson combinations denoted 
	by X, Y, and Z. The first-generation TDI assumes the distances between spacecrafts are 
	constant, just like a rigid body \cite{tinto_time-delay_2014}. 
	
	We use data streams whose characteristic is that instrumental noises are 
	uncorrelated \cite{blaut_mock_2010} by recombining the three channels XYZ into AET 
	\cite{armstrong_timedelay_1999, tinto_time_2004, tinto_time-delay_2014}:
	\begin{equation}
		\begin{gathered}
			A=\frac{Z-X}{\sqrt{2}}, \\
			E=\frac{X-2 Y+Z}{\sqrt{6}}, \\
			T=\frac{X+Y+Z}{\sqrt{3}}.
		\end{gathered}
	\end{equation}
	
\begin{figure}
		\centering
		\includegraphics[width=1\linewidth]{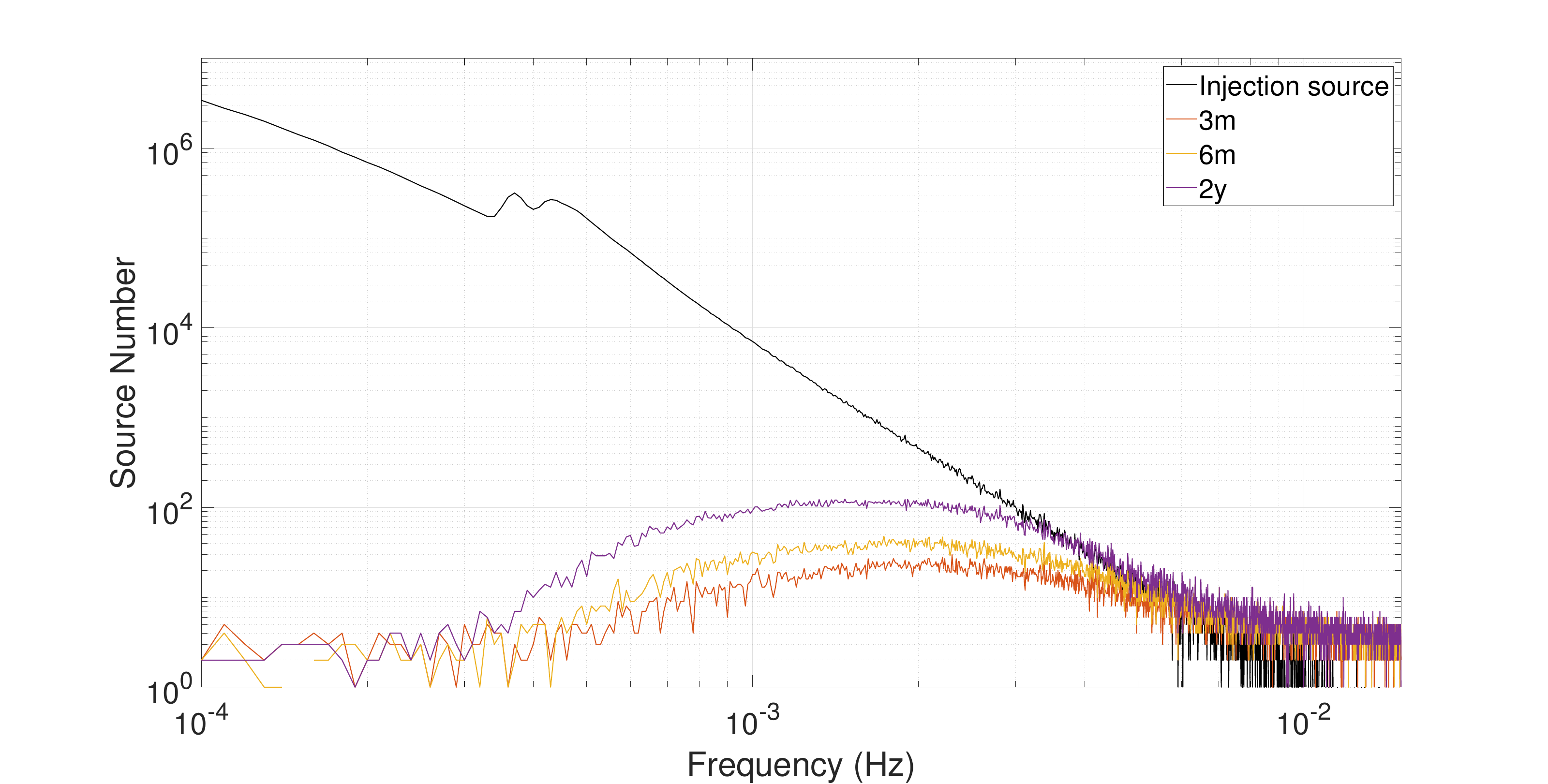}
		\caption{The black line presents the number of injection sources in LDC1-4 per frequency 
			bin from $1\times10^{-4}$ Hz to $1.5\times10^{-2}$ Hz. The other three coloured lines 
			represent the reported sources from direct-search to different 
			data lengths. The number of accelerate-search to the two-year data reported sources 
			(not shown in this figure) is the same as direct-search to the first three-month data. 
			The width of each frequency bin is 0.01mHz.}
		\label{fig:psignal-number}
	\end{figure}

	There are 29857650 DWD binaries in LDC1-4 data, adding stationary gaussian simulated 
	instrumental noise. Each series in LDC1-4 data is uniformly sampled every 15 seconds. The data 
	is given by: 
	\begin{equation}
		\bar{y}^{I}=\bar{S}^{I}+\bar{n}^{I},
	\end{equation}
	where $\bar{S}^{I}$ denotes the collective GW signal from all the
	DWD binaries with $I \in \{A,E,T\}$, $\bar{n}^{I}$ denotes the instrumental noise realization, 
	and $\bar{x}$ denotes a row vector.
	For LDC1-4 data, sampling number N $= 4194304$ corresponds to the observational time $T_{obs} 
	=2$ years. Each 
	binary system is described by eight parameters, which are Frequency $f$, 
	FrequencyDerivative $\dot{f}$, EclipticLatitude $\beta$, EclipticLongitude $\lambda$, Amplitude 
	$\mathcal{A}$, Inclination $\iota$, Polarization $\psi$ and InitialPhase $\phi_{0}$. 
	Almost all DWD binaries in LDC1-4 data have frequencies in the range of (0.1,15) mHz. The 
	number of these sources overgrows as the frequency decreases, as shown in the solid black line 
	in Fig.~\ref{fig:psignal-number}. The other lines present the number of reported sources in 
	different frequency bins from three search situations.
	
\begin{figure}
		\centering
		\includegraphics[width=1\linewidth]{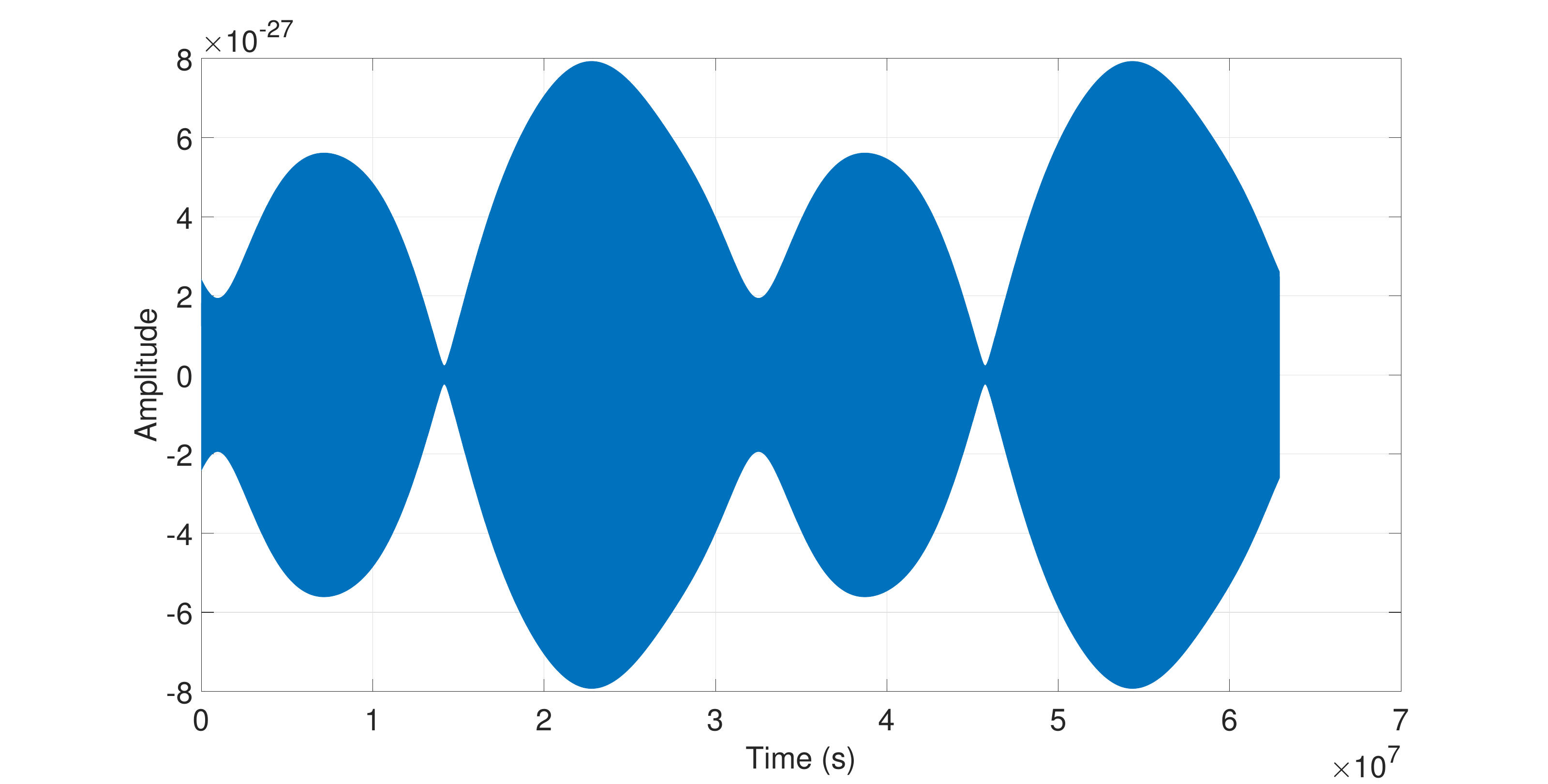}
		\caption{One DWD binary gravitational wave is modulated by LISA with a 
			one-year periodicity. The data length is two years.}
		\label{fig:psignal}
	\end{figure}

	The LISA orbital motion produces the Doppler effect. The earth moves around the sun at about 30 
	kilometres per second, and LISA has a similar speed. By the doppler effect formula, 
	$f_1=f\times\frac{c\pm v}{c}$ (c is the speed of light), the GW frequency will move 
	$\approx 0.02\%$ of itself back and forth. It disperses the energy of the 
	gravitational wave signal at a single frequency and overlaps signals with different 
	frequencies, making the problem of resolving them more challenging. As LISA orientation 
	changes, its sensitivity to different sky areas will change. For the same signal, its signal 
	strength induced by LISA periodically changes with LISA sensitize in different directions, as 
	shown in Fig.~\ref{fig:psignal}. In other words,  
	a GW signal has different SNRs detected at different periods of one year. 

	\section{Search method}
	\label{sec:snglsrcstat}
	\subsection{Search pipeline: GBSIEVER}
	A single DWD binary gravitational wave signal can be expressed as:
	\begin{equation}
		\bar{s}^{I}(\theta)=\bar{a} \mathbf{X}^{I}(\kappa),
	\end{equation}
	where $\bar{a}=\left(a_{1}, a_{2}, a_{3}, a_{4}\right) \in {R}^{4}$ and 
	$\mathbf{X}^{I}(\kappa)$ is a matrix of template waveforms. $\bar{a}$ is the
	reparametrization of four extrinsic parameters $\mathcal{A}$, $\phi_{0}$, $\psi$ 
	and $\iota$. $\mathbf{X}^{I}(\kappa)$ depends on the 
	four remaining intrinsic parameters: $f$, $\dot{f}$, $\beta$ and $\lambda$. GBSIEVER 
	uses the $\mathcal{F}$-statistic proposed by Jaranowski in 1998 \cite{jaranowski_data_1998} to 
	search for monochromatic gravitational waves. 
	It maximizes the log-likelihood ratio $\log \Lambda$ (combing AET  channels  data) concerning 
	parameters $a_{i}$, by solving: 
	\begin{equation}
		\frac{\partial \log \Lambda}{\partial a_{i}}=0.
	\end{equation}
	Then it substitutes the maximum likelihood estimators $\hat{a}_i$ in  
	$\log \Lambda$ yielding the reduced log-likelihood ratio denoted by $\mathcal{F}$-statistic.  
	In the intrinsic parameters search space, one finds the $\mathcal{F}$-statistic maximum
	and then obtains four extrinsic parameters through analytic expression (see 
	\cite{blaut_mock_2010} for more detail).
	
	In GBSIEVER, the template waveforms are generated by using the expressions in 
	Ref.~\cite{blaut_mock_2010}, while the DWD binaries GW signals in LDC1-4 are generated by the 
	FastGB code 
	\cite{petiteau_lisacode_2008}. We have verified that they are in good 
	agreement for the A and E combinations in the frequency range we are interested in. It divides 
	GWs in the frequency range from $9\times 10^{-5}$ Hz to $1.501\times 10^{-2}$ Hz into 1491 
	frequency bins to search separately. 
	The width of each bin is $2\times 10^{-5}$ Hz. There is a 50\% overlap between two adjacent 
	frequency bins. In each bin, considering spectrum leakage due to the Doppler effect, the
	Tukey window is used to weaken the effect of the edge of each frequency bin, and we take the 
	results in the range with central $1\times 10^{-5}$ Hz (acceptance zone), as shown in the black 
	section of Fig.~\ref{fig:frequency_range}. We search each bin until it finds all GW signals 
	with SNR less than 7 for five consecutive times. We refer to the GW parameter sets used to 
	generate the LDC1-4 data as injection sources, and the acceptance zone search results come from 
	GBSIEVER as reported sources.
	
	Particle swarm optimization (PSO) \cite{kennedy_particle_nodate} performs well in 
	finding the optimal value of high-dimensional, nonlinear, and nonconvex optimization problems.
	It has been proven effective in previous experiments to obtain four intrinsic parameters 
	of DWD binaries \cite{zhang_resolving_2021-1}. In GBSIEVER, 40 particles are randomly scattered 
	in the 4-dimensional search space ($f$, $\dot{f}$, $\beta$ 
	and $\lambda$) and then we calculate the $\mathcal{F}$-statistic for each particle. 
	The particles explore the search space randomly for better $\mathcal{F}$-statistic values 
	following iterative rules.
	If a particle changes upon a good $\mathcal{F}$-statistic value, other particles eventually 
	converge to its location and refine the $\mathcal{F}$-statistic value further. 
	According to experience, it takes 2,000 iterations per search to ensure these particles 
	converge. Every time a signal is found, it is removed from the 
	original data, and we search for the next signal in the residual. 
	
	Cross-validation expands the search ranges for $\dot{f}$ in the secondary run when $f\le4\times 
	10^{-3}$ Hz. Comparing the search results in the primary run can eliminate some false 
	sources in the low-frequency range and improve the detection rate. 
	In this paper, our attention focused on confirmed sources screened by the correlation 
	coefficient R (Section \ref{sec:Detection Criteria}). Therefore, we did not use 
	cross-validation in this paper.
	
	Most of the search time is spent computing the 
	$\mathcal{F}$-statistic. In particular, the longer the data length is, the more time it 
	takes to calculate the $\mathcal{F}$-statistic for each particle in PSO. We refer to the method 
	of searching directly with GBSIEVER as direct-search. 
	We estimate that for direct-search to the two-year data of LDC1-4,
	already calculated in Ref.~\cite{zhang_resolving_2021-1}, it will take about 150 
	days to rerun it on the server we are currently using.
	More details on GBSIEVER can be found in Ref.~\cite{zhang_resolving_2021-1}.
	
	\subsection{Accelerate the individual detection process}
	The direct-search to the two-year data is time-consuming.
	Moreover, it is hard to imagine processing the data after LISA has completed all detection 
	missions.
	We can start a search when LISA has been working for a while.
	It can be processed faster and is enough to reveal 
	preliminary information about DWD binaries. (Results are shown in the section \ref{sec:result}.)
	
\begin{figure}
		\centering
		\includegraphics[width=1\linewidth]{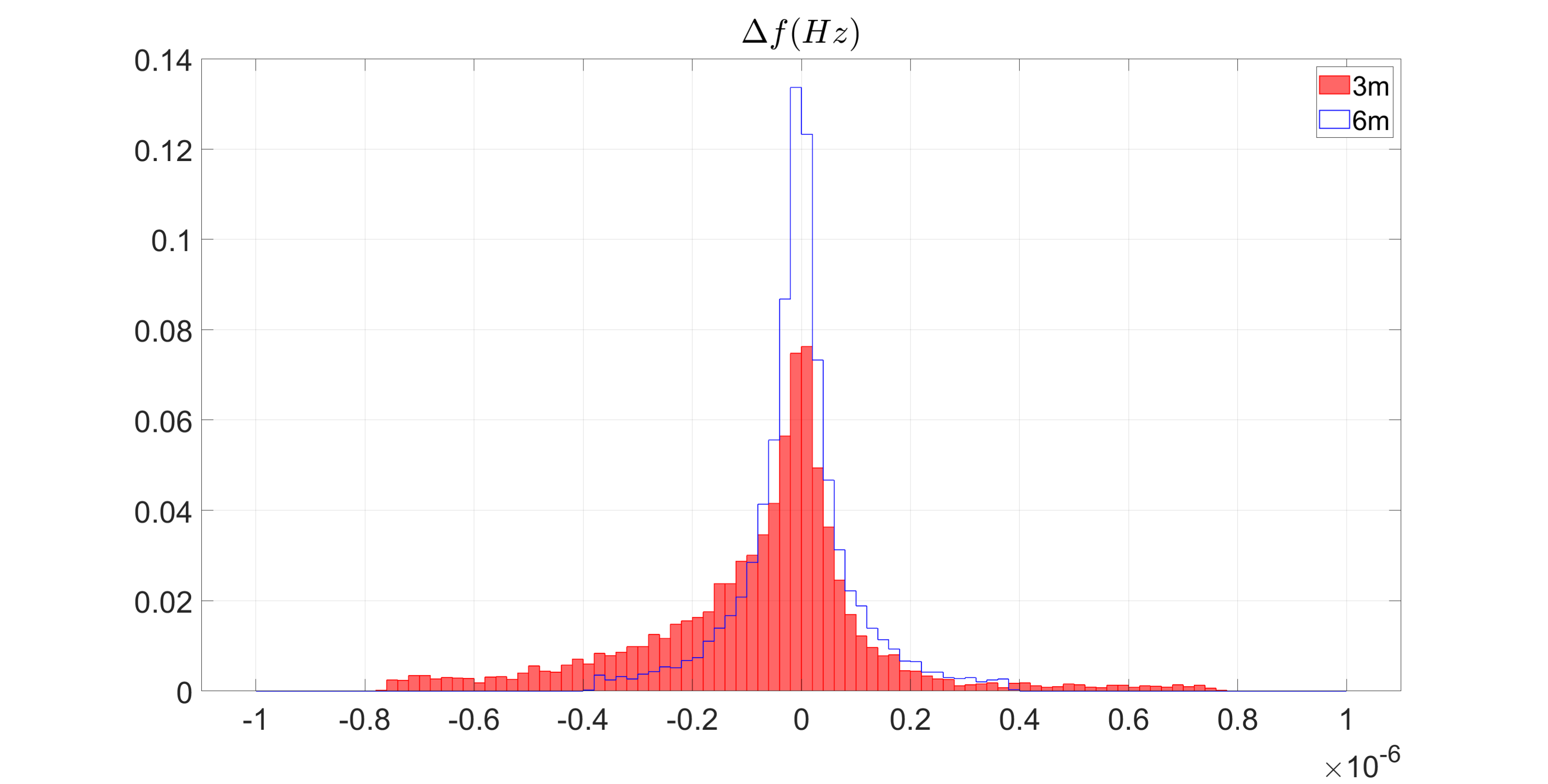}
		\caption{The red blocks and the blue line correspond to the absolute frequency error of 
			direct-search to the first three-month data (3m) and the first six-month data (6m), 
			respectively. Y-coordinate is the number of histograms to all reported 
			sources ratio. Most errors are within the range $|\Delta f|\le1\times 10^{-6}$ 
			Hz. Based on this, we narrow the frequency search range in the following search and 
			reduce the number of iterations.}
		\label{fig:p3maerror}
	\end{figure}

\begin{figure}
		\centering
		\includegraphics[width=1\linewidth]{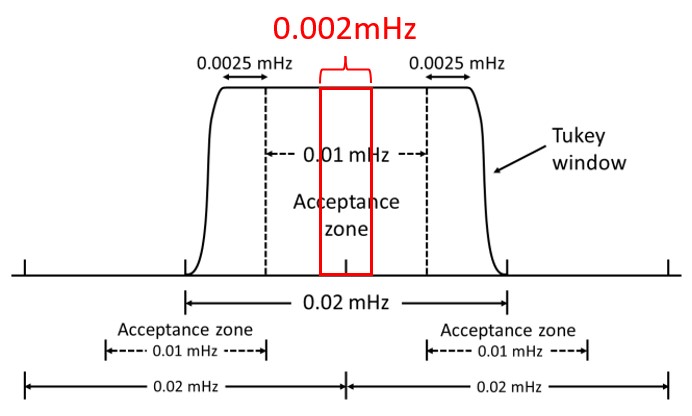}
		\caption{Direct-search using a Tukey window bandlimits 
			the data in the Fourier domain to 0.02 mHz. Only the search results that belong to the 
			central 0.01 mHz interval, called the acceptance zone, are retained as reported 
			sources. The adjacent bin acceptance zone is contiguous. Accelerate-search uses the $f$ 
			results searched from the first three-month data, narrowing the $f$ search range of 
			the two-year data for each source from 0.02 mHz to 0.002 mHz.}
		\label{fig:frequency_range}
	\end{figure}

	We first direct-search to a short period of LDC1-4 data (such as the first three-month data),
	and then use $f$ results from this, limit the search
	space of $f$ in the following search (such as the two-year data).
	In GBSIEVER, the width of the frequency range of each bin is 0.02 mHz. 
	For the first three-month data and the first six-month data, most report sources 
	have absolute frequency errors $|\Delta f|\le1\times 10^{-6}$ Hz (See 
	Fig.~\ref{fig:p3maerror}). 
	So we center on the $f$ value from the direct-search to the first three-month data, extending 
	it on 
	both sides with 0.001 mHz, to obtain the new range of frequency in searching the two-year data 
	whose width is 0.002 mHz (the red line in Fig.~\ref{fig:frequency_range}), which is one-tenth 
	of the previous search space. As the search space decreases, it 
	needs lower iterations for search points to converge. We reduced the number of 
	iterations from 2000 to 400. Therefore the time spent per search was also reduced by five 
	times. We call this method accelerate-search. It limits the number of reported 
	sources in each frequency bin and narrows the frequency search range of each source based on 
	reported sources from the direct-search to the first acquired data. In this paper, 
	accelerate-search is based on the $f$ results from direct-search to the first three-month data.
	
	\subsection{Detection Criteria}
	\label{sec:Detection Criteria}
	We have two criteria for judging the reliability of these reported sources. Firstly, we 
	calculate their optimal SNR:
	\begin{equation}
		\label{equation_snr}
		SNR = 
		\sqrt{T\sum_{I}(\bar{s}^{I}(\theta^{\prime})*\bar{s}^{I}(\theta^{\prime}))/S_n^I(f)},
	\end{equation}
	where $\theta^{\prime}$ represents reported source
	parameter sets. $I \in \{A,E,T\}$. $T = 15 s$ represents the sampling interval. Because each 
	DWD binary signal locates a tiny 
	frequency band, the noise PSD $S_n(f)$ in this band is treated as a constant approximately. 
	
\begin{figure}
		\centering
		\includegraphics[width=1\linewidth]{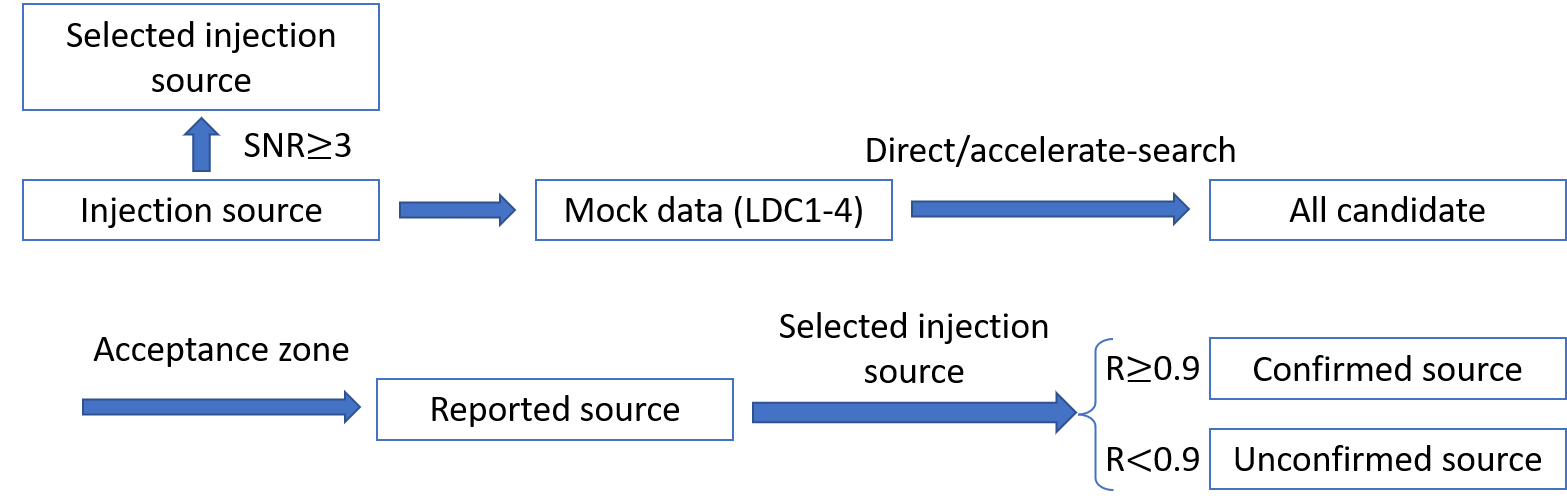}
		\caption{The search diagram in this paper. Injection sources are used to generate mock LISA 
			data LDC1-4, among which SNR $\ge3$ in two years of observation is the selected 
			injection source. The candidates in the acceptance zone are reported sources. The 
			correlation coefficient R of each reported source calculate with the selected 
			injection sources. Moreover, the reported source with R $\ge0.9$ is a confirmed source. 
			Otherwise, it is considered an unconfirmed source. Detection rate D is the 
			ratio of the confirmed sources number $N_c$ to the reported sources number $N_r$. 
			The results  of those sources in different search strategies are shown in Tab. 
			\ref{table:1}.}
		\label{fig:flow_chart}
	\end{figure}

	In addition to the SNR, we also calculate the correlation coefficient R between the reported 
	source parameter sets $\theta^{\prime}$ and the injection source parameter sets $\theta$ to 
	assess the similarity between their GW signal waveforms:
	\begin{equation}
		\label{1}
		\begin{aligned}
			R\left(\theta, \theta^{\prime}\right) &=\frac{C\left(\theta, 
				\theta^{\prime}\right)}{\left[C(\theta, \theta) C\left(\theta^{\prime}, 
				\theta^{\prime}\right)\right]^{1 / 2}},\\
			C\left(\theta, \theta^{\prime}\right) &= 
			{\sum_{I}(\bar{s}^{I}(\theta)*\bar{s}^{I}(\theta^{\prime}))/S_n^I(f)}.
		\end{aligned}
	\end{equation}
	We pick up parameter sets with SNR $\ge3$ from all injection sources in two years of 
	observation, as $\theta$ in Eq.~\ref{1}. They named selected injection sources, and the number 
	of them is 66696. 
	Other injection sources with lower SNR are not considered to exist in 
	our reported sources. For each reported source, the corresponding R takes the 
	maximum value calculated with selected injection source parameter sets which have $\Delta f = 
	\lvert f(\theta) - f(\theta^{\prime})\rvert < 6/T_{obs}$. $T_{obs}$ is the duration of 
	detection. Generally, a reported source with R $\ge0.9$ is seen as a confirmed source. 
	And the reported source with R $<0.9$ is seen as an unconfirmed source or a false source.
	The detection rate D of DWD is defined as the number of confirmed sources $N_c$
	relative to the reported sources $N_r$ in one search, 
	$D = N_c/N_r$ as did in previous literature \cite{zhang_resolving_2021-1} 
	\footnote{Note that, since there are always DWD GW singals in the data, this detection 
		definition is not the same as the transient signal search.}.
	Fig.~\ref{fig:flow_chart} shows the search diagram relating to the souces defined in this paper.

	\section{Search results}
	\label{sec:result}
	\subsection{Source candidates}
\begin{figure}
		\centering
		\includegraphics[width=1\linewidth]{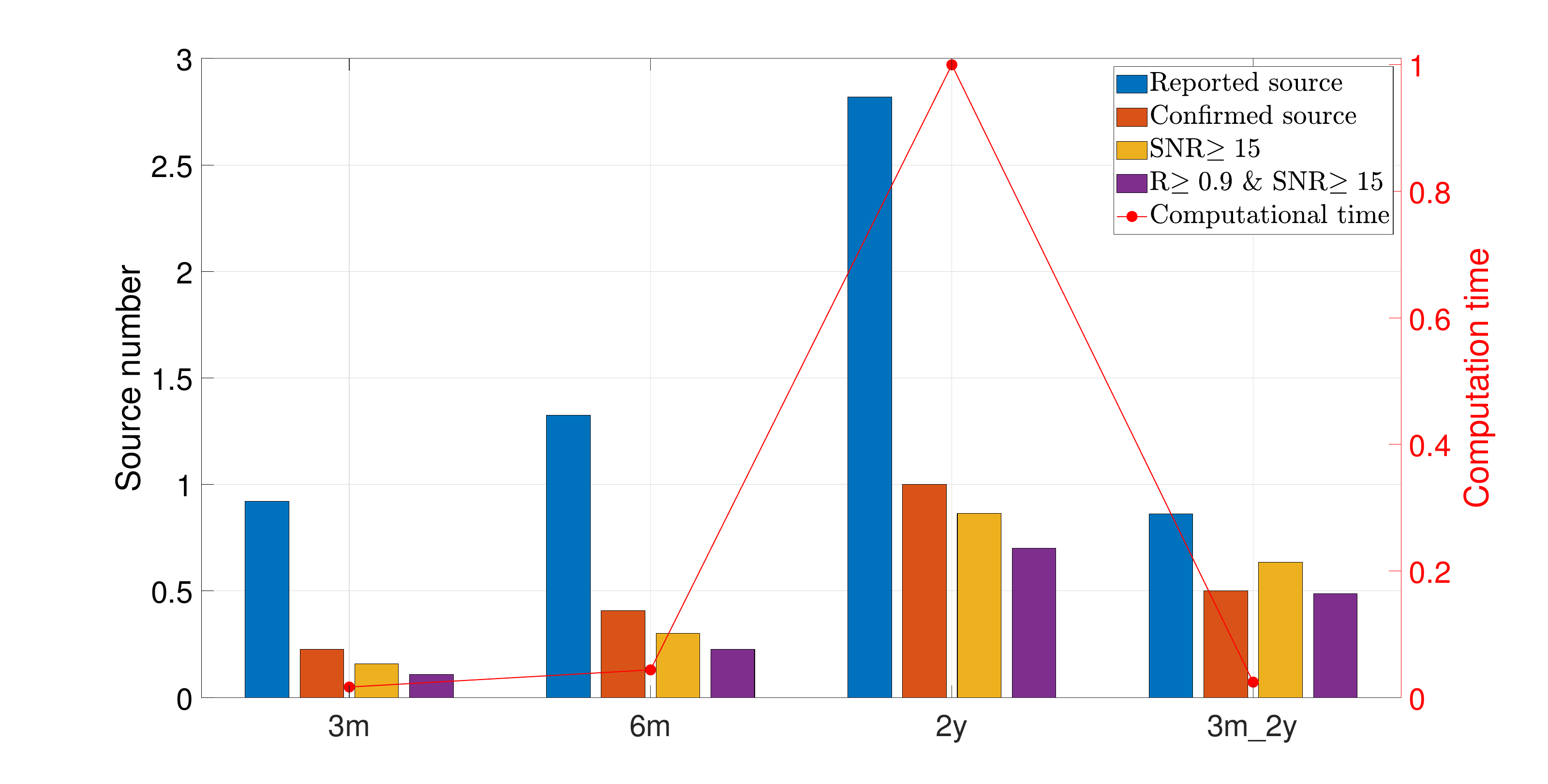}
		\caption{The figure visualizes the data in Tab.~\ref{table:1}. 
			The left y-coordinate unit is the number of confirmed sources from the direct-search to 
			the two-year data (2y). The right y-coordinate takes the computation time of 
			direct-search to the two-year data as a unit.}
		\label{fig:p_time_number}
	\end{figure}

\begin{figure}
		\centering
		\includegraphics[width=1\linewidth]{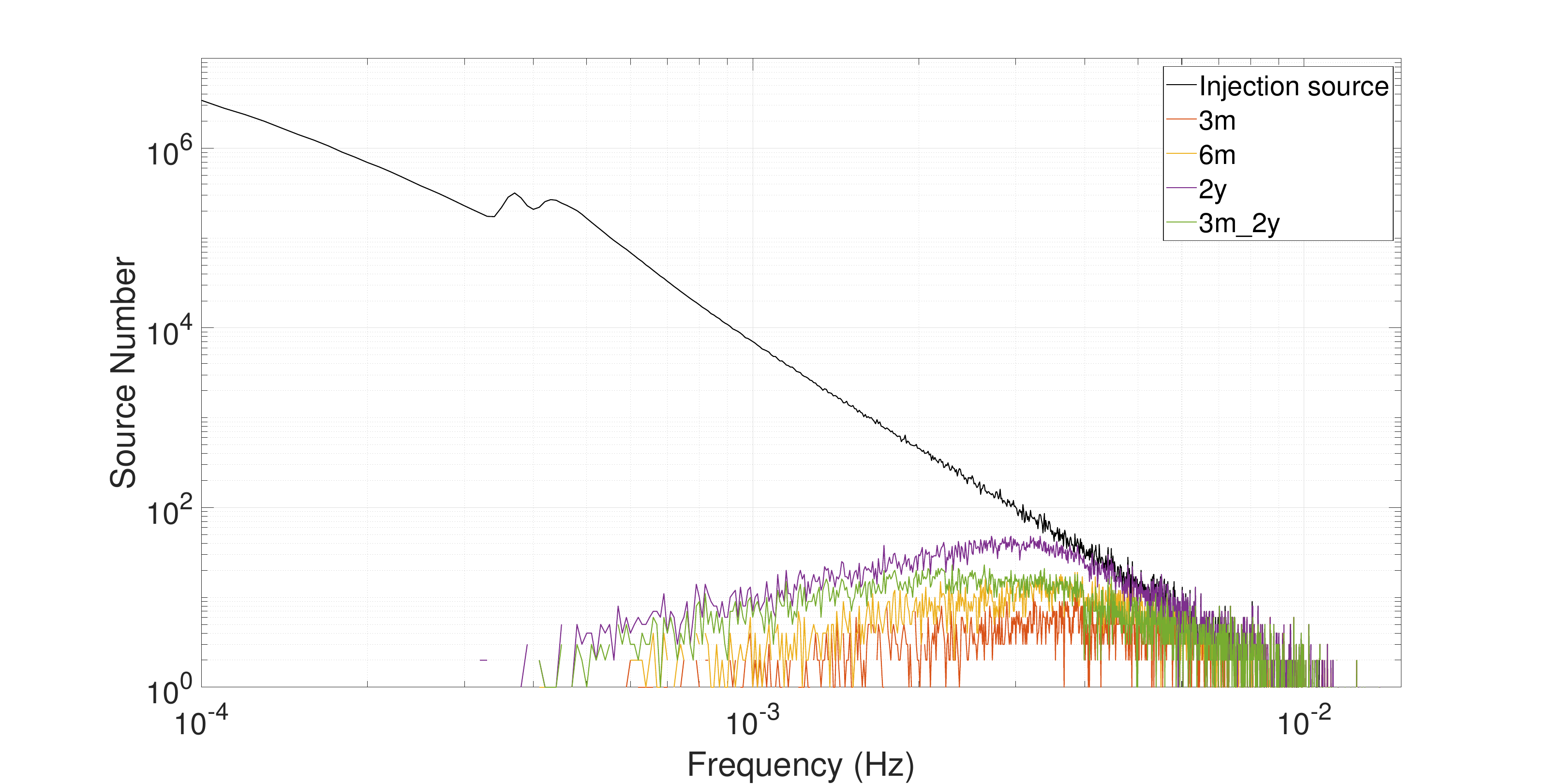}
		\caption{The black line has the same meaning as Fig.~\ref{fig:psignal-number}. The other 
			four colored lines represent all confirmed sources obtained by four search situations. 
			The width of each frequency bin is 0.01mHz.}
		\label{fig:p_signal_number3}
	\end{figure}

	We directly search the first three-month data (3m), the first six-month 
	data (6m) and the two-year data (2y), and apply the above proposed accelerating method to
	the two-year data (3m\_2y) based on the results of direct-search to the first three-month data. 
	Table \ref{table:1} shows the computation time and the number of sources for each search 
	situation.
	Computation time means the time it takes to conduct a complete search, and search efficiency 
	is the normalized number of confirmed sources detected in a unit of time.
	The search efficiency of accelerate-search to the two-year data is twenty times as much as that 
	of direct-search to the two-year data. It is because accelerate-search has a lower number 
	of iterations per search and uses less time for searching in low-frequency locations, 
	where reported sources are often unconfirmed. Most reported sources with $f<4\times10^{-3}$ Hz 
	admit R $<0.9$, which reduces the detection rate (compare Fig.~\ref{fig:psignal} and 
	Fig.~\ref{fig:p_signal_number3}). 
	Direct-search to the two-year data has the most sources and consumes the longest computation 
	time. It finds 37399 reported sources, with 13269 confirmed sources among them.
	This means the detection rate $D_{2y} = \frac{13269}{37399} = 35\%$.
	Accelerate-search to the two-year data has 11440 reported sources and 6651 confirmed 
	sources, with a detection rate $D_{3m\_2y} = \frac{6651}{11440} = 58\%$. The
	detection rates for the direct-search to the first three-month and the first 
	six-month data are 27\% and 31\%, respectively. 
	If we take the reported sources from direct-search to the two-year data as the 
	standard, the normalized detection rate $D^n = N_c/N^{2y}_r$. The $D^n_{3m}$, $D^n_{6m}$, 
	$D^n_{2y}$, $D^n_{3m\_2y}$ are 8\%, 15\%, 35\% and 18\%, respectively.
	We take the computation time consumed in direct-search to the two-year data as a unit in 
	Fig.~\ref{fig:p_time_number}. Therefore, the computation times (red line) for direct-search 
	to the first-three month data and the first six-month data, accelerate-search to the two-year 
	data are 0.016, 0.044, and 0.024, respectively.
	In the end, accelerate-search to the two-year data takes about one-fortieth 
	time of direct-search to the two-year data to obtain about half the number of confirmed 
	sources.
	
\begin{figure}
		\centering
		\includegraphics[width=1\linewidth]{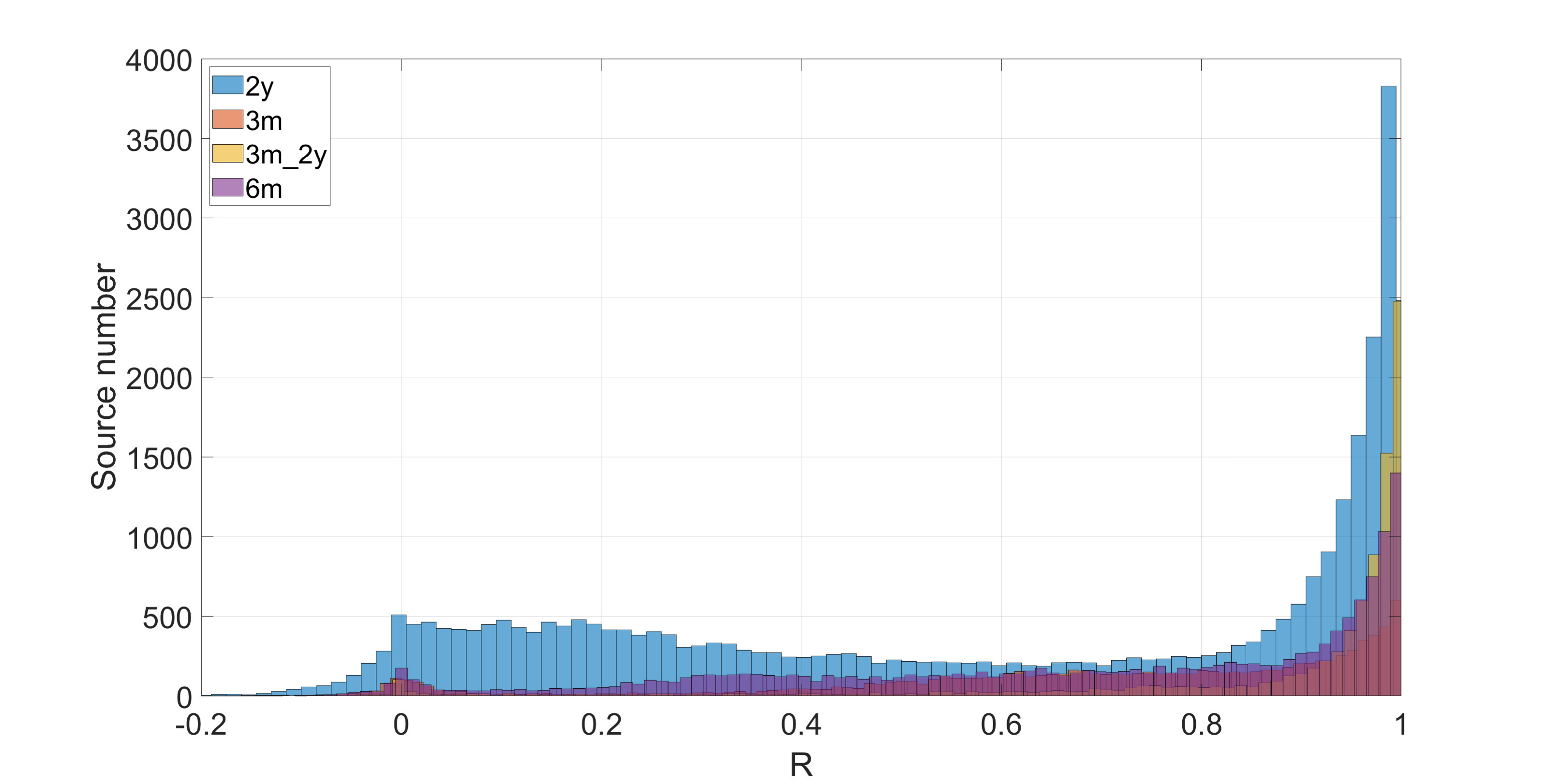}
		\caption{The histogram distribution of R of reported sources, which is obtained in four 
			search situations.}
		\label{fig:p_hist_R}
	\end{figure}

\begin{figure*}
		\centering  
		\subfigbottomskip=2pt 
		\subfigcapskip=-5pt 
		\subfigure[\label{fig:p_hist_SNR1}]{
			\includegraphics[width=0.48\linewidth]{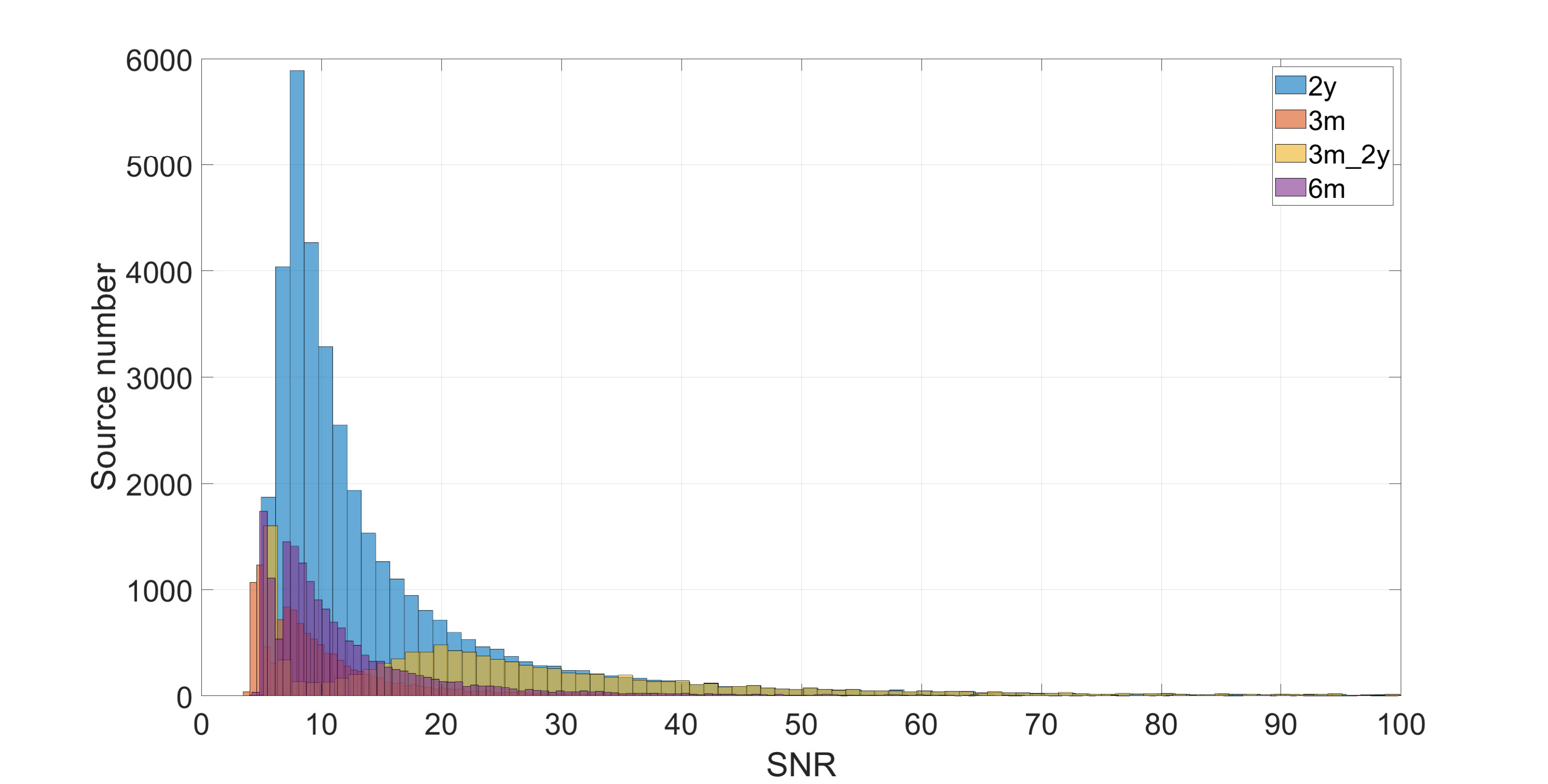}}
		\subfigure[\label{fig:p_hist_SNR2}]{
			\includegraphics[width=0.48\linewidth]{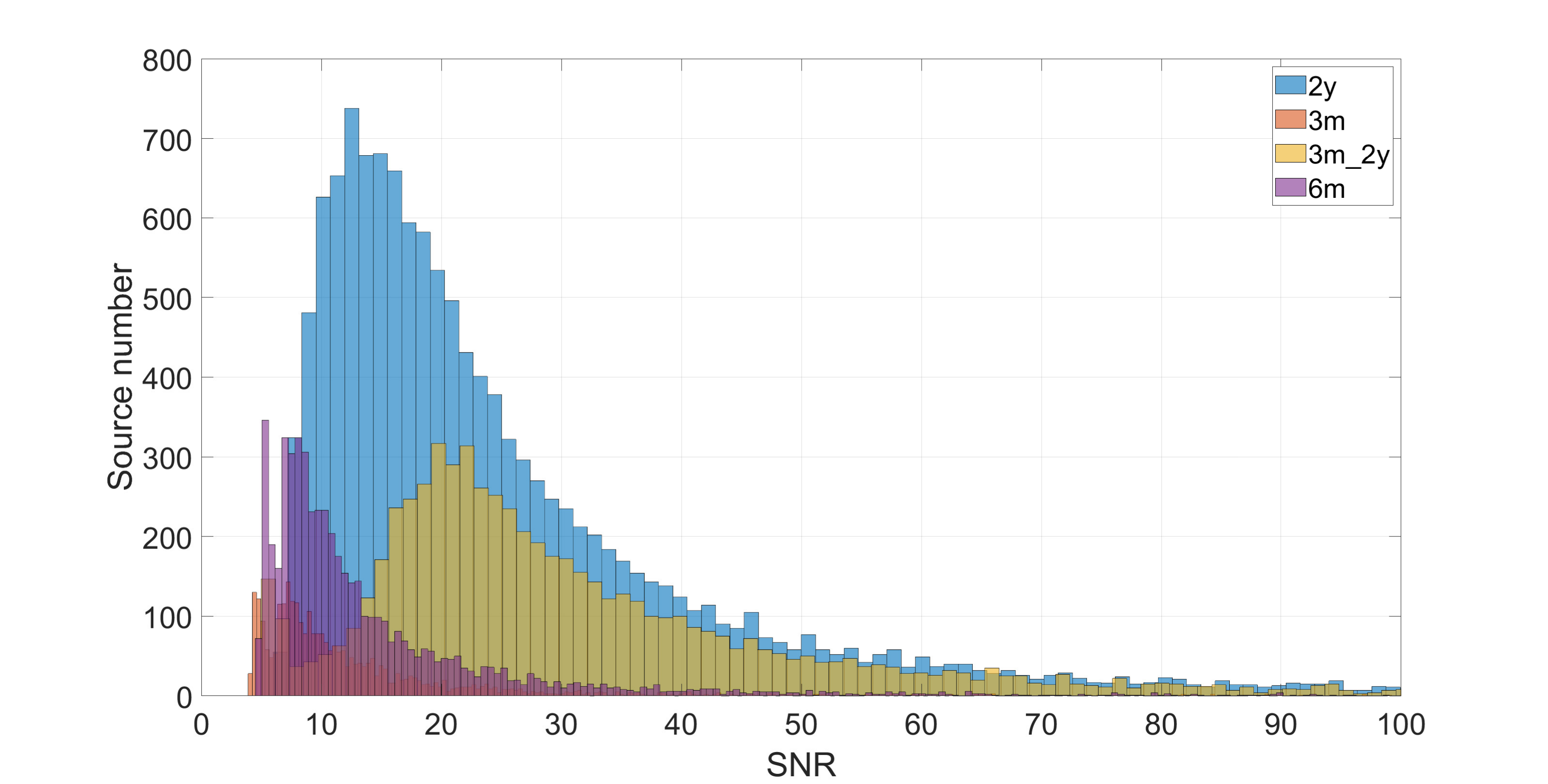}}
		\caption{The histogram distribution of SNR of reported sources 
			(Fig.~\ref{fig:p_hist_SNR1}), and confirmed sources (Fig.~\ref{fig:p_hist_SNR2}).}
	\end{figure*}

	Fig.~\ref{fig:p_hist_R} shows the correlation coefficient R histogram distribution of the 
	reported sources. A higher R value means the reported source is more similar to the injection 
	source. 
	The SNR histogram distribution of reported sources and confirmed sources are 
	shown in Fig.~\ref{fig:p_hist_SNR1} and Fig.~\ref{fig:p_hist_SNR2}. The confirmed sources 
	in accelerate-search are the sub-population in direct-search with higher SNRs.
	A considerable proportion of reported sources from direct-search to the two-year data (2y) 
	with SNR $\sim 10$ does not meet our screening criteria (R $\ge 0.9$), and it does not 
	belong to the list of confirmed sources.
	For space-based gravitational wave detection, source with SNR $\ge15$ are usually selected as a
	reliable reported sources.
	Direct-search to the two-year data finds 11453 reported sources with SNR 
	$\ge15$, and the accelerate-search finds 8411 ones, which is 73\% of the direct-search 
	result. For those sources with SNR 
	$\ge20$, two search situations have a similar SNR histogram distribution. It means 
	accelerate-search has a similar ability in searching for GW with a slightly bigger SNR. 
	In actual GW detection, we cannot calculate R because the injection source parameter sets are 
	unknown, so the main criterion is SNR.
	We checked out the reported sources that satisfy R $\ge0.9$ and SNR $\ge15$. The higher the 
	proportion of these sources in SNR $\ge15$, the better SNR $\ge15$ as the criterion for LISA to 
	detect DWD binaries GWs. For the 
	direct-search to the first three-month data, the first six-month data, the two-year data, and 
	the accelerate-search to the two-year data, the proportion is 68\%, 75\%, 81\%, and 77\%, 
	respectively.
	Fig.~\ref{fig:p_signal_number3} shows the number of confirmed sources in 
	different frequency bins from four search situations. When the frequency $f$ is lower than 
	$4\times10^{-3}$ Hz, the search capability decreases. Moreover, it cannot even search for some 
	confirmed sources at a frequency below $1\times10^{-3}$ Hz.
	
	\begin{table*}
		\begin{center}   
			\begin{tabular}{|c|c|c|c|c|c|c|c|c|}
				\hline
				& \makecell{Computation\\ time (s)} & All candidate & \makecell{Reported source\\ 
					($N_r$)} & \makecell{Confirmed source\\ ($N_c$)} & \makecell{Detection\\ rate 
					(D)} 
				& \makecell{Search\\ efficiency} & $SNR \ge 15$ & 
				\makecell{$R \ge 0.9$ \&\\ $SNR \ge 15$}\\
				\hline
				3m & 222,806 & 20737 & 12,213 & 3010 & 27\% & 13.79 & 2085 & 1424
				\\
				\hline
				6m & 590,209 & 30220 & 17,571 & 5418 & 31\% & 9.37 & 3993 & 3004
				\\	
				\hline
				2y & 13,525,967 & 100783 & 37,399 & 13269 & 35\% & 1.00 & 11453 & 9300 \\
				\hline	
				3m\_2y & 327,818 & 12213 & 11,440 & 6651 & 58\% & 20.70 & 8411 & 6477 \\	
				\hline	
			\end{tabular}
			\caption{This table shows the source number of direct-search to the first three-month 
				data, the first six-month data, the two-year data, and accelerate-search to 
				the two-year data, as well as their computation time. All source contains search 
				results that are not in the acceptance zone. We take the search efficiency of 
				direct-search to the two-year data as the unit.} 
			\label{table:1} 
		\end{center}   
	\end{table*}	
	
	
	\subsection{Residual data}
\begin{figure*}
		\centering  
		\subfigbottomskip=2pt 
		\subfigcapskip=-5pt 
		\subfigure[\label{fig:p_residual_3months1}3m]{
			\includegraphics[width=0.48\linewidth]{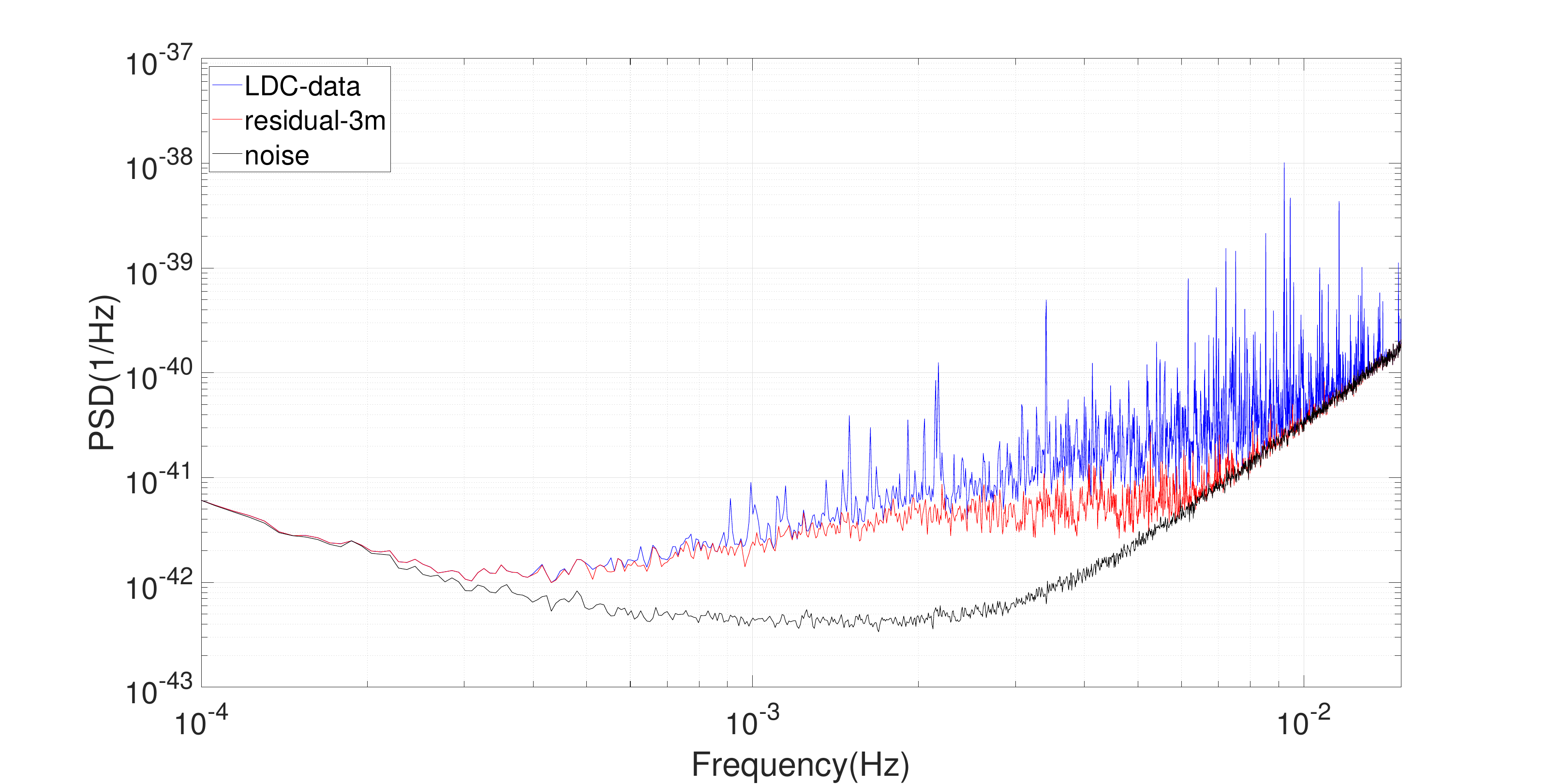}}
		\subfigure[\label{fig:p_residual_6months1}6m]{
			\includegraphics[width=0.48\linewidth]{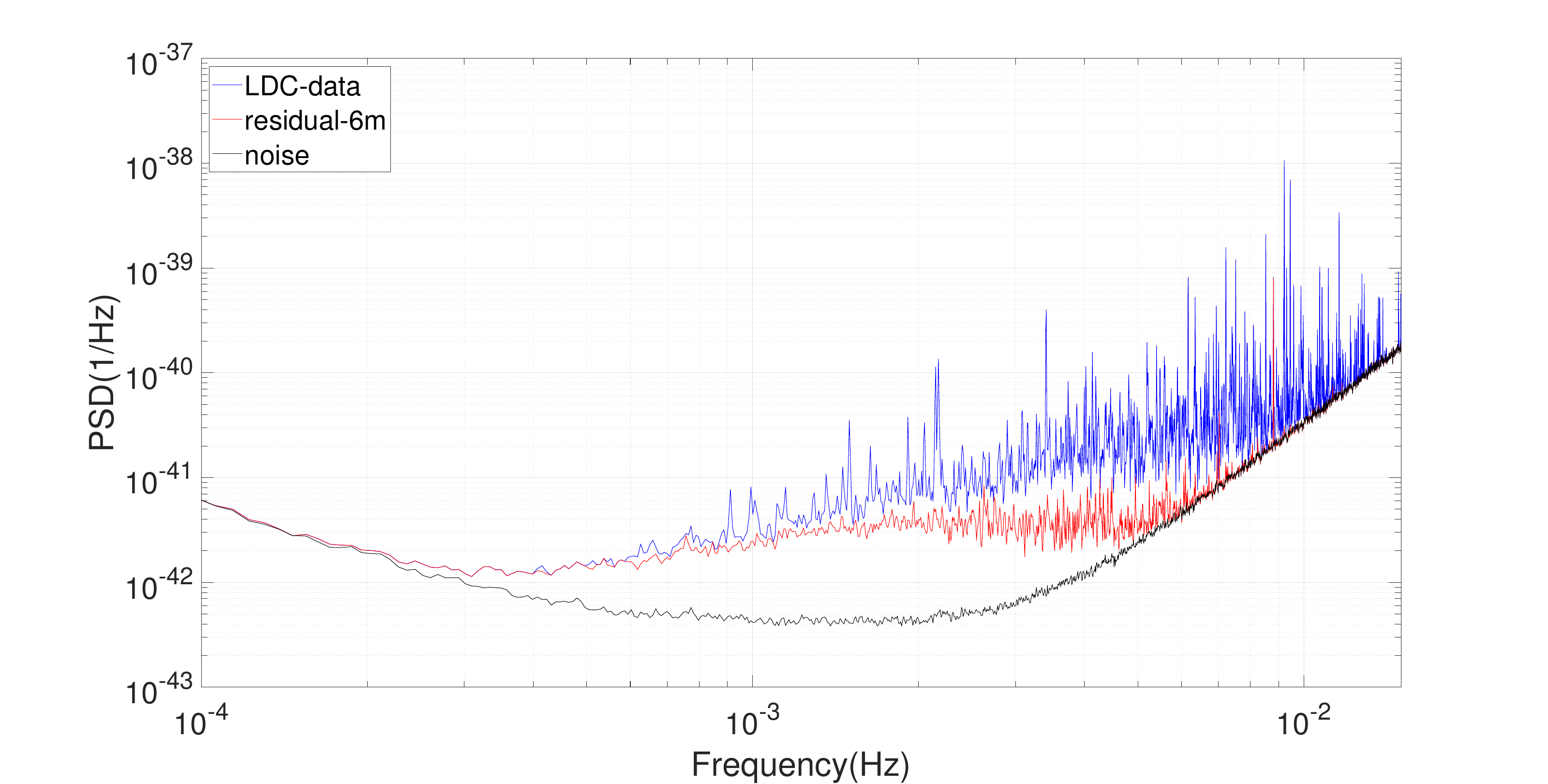}} \\
		\subfigure[\label{fig:p_residual_2years1}2y]{
			\includegraphics[width=0.48\linewidth]{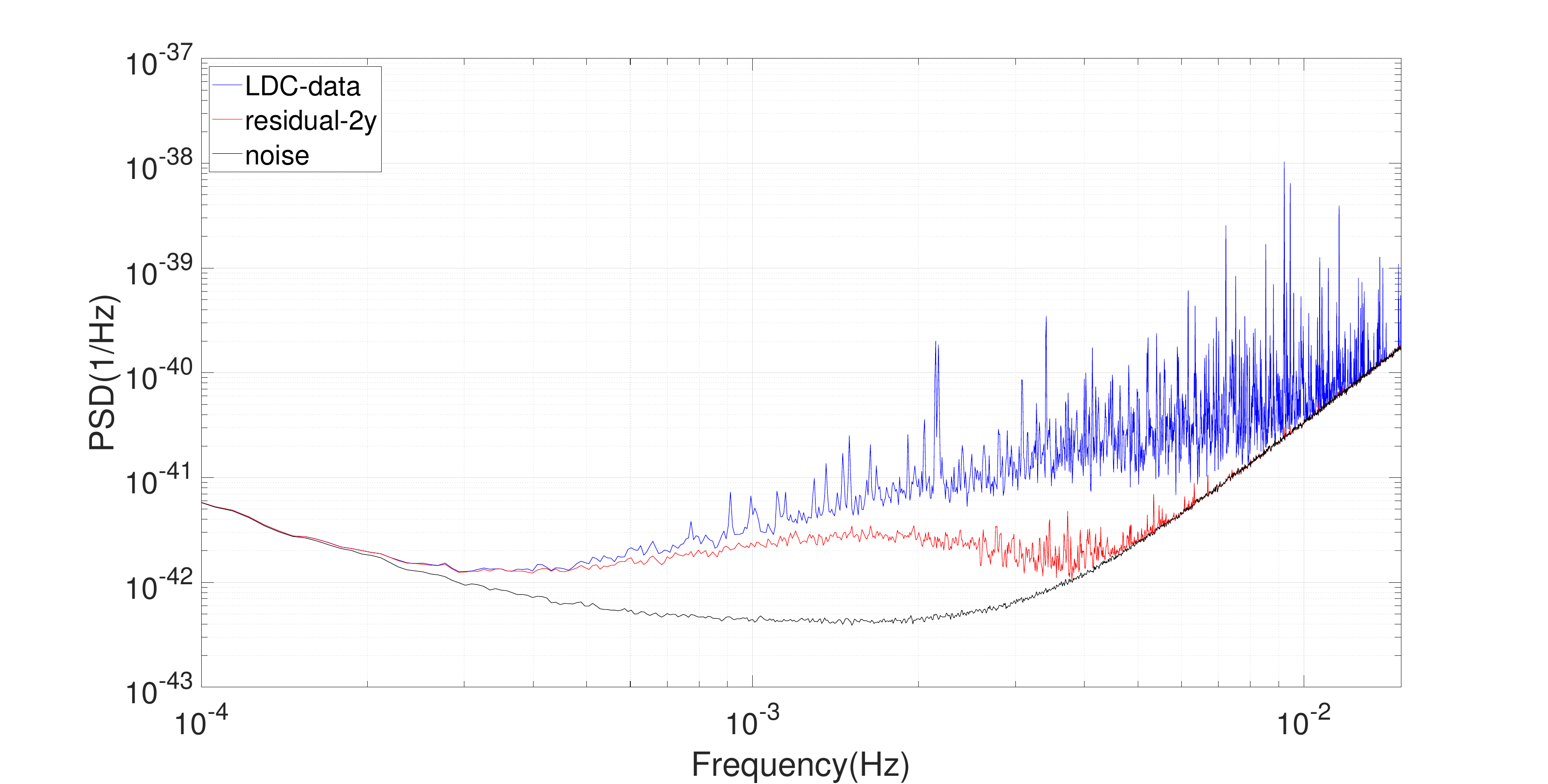}}
		\subfigure[\label{fig:p_residual_3months_2years1}3m\_2y]{
			\includegraphics[width=0.48\linewidth]{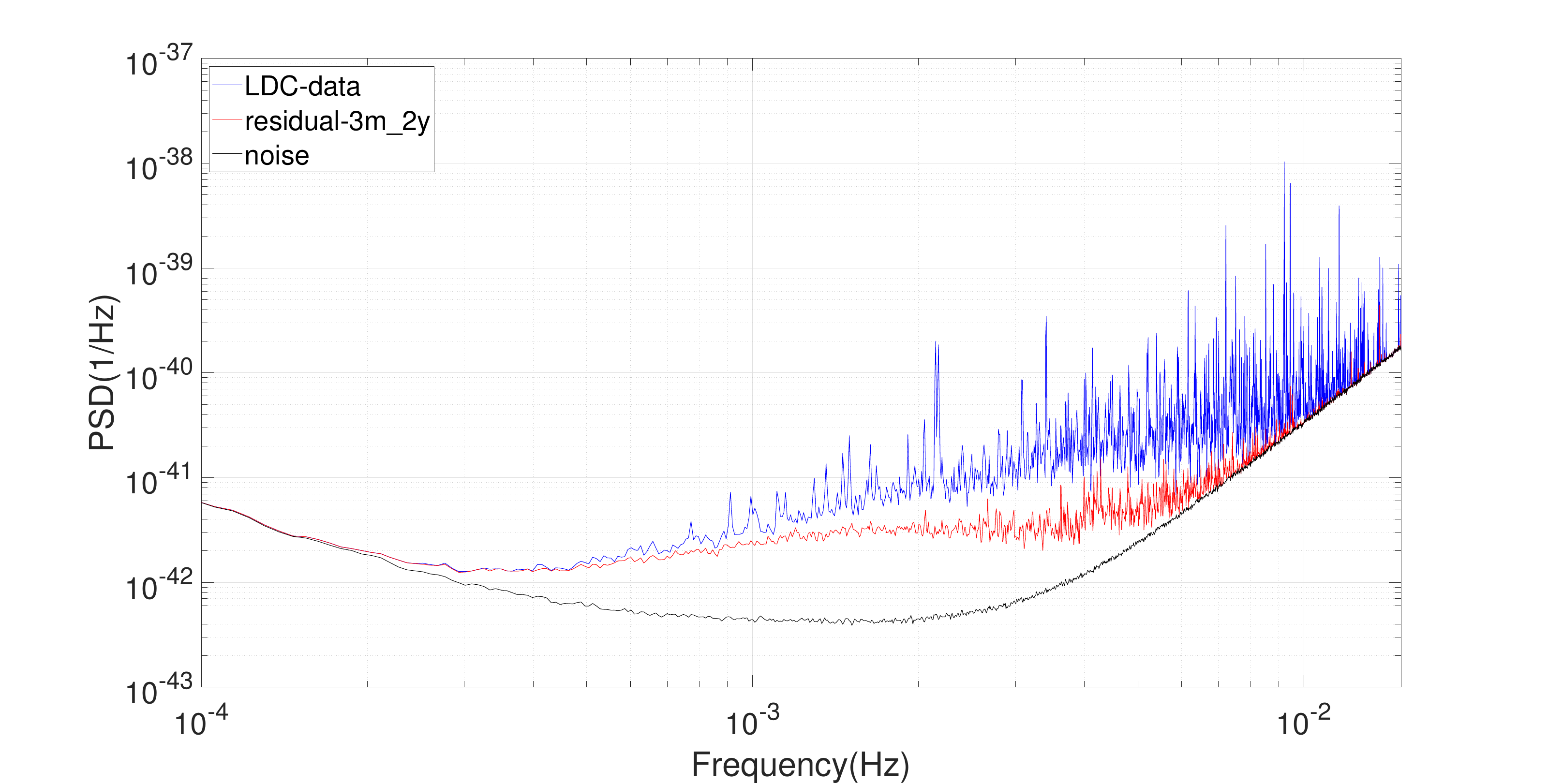}}
		\caption{\label{fig:p_residual1}These figures show the PSD comparison of the LDC1-4 data, 
			the residuals from four search situations, and the LISA instrumental noise. The 
			noise and data length is the same as the residual. All data is calculated on the TDI A.}
	\end{figure*}

	We calculate the LISA response signals of all confirmed sources in four search situations and 
	subtract these signals from the LDC1-4 TDI A data to obtain the residual data. 
	Fig.~\ref{fig:p_residual1} compares the LDC1-4 data power spectral densities 
	(PSDs), the residuals PSDs and the instrumental noise PSDs in different search situations. The 
	time length of LDC data and noise is consistent with that of the residual.
	In space-based gravitational wave detection, DWD binary signals in the 
	Milky Way constitute a strong foreground noise.
	With the increase in data length, more confirmed sources were found, and the residual of DWD 
	binary signals is more similar to LISA instrumental noise. However, confirmed 
	sources are mainly concentrated in the high-frequency domain ($f\ge4\times10^{-3}$ Hz) and the
	medium-frequency domain ($1\times10^{-3}$ Hz $\le f<4\times10^{-3}$ Hz). In the low-frequency 
	domain ($f<1\times10^{-3}$ Hz), there is little difference in the residuals obtained in the 
	four search situations.
	Comparing the direct-search and the accelerate-search to the two-year 
	data, the main difference between their residuals is   
	because three-month data is too short, so the search cut-off according to SNR $\ge7$ 
	miss some signals, which limits the search number in accelerate-search to the two-year data. 
	Suppose we use longer data search results, such as the direct-search to the first six-month 
	data, which has 17571 
	reported sources. We can theoretically have more confirmed sources for 
	accelerate-search to the two-year data, and its residual PSD will be more similar to the 
	residual PSD of direct-search to the two-year data.
	
	\subsubsection{MBHB's SNR}
\begin{figure}
		\centering
		\includegraphics[width=1\linewidth]{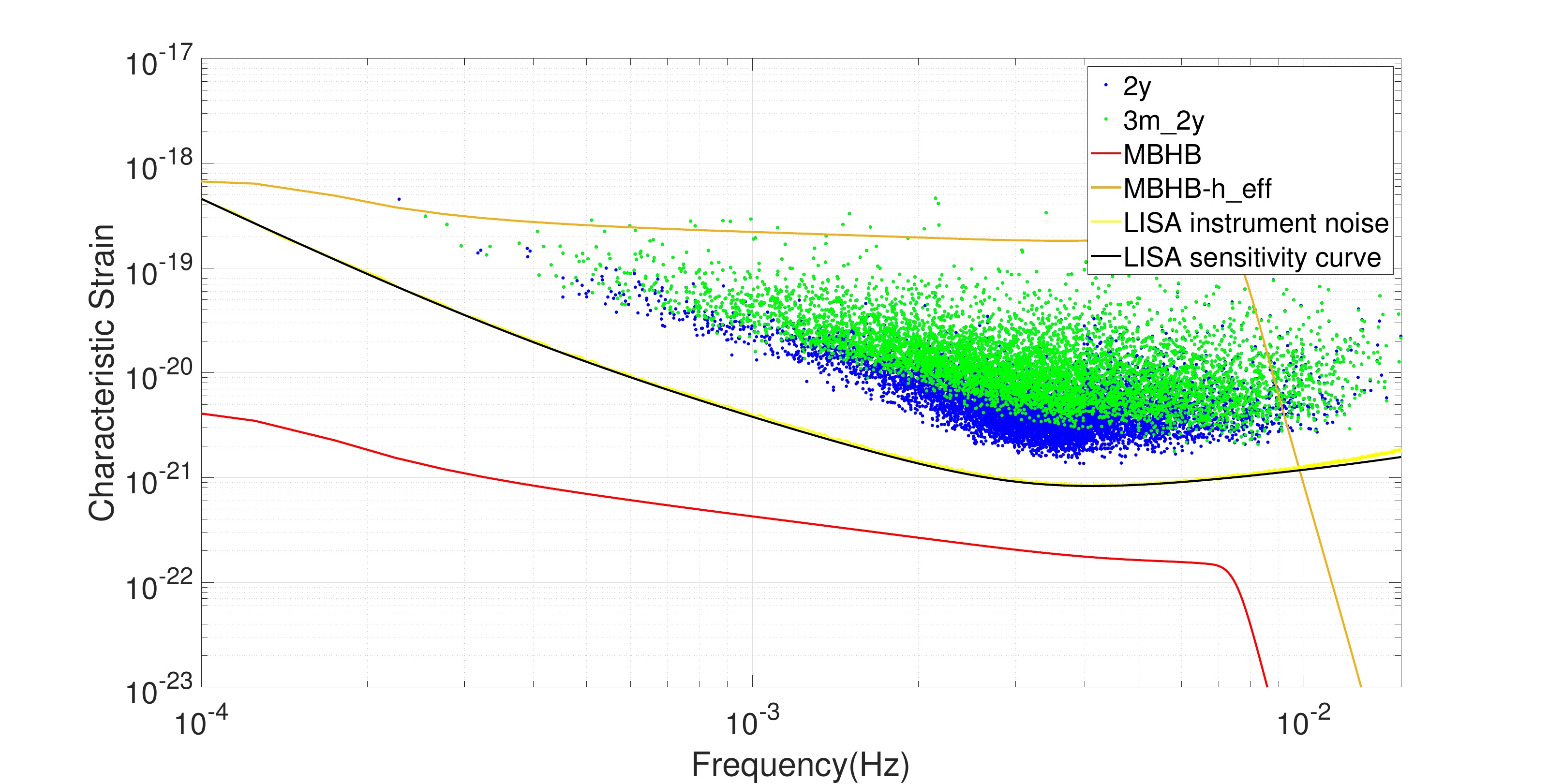}
		\caption{The blue and green points correspond to confirmed sources from 
			direct-search and accelerate-search to the two-year data, respectively. The LISA 
			sensitivity curve is consistent with mock LISA instrumental noise. And the MBHB signal 
			comes from LDC1-1 data, whose detector frame masses are $2.8\times10^{5} M_{\odot}$ and 
			$2.8\times10^{6} M_{\odot}$.}
		\label{fig:p_sensitivity}
	\end{figure}

	We use LDC1-1 data, the LISA response signal of a single gravitational wave generated by a 
	massive black hole binary (MBHB), to test the influence of different noise assumptions on 
	MBHB's SNR. The two 
	black hole detector frame masses are $2.8\times10^{5} M_{\odot}$ and $2.8\times10^{6} 
	M_{\odot}$, 
	and their redshift $z=6.1$. LDC1-1 sample frequency is 1/10, and we interpolate its 
	signal at a sampling frequency of 1/15, which is consistent with LDC1-4. The signal length of 
	LDC1-1 is 1.33 years. We calculate the SNR of MBHB by:
	\begin{equation}
		SNR = \left|\sqrt{\sum{\frac{1}{Nf_{s}}(\tilde{h}_{MBHB}/\bar{S}_{n})
				\tilde{h}_{MBHB}^{\dag}}}\right|,
	\end{equation}
	where $N$ is the sampling number and $f_{s}$ is the sampling frequency. $\tilde{h}_{MBHB}$ 
	represents the discrete Fourier transform (DFT) of time-domain signal ${h}_{MBHB}$ in LDC1-1.  
	${\dag}$ is a transposed conjugate symbol. And $\bar{S}_{n}$ 
	represents the PSD with different noise assumptions.
	All calculations are based on the TDI A. Tab.~\ref{table:2} presents the SNR 
	for this MBHB signal on different noise assumptions, which are 
	LISA instrumental noise, LDC1-4 data, and residuals from four search situations 
	(Fig.~\ref{fig:p_residual1}). 
	It can be seen that the foreground noise from DWD binaries does not significantly affect the 
	SNR of the MBHB signal.
	Fig.~\ref{fig:p_sensitivity} shows the MBHB signal, confirmed sources with two search 
	situations, and the LISA sensitivity curve from Ref.~\cite{robson_construction_2019}, which is 
	consistent with LISA instrumental noise. The vertical coordinate is dimensionless 
	\cite{moore_gravitational-wave_2015}. Rather than plotting the signal directly, it often plots 
	$h_{eff}^2=16f(2fT)S_h(f)/5$ to account for the influence of coherent signal extraction 
	\cite{robson_construction_2019}. Accelerate-search to the two-year data has fewer confirmed 
	sources than direct-search to the two-year data. The missing parts are the blue dots whose 
	characteristic strains are closer to the LISA sensitivity curve, which means they have a 
	lower SNR.
	\begin{table*}
		\begin{center}   
			\begin{tabular}{|c|c|c|c|c|}
				\hline
				& Instrument noise & LDC1-4 & 2y-residual & 
				3m\_2y-residual 	
				\\
				\hline
				MBHB & 135.7084 & 99.7673 & 121.3335 & 112.0379  \\
				\hline
			\end{tabular}
			\caption{For a certain MBHB signal, SNR is obtained under different noise 
				assumptions. The different noise assumptions do not have a tremendous influence on 
				the MBHB's SNR.} 
			\label{table:2} 
		\end{center}   
	\end{table*}
	
	\subsubsection{Contamination effects on the subtraction of DWDs}
	Inaccurate subtraction of individual DWD can lead to contamination on the residual data 
	that becomes more pronounced as detection continues. We explore this effect by a case study, 
	using a frequency bin ranging from $4.18\times{10}^{-3}$ Hz to $4.20\times{10}^{-3}$ Hz 
	(the 410th bin in our data). To evaluate the impact of inaccurate detections, we consider 
	reported sources as ``inaccurate" sources, while injection sources were deemed ``accurate". 
	The A channel data is used to show this effect: 
	\begin{equation}
		\begin{aligned}
			\bar{r}^{A}_{inj-t} &= \bar{y}^{A}_{t}-\bar{s}^{A}_{inj-t}, \\
			\bar{r}^{A}_{rep(K)-t} &= \bar{y}^{A}_{t}-\bar{s}^{A}_{rep(K)-t}, \\
		\end{aligned}
	\end{equation} 
	where $\bar{r}^{A}$ is the residuals in the A channel, $\bar{s}^{A}$ is the GWs signals 
	come from injection or reported sources. We 
	consider the first three-month ($t=3m$) and the two-year ($t=2y$) data, and compare the 
	reported sources from direct-search to the first three-months data ($K=3m$) and the 
	two-year data ($K=2y$), and accelerate-search to the two-year data ($K=3m\_2y$).
	
\begin{figure*}
		\centering  
		\subfigbottomskip=2pt 
		\subfigcapskip=-5pt 
		\subfigure[\label{fig:1}]{
			\includegraphics[width=0.48\linewidth]{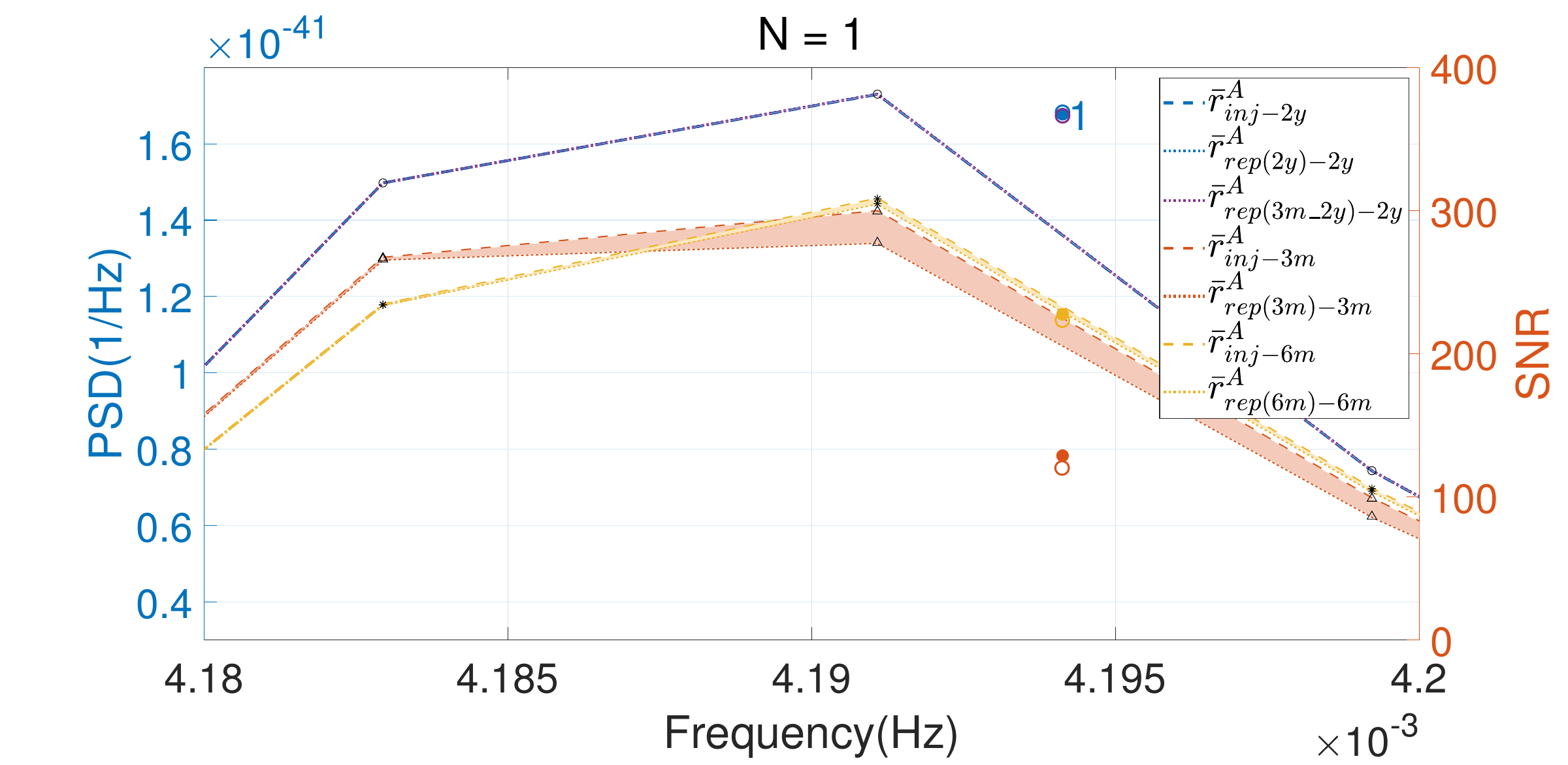}}
		\subfigure[\label{fig:3}]{
			\includegraphics[width=0.48\linewidth]{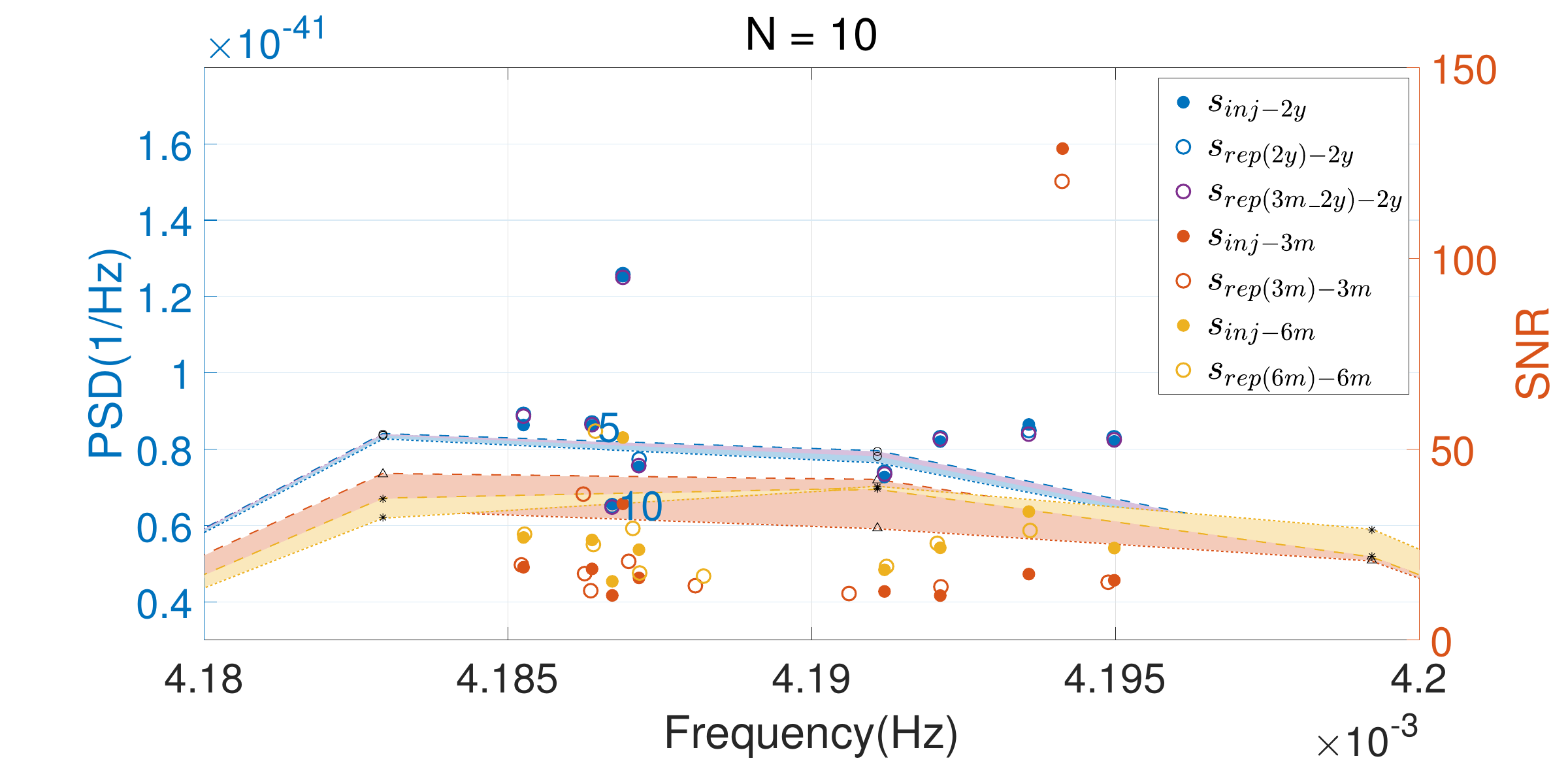}}
		\subfigure[\label{fig:5}]{
			\includegraphics[width=0.48\linewidth]{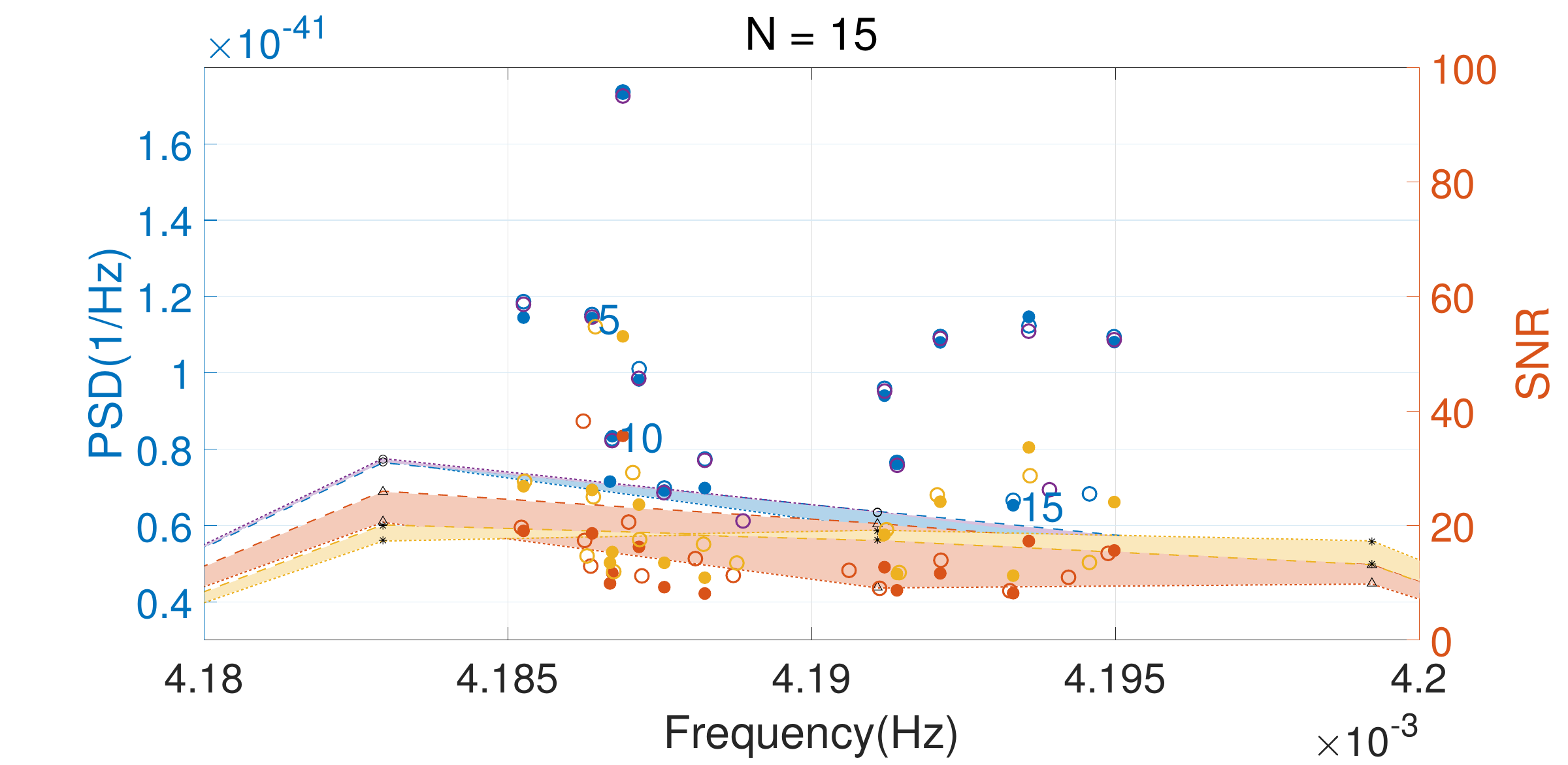}}
		\subfigure[\label{fig:6}]{
			\includegraphics[width=0.48\linewidth]{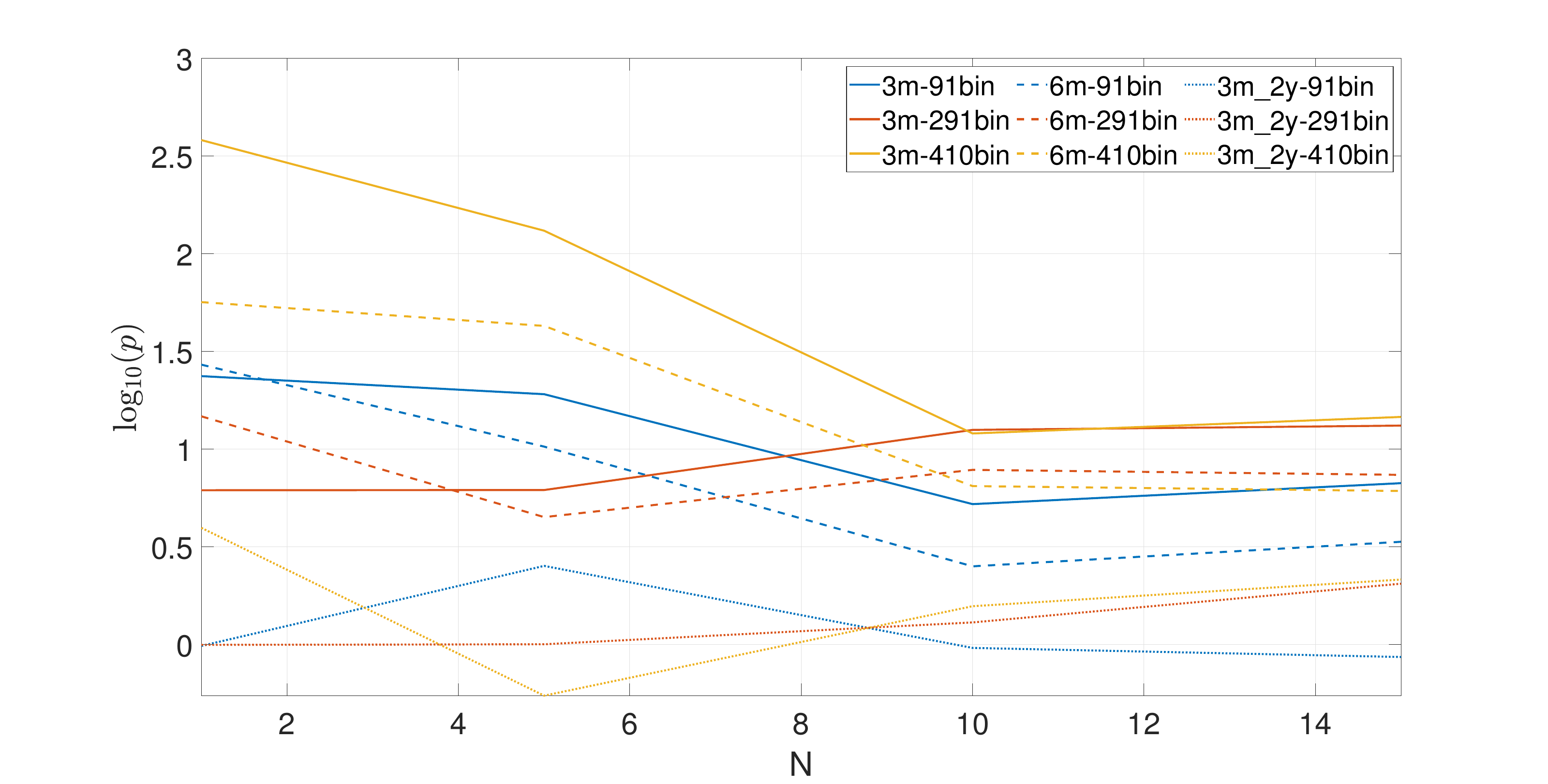}}
		\caption{Figures~\ref{fig:1}, \ref{fig:3}, and 
			\ref{fig:5} display the PSDs of residuals $\bar{r}^{A}_{inj-t}$ and 
			$\bar{r}^{A}_{rep(K)-t}$, and their corresponding subtracted sources whose SNR 
			calculated by Eq.~\ref{equation_snr} for the   
			410th bin in our data. The title indicates the number of subtracted 
			sources (N), and the right y-axis corresponds to their SNRs. In each 
			figure, the larger area ($\int |S_{\bar{r}^{A}_{inj-t}}(f) - 
			S_{\bar{r}^{A}_{rep(K)-t}}(f)| df$) between the lines of the same $t$ is, the 
			greater impact of the contamination caused by inaccurate subtraction. 
			Figure~\ref{fig:6} shows the trend of areas (Eq.~\ref{eqp}) which 
			take the $\int |S_{\bar{r}^{A}_{inj-2y}}(f) - S_{\bar{r}^{A}_{rep(2y)-2y}}(f)| df$ 
			as unit one with the number of subtracted sources increases, and includes case 
			studies in the 91st frequency bin ($9.90\times{10}^{-4}$ Hz -- 
			$1.01\times{10}^{-3}$ Hz) and the 291st frequency bin ($2.99\times{10}^{-3}$ Hz -- 
			$3.01\times{10}^{-3}$ Hz). We have marked the order of subtraction (1, 5, 10, 15) of 
			the sources $s_{inj-2y}$.}
		\label{fig:contimation}
	\end{figure*}

	By comparing the power spectral densities (PSDs) of the residuals 
	$\bar{r}^{A}_{inj-t}$ and $\bar{r}^{A}_{rep(K)-t}$, we can assess the extent of the 
	inaccurate subtracted contamination. 
	We plot the PSDs of the residuals $\bar{r}^{A}_{inj-2y}$, $\bar{r}^{A}_{rep(2y)-2y}$, 
	$\bar{r}^{A}_{rep(3m\_2y)-2y}$, $\bar{r}^{A}_{inj-3m}$, $\bar{r}^{A}_{rep(3m)-3m}$, 
	$\bar{r}^{A}_{inj-6m}$, $\bar{r}^{A}_{rep(6m)-6m}$ of the 410th frequency bin in 
	Fig.~\ref{fig:1} -- Fig.~\ref{fig:5}, alongside the 
	corresponding subtracted sources whose SNRs calculated by Eq.~\ref{equation_snr}.
	
	A larger area ($\int |S_{\bar{r}^{A}_{inj-t}}(f) - 
	S_{\bar{r}^{A}_{rep(K)-t}}(f)| df$) between two PSDs indicates a greater 
	contamination effect from reported (inaccurate) sources. These areas have been colored in 
	Fig.~\ref{fig:1} -- Fig.~\ref{fig:5}. For the case study in 
	the 410th frequency bin, we subtract 1, 10, and 15 sources, respectively, resulting in 
	areas of $1.72\times10^{-49}$, $9.49\times10^{-48}$, $1.42\times10^{-47}$ for $K=t=2y$,
	and $2.24\times10^{-49}$, $4.90\times10^{-48}$, $7.03\times10^{-48}$ for $K=3m\_2y$, 
	$t=2y$, 
	and $2.17\times10^{-47}$, $3.76\times10^{-47}$, $4.78\times10^{-47}$ for $K=t=3m$,
	and $3.21\times10^{-48}$, $2.02\times10^{-47}$, $2.00\times10^{-47}$ for $K=t=6m$.
	As the number (N) of subtracted sources increases, the areas grow larger, indicating a more 
	significant contamination effect from reported (inaccurate) sources.
	
	We conduct the same process in two additional frequency bins, the 91st 
	frequency bin ($9.90\times{10}^{-4}$ Hz$ - 1.01\times{10}^{-3}$ Hz) and the 291st frequency 
	bin ($2.99\times{10}^{-3}$ Hz$ - 3.01\times{10}^{-3}$ Hz). Then we plot the trend of areas 
	(Fig.~\ref{fig:6}), which takes the $\int 
	|S_{\bar{r}^{A}_{inj-2y}}(f) - S_{\bar{r}^{A}_{rep(2y)-2y}}(f)| df$ as unit one: 
	\begin{equation}
		\label{eqp}
		p = \frac{\int |S_{\bar{r}^{A}_{inj-t}}(f) - 
			S_{\bar{r}^{A}_{rep(K)-t}}(f)| df}{\int |S_{\bar{r}^{A}_{inj-2y}}(f) - 
			S_{\bar{r}^{A}_{rep(2y)-2y}}(f)| df}. 
	\end{equation}
	The number of subtracted sources are 1, 5, 10, and 15, respectively.
	In Fig.~\ref{fig:6}, the $p$ vaule is higher when we use the shorter data length, such as 
	3m and 6m. This is because the shorter data has a greater potential for inaccurate 
	removal of sources compared to the longer data, resulting in more contamination on the 
	residual, which may lead to fewer confirmed sources when search the shorter data.
	
	We repeated the same process for DWD as we did for MBHB. We calculate the 
	SNR for the brightest injection source in each frequency bin in the A channel by 
	Eq.~\ref{equation_snr}, based on various noise assumptions: LISA instrumental noise, LDC1-4 
	data, and residuals from two search situations (Fig.~\ref{fig:p_residual1}). The SNRs are 
	presented in the Tab.~\ref{table:4}. Due to the monochromatic nature of DWDs, they are more 
	sensitive to noise at a single frequency. 
	As a result, different noise assumptions can have a greater impact on the SNR of DWDs 
	compared to that of MBHBs. 
	The SNRs of a DWD can differ by a factor of 5 to 9 under two noise assumptions of 
	instrument noise or LDC1-4 data.
	This highlights the potential for subtracting out bright DWDs to enhance the detection 
	sensitivity for weaker ones when searching in a specific frequency bin.
	\begin{table*}
		\begin{center}   
			\begin{tabular}{|c|c|c|c|c|}
				\hline
				& Instrument noise & LDC1-4 & 2y-residual & 
				3m\_2y-residual 	
				\\
				\hline
				91bin & 58.9001 & 11.7234 & 18.0358 & 17.7068  \\
				\hline
				291bin & 62.5891 & 10.9944 & 28.2086 & 21.3617  \\
				\hline
				410bin & 264.4705 & 31.2997 & 138.5068 & 92.9885  \\
				\hline
			\end{tabular}
			\caption{For the brightest DWD source in each bin, SNRs are obtained in 
				the A channel under different noise assumptions. The influence of different 
				noise assumptions on the SNR of DWD is much greater than that of MBHB.} 
			\label{table:4} 
		\end{center}   
	\end{table*}
	
\begin{figure*}
		\centering  
		\subfigbottomskip=2pt 
		\subfigcapskip=-5pt 
		\subfigure[\label{fig:7}]{
			\includegraphics[width=0.48\linewidth]{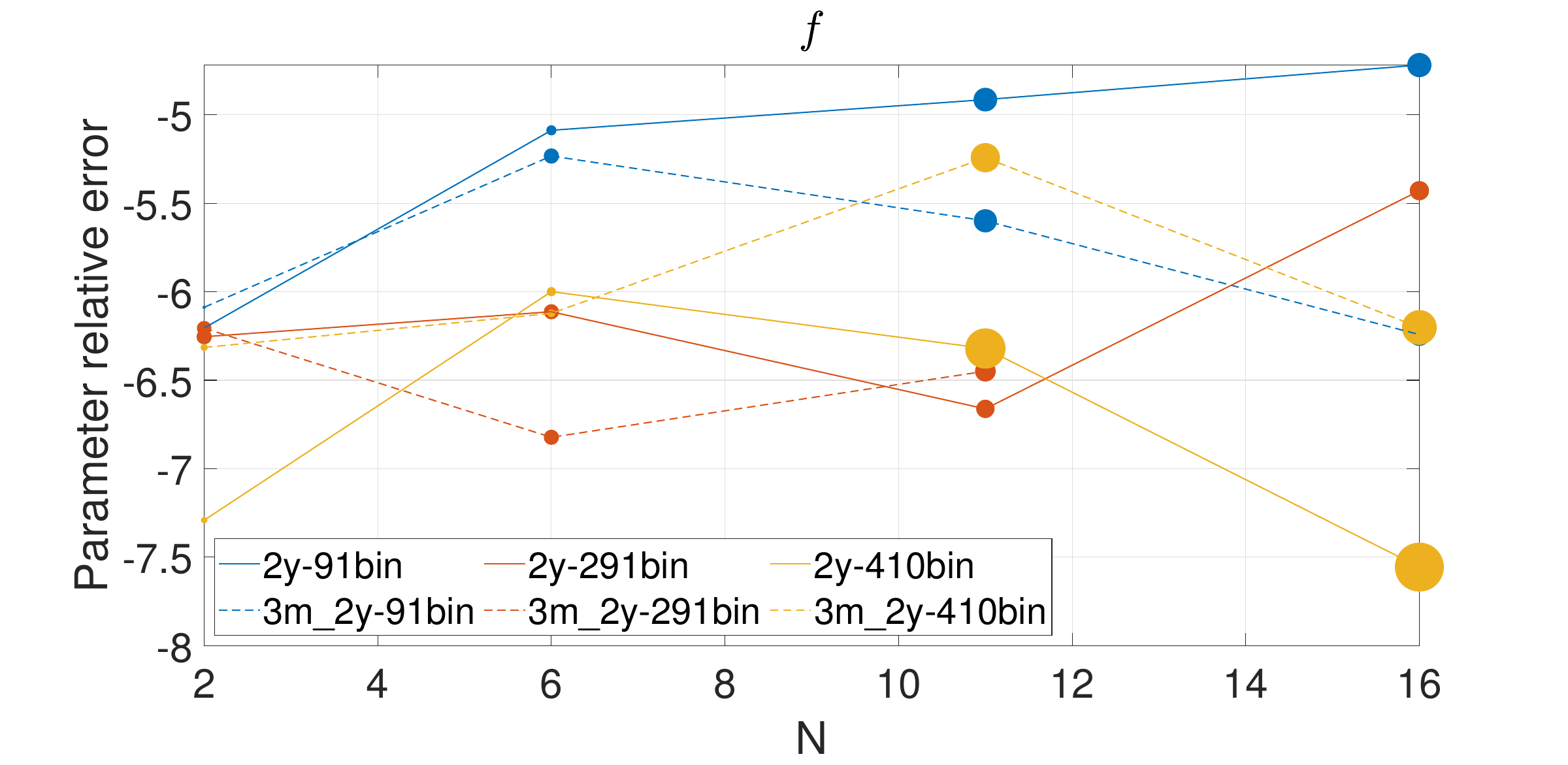}}
		\subfigure[\label{fig:8}]{
			\includegraphics[width=0.48\linewidth]{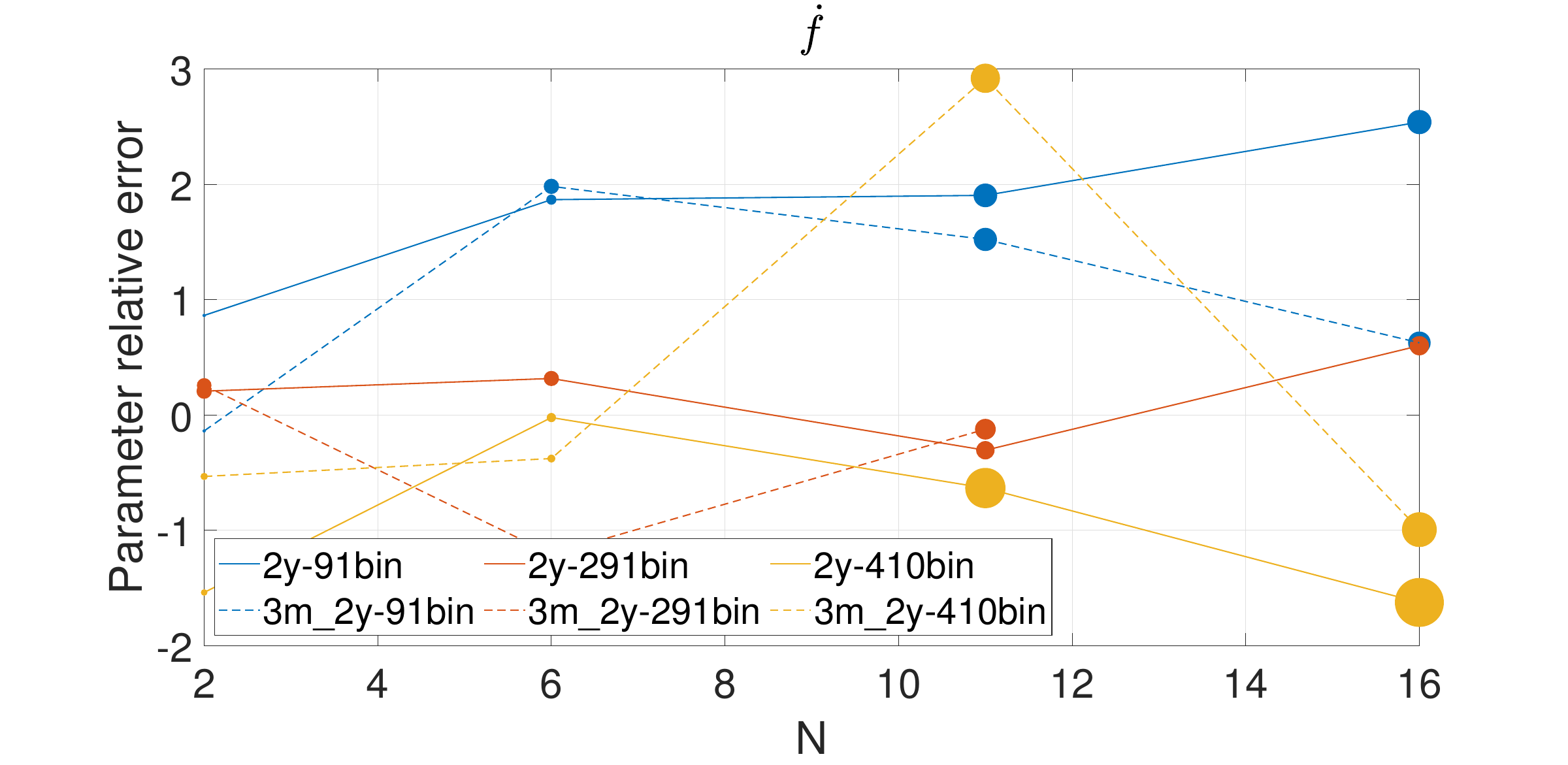}}
		\subfigure[\label{fig:9}]{
			\includegraphics[width=0.48\linewidth]{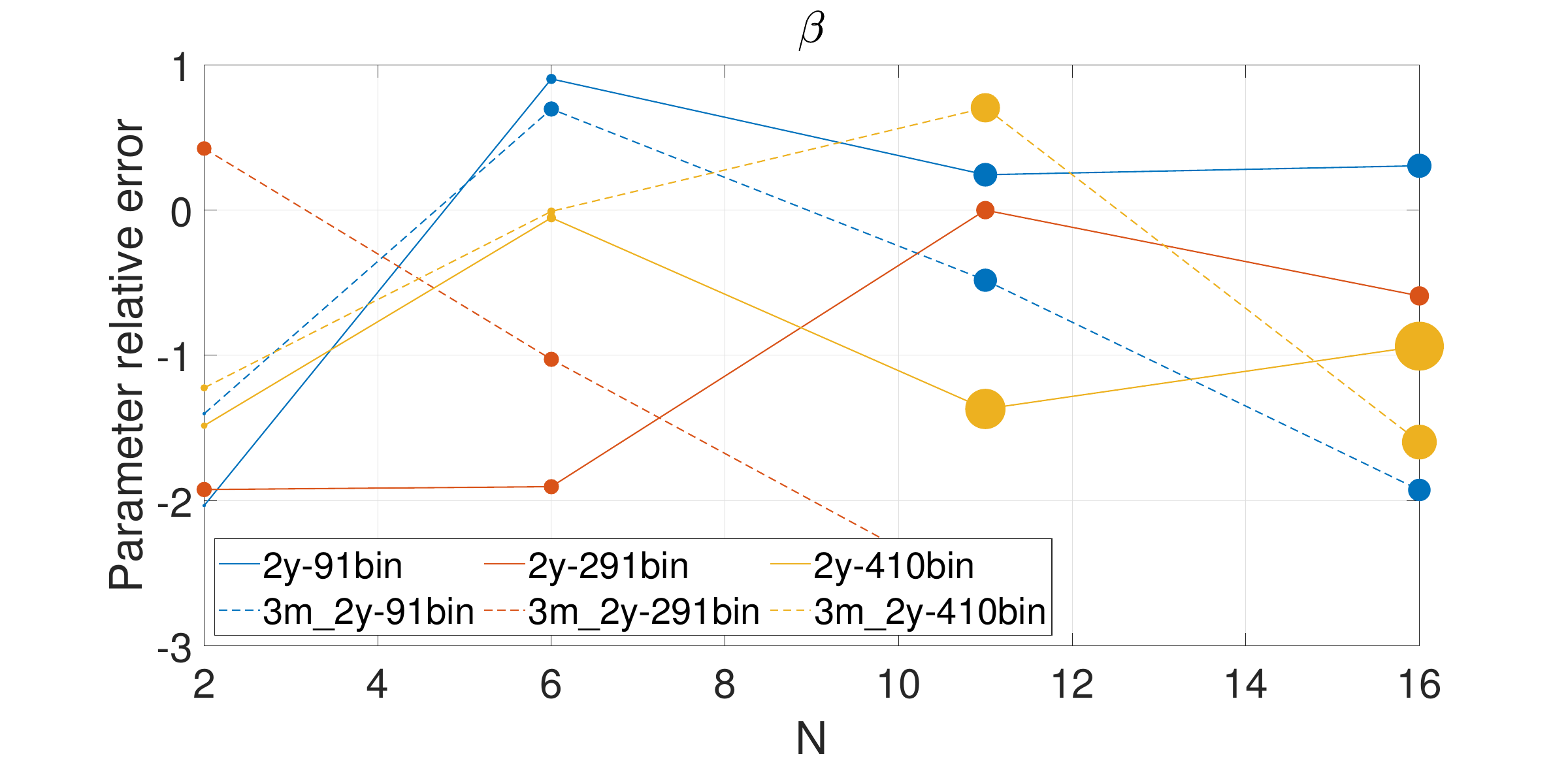}}
		\subfigure[\label{fig:10}]{
			\includegraphics[width=0.48\linewidth]{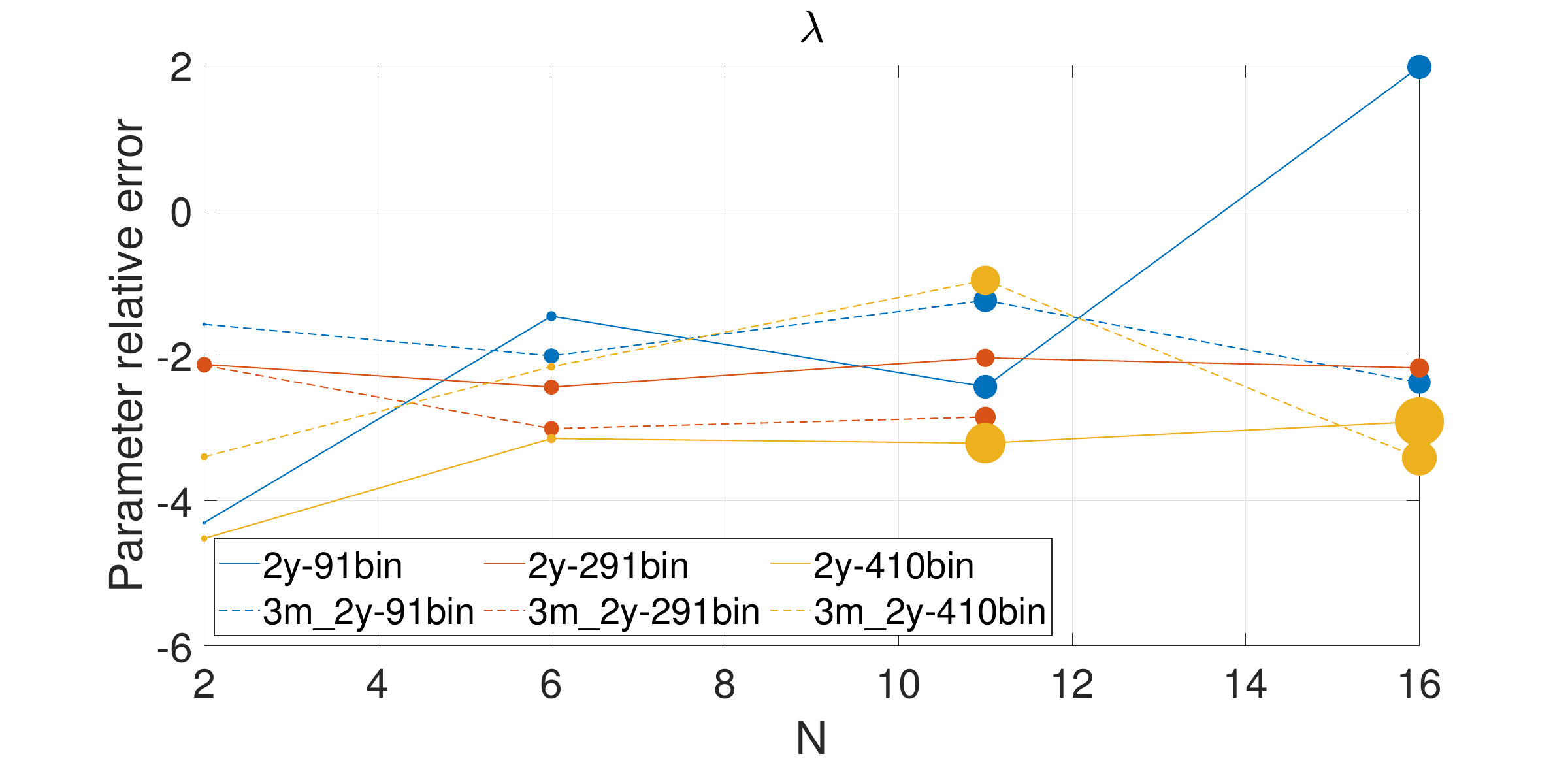}}
		\caption{The X-axis represents the number (N) of subtracted sources, and 
			the Y-axis represents the four intrinsic parameter ($f$, $\dot{f}$, $\beta$, 
			$\lambda$) relative error in logarithmic base 10. 
			The diameter of each circle represents the 
			area value obtained by subtracting all preceding sources (1, 5, 10, 15). To display 
			the area value, we use $1\times10^{-51}$ as unit one and take the square root to 
			obtain the diameter of the circle. For example, the circle in ``3m\_2y-410bin, 
			N=16" has an area value of $1.42\times10^{-47}$, and its diameter is
			$\sqrt{\frac{1.42\times10^{-47}}{1\times10^{-51}}} \approx 119$.)}
		\label{fig:influpara}
	\end{figure*}

	The impact of contamination on the accuracy of the next subtracted source's 
	parameters is a complex issue. In this study, we present the four intrinsic parameter ($f$, 
	$\dot{f}$, $\beta$, $\lambda$) accuracies for the next reported sources (2, 6, 11, 16) 
	based on the respective areas ($\int |S_{\bar{r}^{A}_{inj-t}}(f) - 
	S_{\bar{r}^{A}_{rep(K)-t}}(f)| df$) obtained from subtracting 1, 
	5, 10, 15 sources in three frequency bins. To ensure that the data length does not affect 
	the parameters' accuracy, we only consider $t=2y$ and $K=2y~\&~3m\_2y$. In 
	Fig.~\ref{fig:influpara}, the diameter of the circle represents the area value obtained by 
	subtracting all preceding sources (1, 5, 10, 15). 
	While the areas increase as the number (N) of subtracted sources grows, the parameter 
	relative error has no clear correspondence with N. This indicate that the accuracy of 
	estimated source parameters may be influenced by multiple factors, such as the SNR of the 
	source itself, the distance in parameter space between this source and other subtracted 
	sources, the density of all sources distribution in the frequency bin, and the bin in which 
	this source resides. Further studies are needed to fully understand these factors.
	
	\subsection{Detection parameter error}
\begin{figure}
		\centering
		\includegraphics[width=1\linewidth]{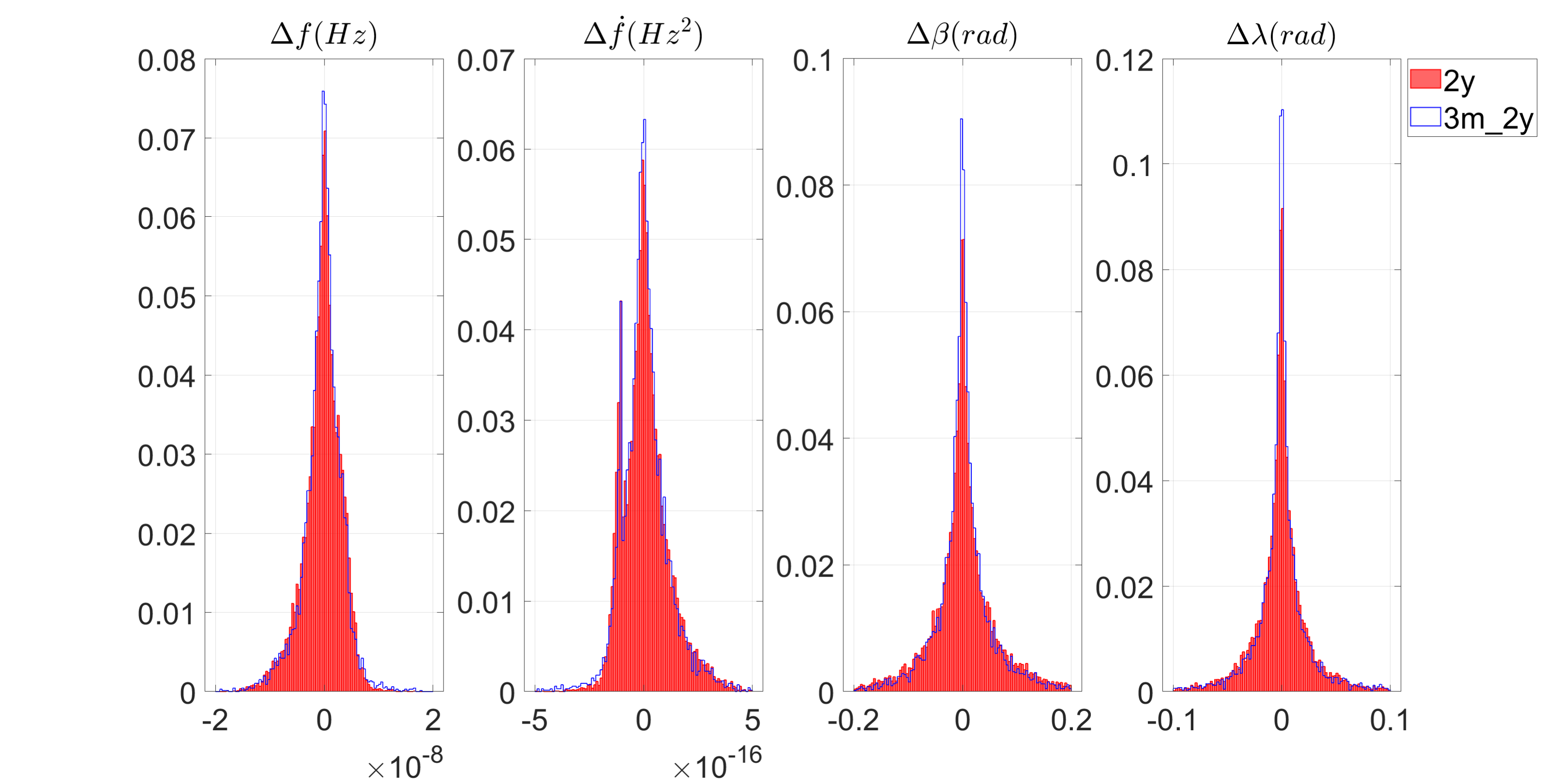}
		\caption{The absolute error histogram for four intrinsic parameters was obtained from 
			direct-search and accelerate-search to the two-year data. The red zone represents 
			direct-search, and the blue line represents accelerate-search. 
			The second histogram has a bimodality due to some sources with $f\le1\times10^{-3}$ Hz 
			has $\dot{f}$ fall on the negative boundary of the search range, making they have a 
			similar absolute error with $-1\times10^{-16}$.}
		\label{fig:paerror}
	\end{figure}

	What are the errors between confirmed source parameters and their corresponding injection 
	source parameters? Fig.~\ref{fig:paerror} shows the absolute error histogram distribution of 
	four intrinsic parameters for confirmed sources in direct-search to the two-year data and 
	accelerate-search to the two-year data. In direct-search to the two-year data, the fraction of 
	$|\Delta f|\le0.5\times 10^{-8}$ Hz is 86.4\%, the fraction of $|\Delta\dot{f}|\le2.5\times 
	10^{-16}Hz^{-1}$ is 96.3\%, the fraction of $|\Delta\beta|\le0.1$ rad is 87.5\%, and the 
	fraction of $|\Delta\lambda|\le0.05$ rad is 91.9\%.
	In accelerate-search to the two-year data, the fraction of $|\Delta f|\le0.5\times 10^{-8}$ Hz 
	is 87.5\%, the fraction of $|\Delta\dot{f}|\le2.5\times 10^{-16}Hz^{-1}$ is 96.0\%, 
	the fraction of $|\Delta\beta|\le0.1$ rad is 91.7\%, and the fraction of 
	$|\Delta\lambda|\le0.05$ rad is 92.4\%. Note that, although the fraction of 
	$|\Delta f|\le0.5\times 10^{-8}$ Hz in accelerate-search is slightly larger than in
	direct-search, the confirmed sources in accelerate-search are the sub-population in  
	direct-search with higher SNRs as shown in Fig.~\ref{fig:p_hist_SNR2}.

	\section{LISA's ability to detect the Milky Way structure}
	\subsection{Source distribution in ecliptic coordinate system}
\begin{figure*}
		\centering  
		\subfigbottomskip=2pt 
		\subfigcapskip=-5pt 
		\subfigure[\label{fig:p_3months_results}3m]{
			\includegraphics[width=0.48\linewidth]{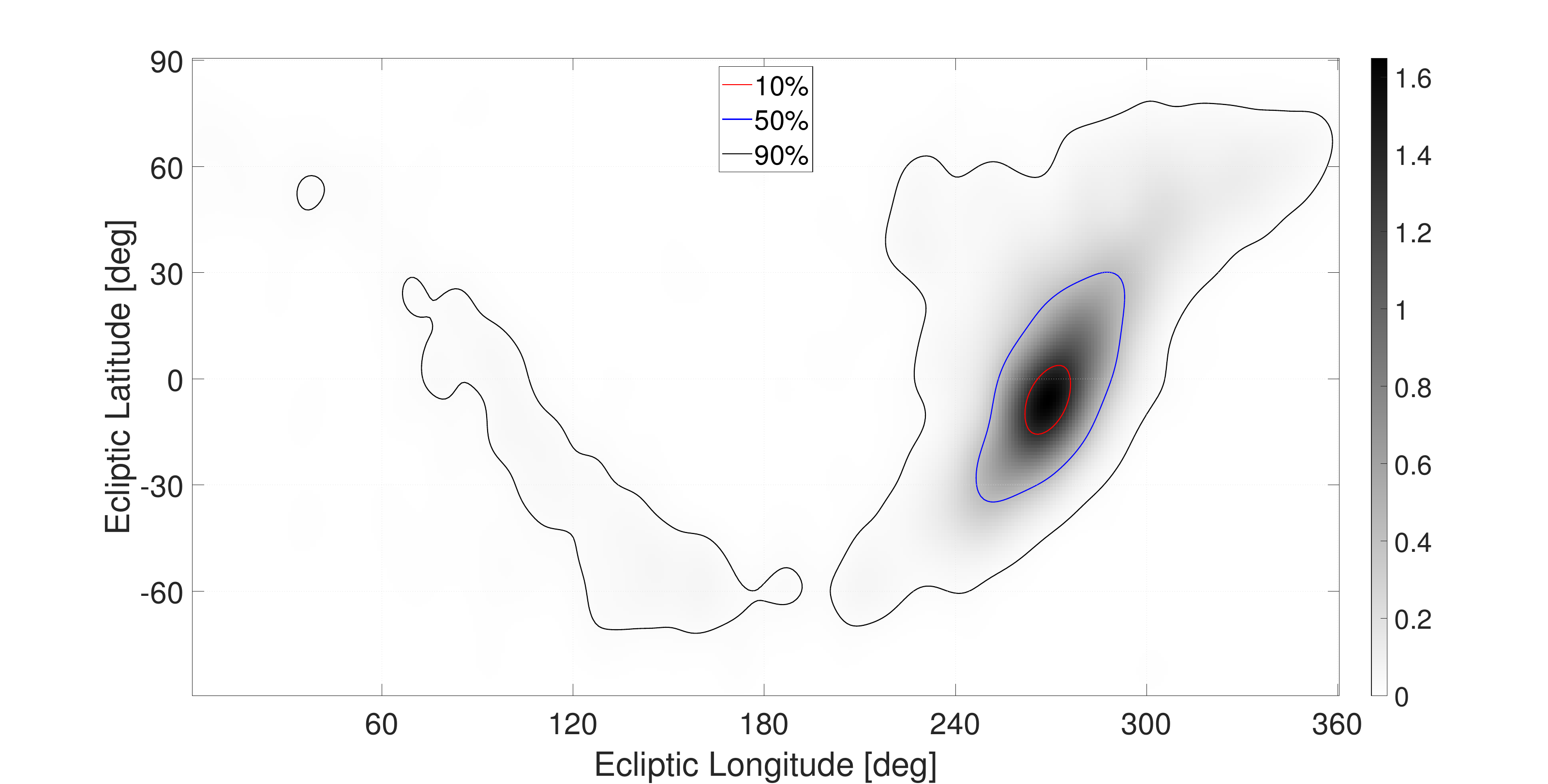}}
		\subfigure[\label{fig:p_6months_results}6m]{
			\includegraphics[width=0.48\linewidth]{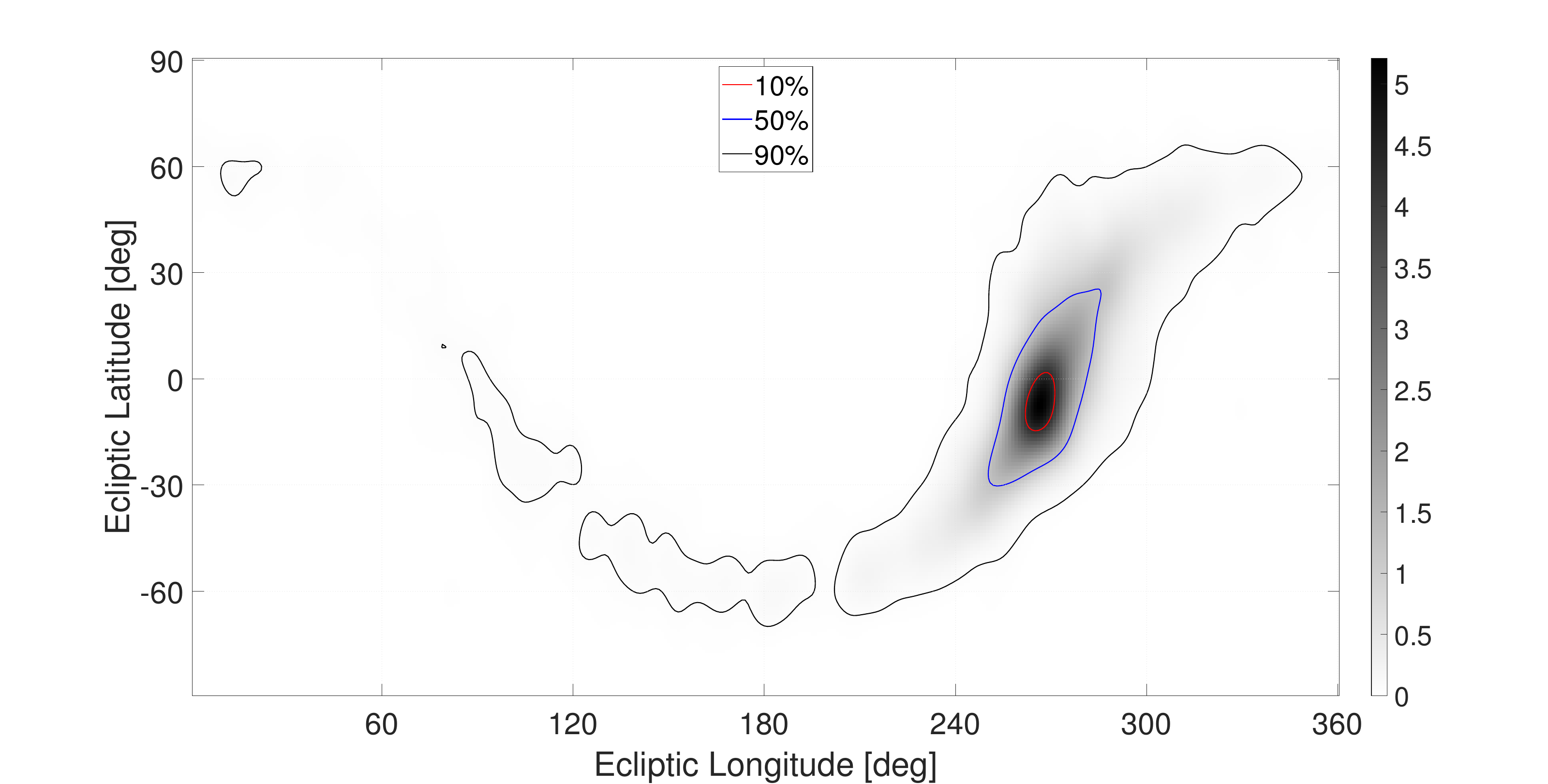}} \\
		\subfigure[\label{fig:p_2years_results}2y]{
			\includegraphics[width=0.48\linewidth]{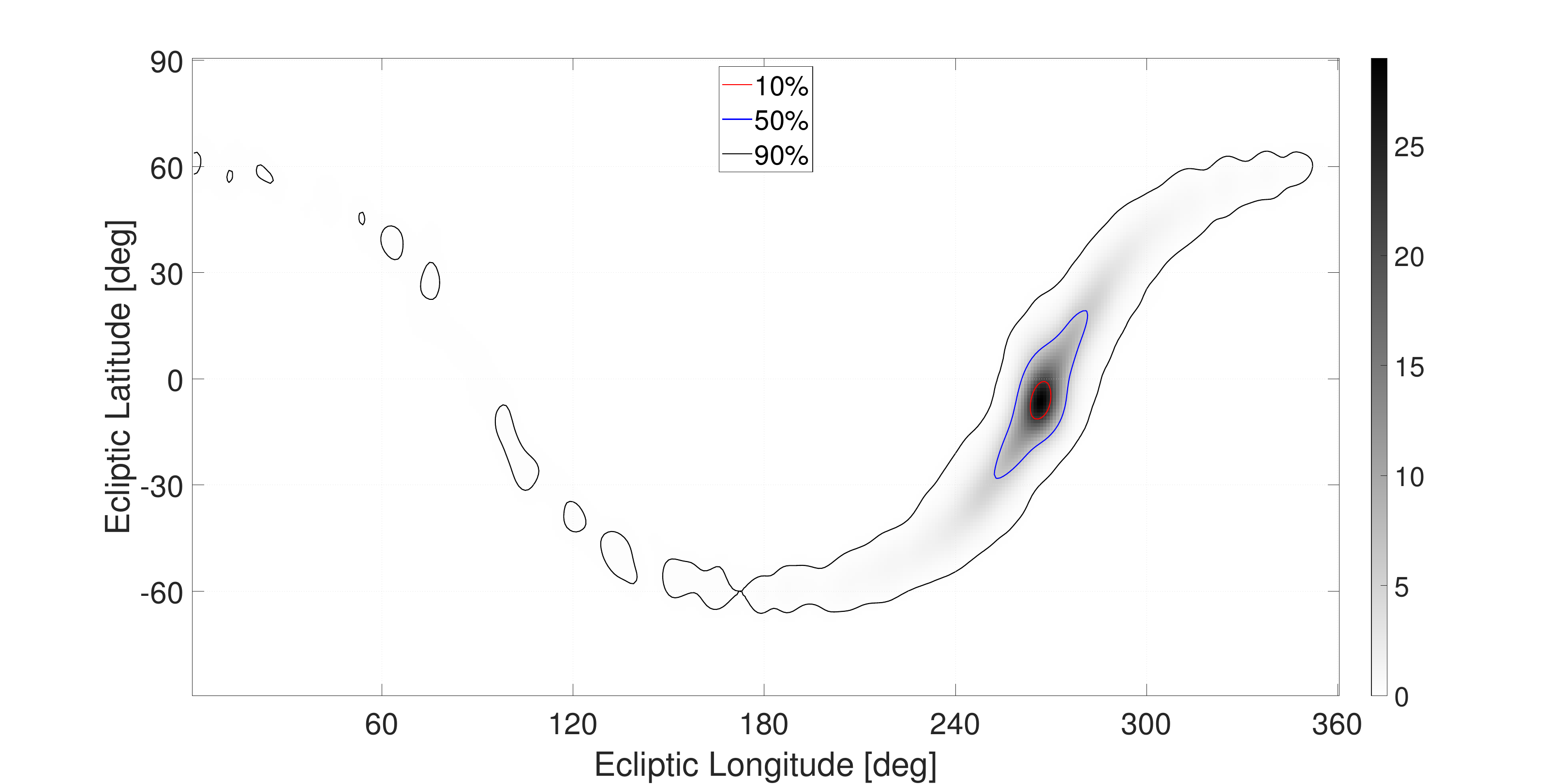}}
		\subfigure[\label{fig:p_3months_limit_2years_results}3m\_2y]{
			\includegraphics[width=0.48\linewidth]{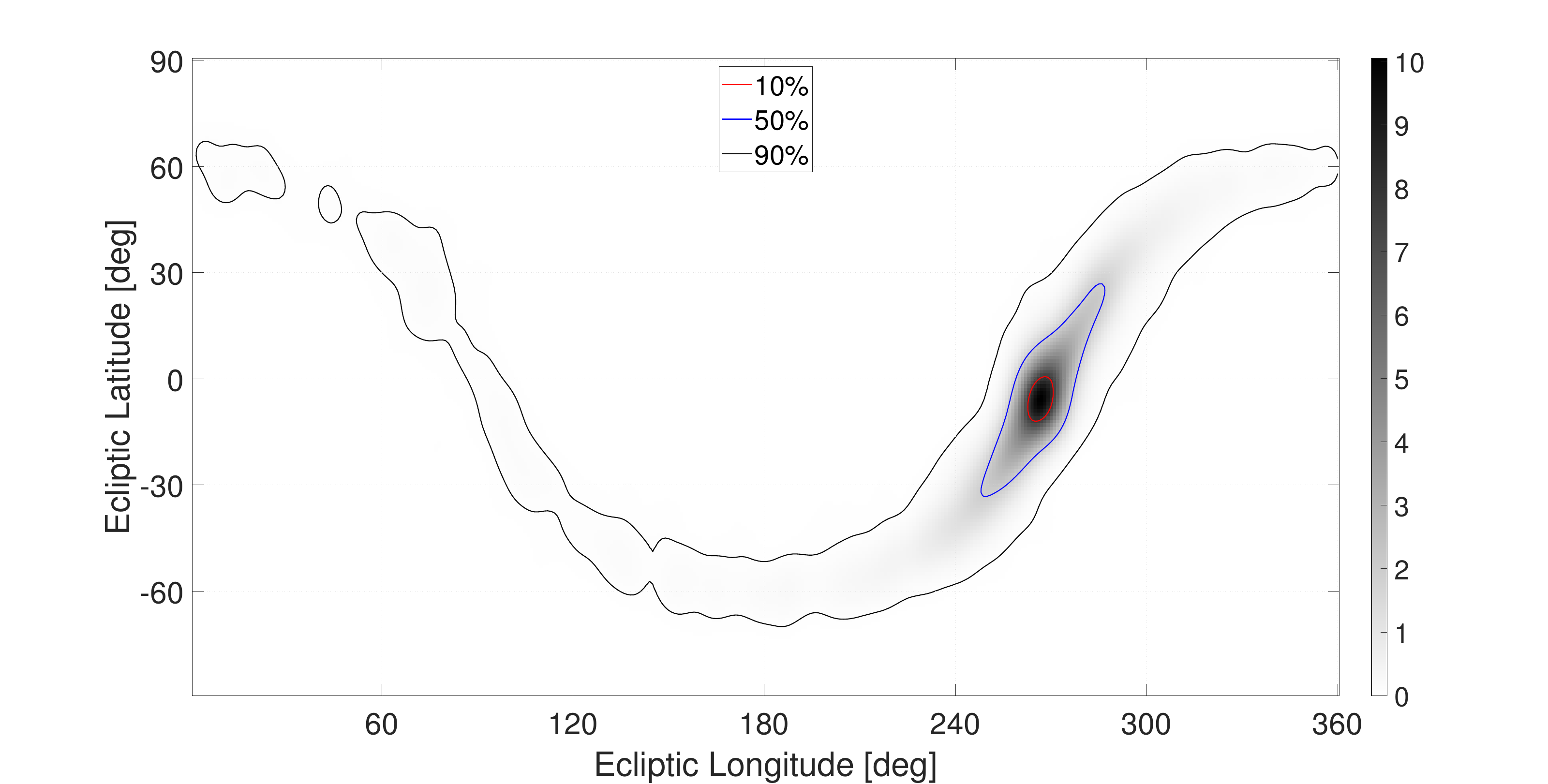}}
		\caption{\label{fig:p_results}The distribution of confirmed sources in the ecliptic 
			coordinate system. According to the number of sources per square degree, the sky is 
			divided into several areas. Specifically, the red, blue and black lines represent the 
			boundaries of the top 10\%, 50\% and 90\% distribution area of all confirmed sources, 
			respectively.}
	\end{figure*}

	LISA's detecting gravitational waves generated by DWD binaries in the Milky Way may be 
	a powerful astronomical tool \cite{korol_prospects_2017, korol_detectability_2018, 
		lamberts_predicting_2019, korol_multimessenger_2019, korol_populations_2020, 
		georgousi_gravitational_2022}. Specifically, we use the above confirmed sources to discuss 
	the 
	possible research of the Milky Way structure through LISA. Fig.~\ref{fig:p_results} presents 
	the distribution of 
	confirmed sources in the ecliptic coordinate system. Kernel density estimation is used to 
	estimate the density of confirmed sources per square degree of sky. Most sources are centered 
	at the galactic center. We draw three contour lines from high to low density that 
	account for different proportions of all confirmed sources, 10\%, 50\%, and 90\%, respectively. 
	Fig.~\ref{fig:p_3months_results} shows the result of direct-search to the first three-month 
	data, the top 10\% region enclosed by the red circle corresponds to the density more than 1.3 
	sources per square degree, and the top 50\% more than 0.4. 
	Fig.~\ref{fig:p_6months_results} shows the result of direct-search to the first six-month data, 
	the top 10\% region corresponds to the density 
	more than 4.2 sources per square degree, and the top 50\% more than 1.3. 
	Fig.~\ref{fig:p_2years_results} shows the result of direct-search to the two-year data, 
	the top 10\% region corresponds to the density 
	more than 22.2 sources per square degree, and the top 50\% more than 6.1. 
	Fig.~\ref{fig:p_3months_limit_2years_results} shows the result of accelerate-search to
	the two-year data, the top 10\% region corresponds to the density more 
	than 7.7 sources per square degree, and the top 50\% more than 1.9. 
	All of the confirmed sources from the direct-search to the first three-month data and the 
	first six-month data 
	are scattered in the sky. This is because their data is not long enough, which enlarges the 
	position error of LISA for each DWD binary. However, both direct-search and 
	accelerate-search to the two-year data have less position error.
	Their average latitude error is about $2.3^{\circ}$, and the average longitude error is about 
	$1.6^{\circ}$. When we only consider sources with $f\ge4\times10^{-3}$ Hz, the average latitude 
	and longitude error are $1.3^{\circ}$ and $0.3^{\circ}$, respectively.
	
	\subsection{Source distribution in galactic coordinate system}
\begin{figure*}
		\centering  
		\subfigbottomskip=2pt 
		\subfigcapskip=-5pt 
		\subfigure[\label{fig:p_gala_all_inj0}Injection sources]{
			\includegraphics[width=0.48\linewidth]{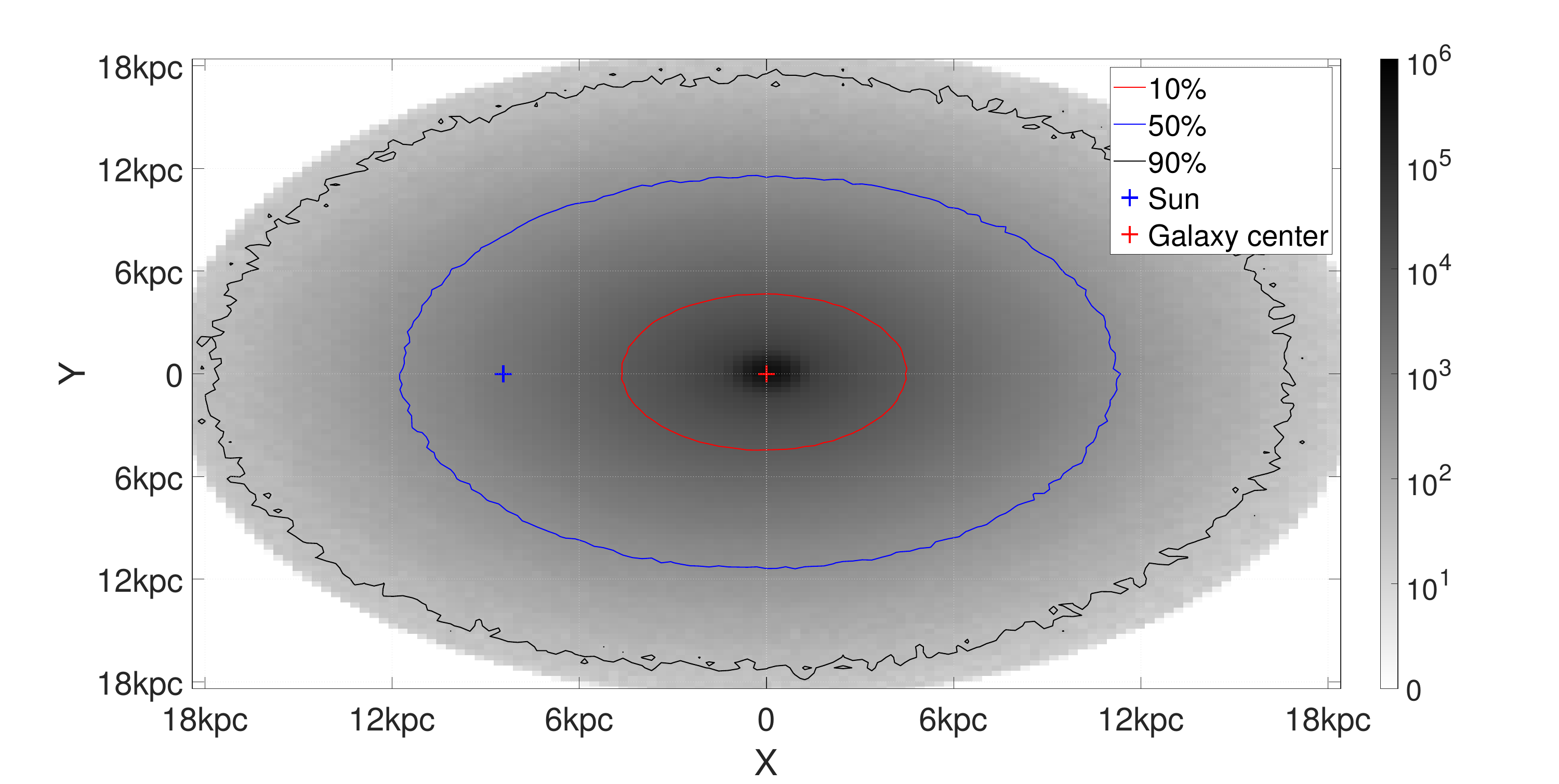}}
		\subfigure[\label{fig:p_gala_all_sel0}Selected sources]{
			\includegraphics[width=0.48\linewidth]{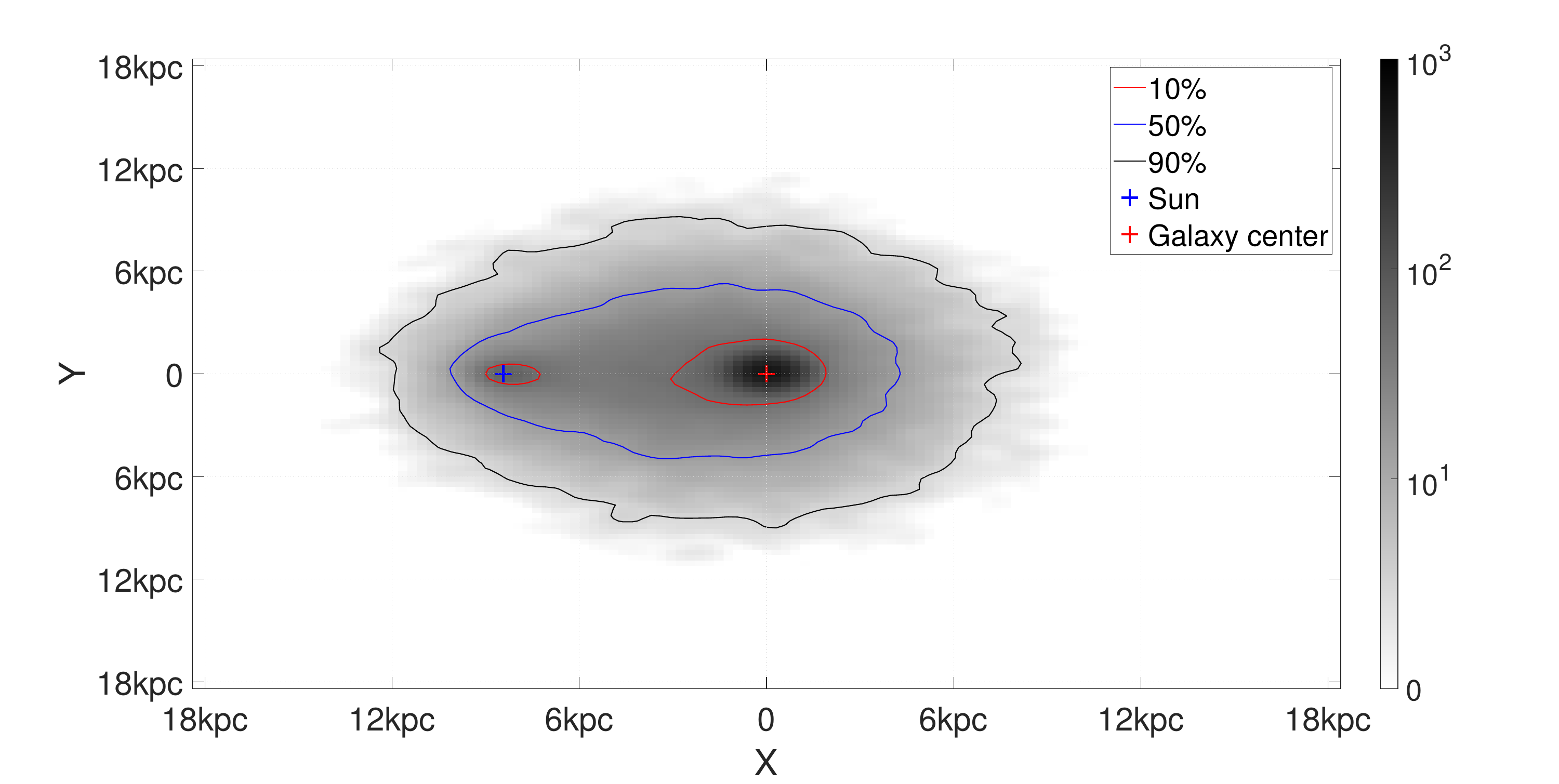}}\\
		\subfigure[\label{fig:p_gala_2yR0.9}2y]{
			\includegraphics[width=0.48\linewidth]{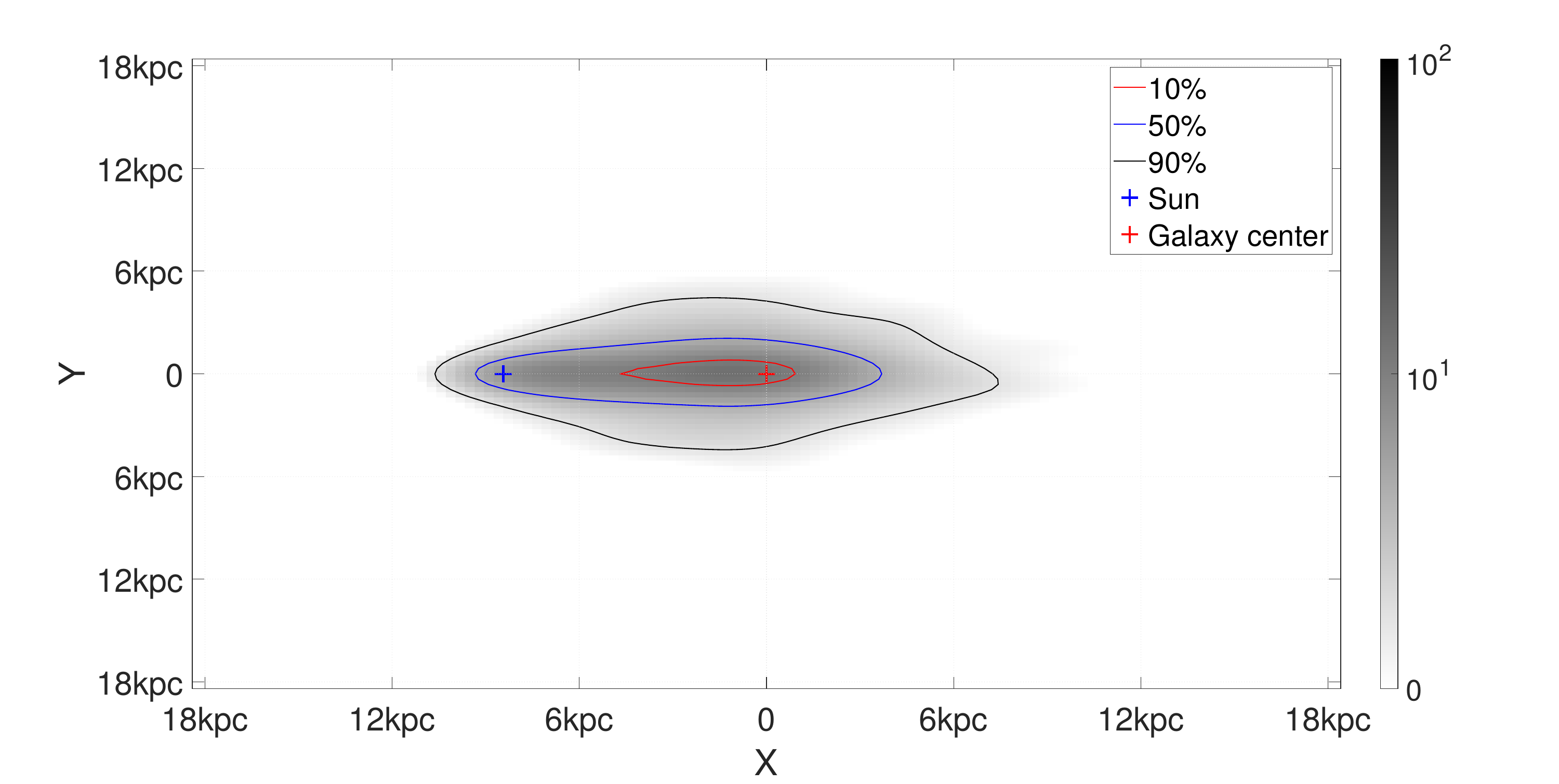}}
		\subfigure[\label{fig:p_gala_3m_2yR0.9}3m\_2y]{
			\includegraphics[width=0.48\linewidth]{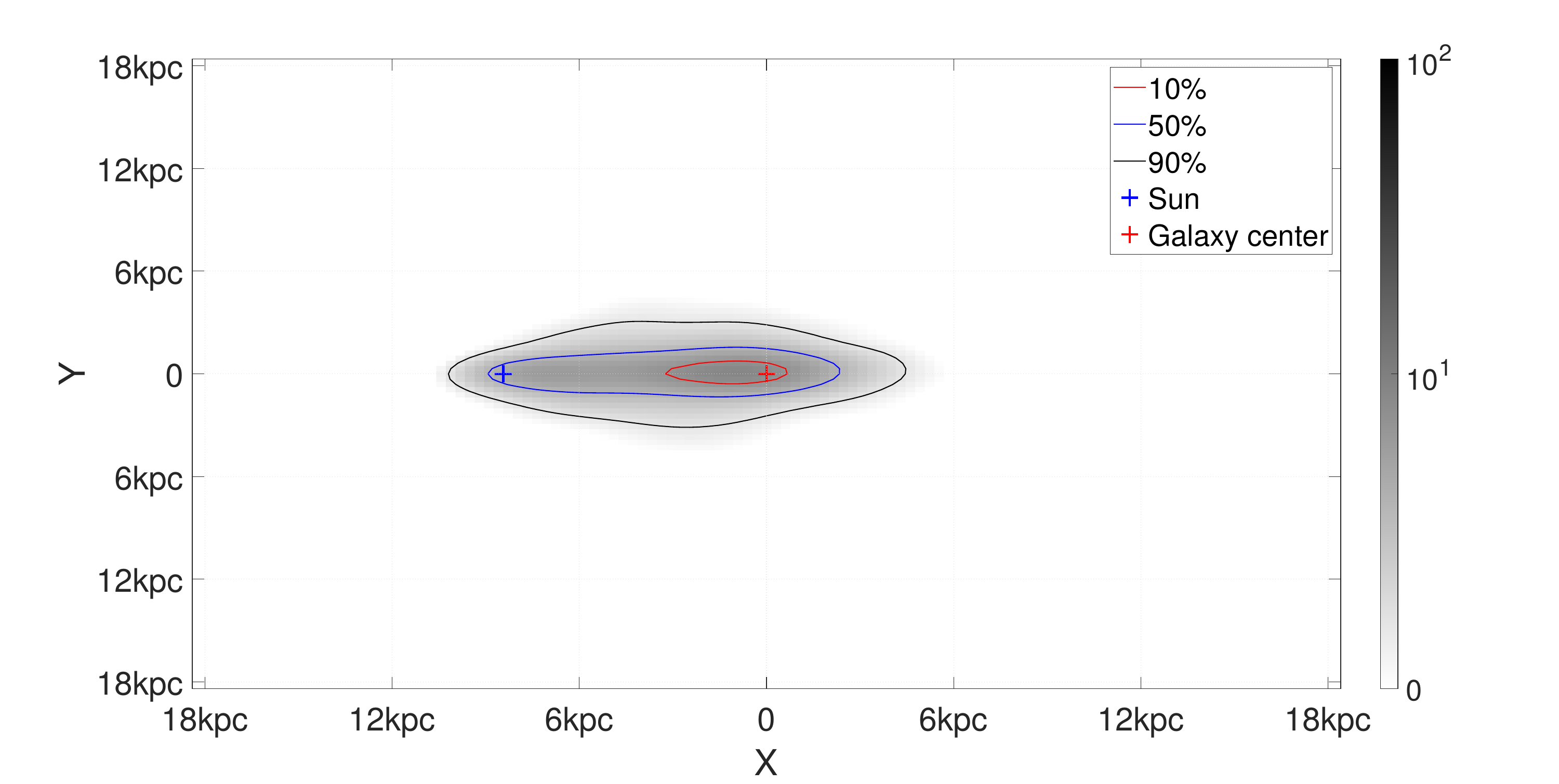}}
		\caption{\label{fig:p_gala}The distribution of confirmed sources with $\dot{f}\ge0$ in the 
			galactic coordinate system. The view is equivalent to looking down at the entire disk 
			from the galactic north pole. The thickness of the disk is not shown.
			The color indicates the number of sources per square thousand light years. The red, 
			blue and black lines represent the top 10\%, 50\% and 90\% distribution area of all 
			confirmed sources, respectively.}
	\end{figure*}

	It is assumed that all confirmed sources with $\dot{f}\ge0$ come from DWD binary systems 
	without mass transfer. Therefore, the distances $D$ and chirp masses $\mathcal{M}$ of DWD 
	binaries can be calculated by Eqs.~\eqref{format1} and \eqref{format2}.
	With the known distance, we convert the source distribution in ecliptic coordinate system to 
	that in galactic coordinate system and plot the density distribution of DWD 
	binaries on 
	the galactic disk. We also use kernel density estimation to access the number of DWD binaries 
	per square thousand light years. In fact, we have ignored the thickness of the galactic disk. 
	Fig.~\ref{fig:p_gala} shows the DWD binaries distribution of four situations, and 
	the location of the sun and the Galactic center.
	Fig.~\ref{fig:p_gala_all_inj0} shows all injection sources with $\dot{f}\ge0$, whose number is 
	26433149, forming a disk with a radius of about 60,000 light years ($\sim$18kpc).
	In the top 10\% density region, there are more than 7413.1 DWD binaries per 
	square thousand light years, and the top 50\% more than 426.6. 
	The density decreases with the increase of distance between the source and the galactic center.
	Fig.~\ref{fig:p_gala_all_sel0} presents the selected injection sources with $\dot{f}\ge0$, 
	whose number is 65510. All selected injection sources were selected by SNR $\ge3$ for LISA over 
	two years of observation.
	These sources are expected to have SNR $\ge6$ during the four years of LISA observation. In 
	the top 10\% density region, there are more than 50.1 DWD binaries per square thousand light 
	years, and the top 50\% more than 14.1.
	The densest parts are concentrated near the solar system and the galactic 
	center. We expect to eventually search all of these selected injection sources throughout the 
	life of LISA.
	Fig.~\ref{fig:p_gala_2yR0.9} presents confirmed sources with $\dot{f}\ge0$ from the 
	direct-search to the two-year data, whose number is 9927. In the top 10\% density region, 
	there are more than 10.5 DWD binaries per square thousand light years, and the top 50\% more 
	than 4.4.
	These sources are centralized on the sun-galactic center line, which we think 
	is related to the direction of LISA's sensitivity.
	Fig.~\ref{fig:p_gala_3m_2yR0.9} presents all confirmed sources with $\dot{f}\ge0$ from 
	the accelerate-search to the two-year data, whose number is 5205. In the top 10\% density 
	region, there are more than 7.1 DWD binaries per square thousand light years, and the top 50\% 
	more than 3.9.
	Accelerate-search to the two-year data is less 
	capable of detecting sources on the other side of the Galactic center than direct-search to 
	the two-year data because they are so far away and tend to be accompanied by a lower SNR.
	
	\subsection{Distance and chirp mass error}
\begin{figure*}
		\centering  
		\subfigbottomskip=2pt 
		\subfigcapskip=-5pt 
		\subfigure[\label{fig:p_2y_m_error}2y]{
			\includegraphics[width=0.48\linewidth]{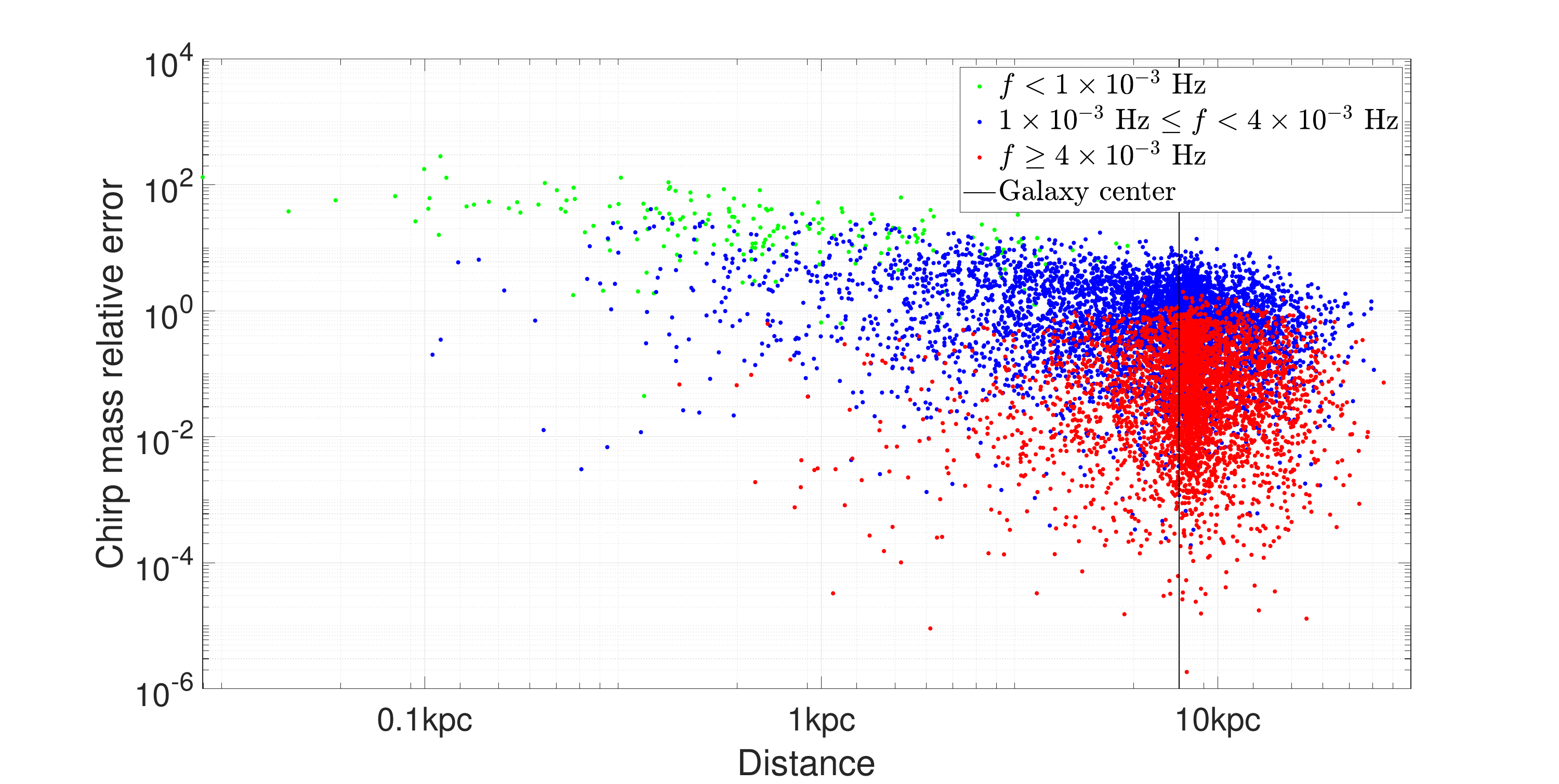}}
		\subfigure[\label{fig:p_3m_2y_m_error}3m\_2y]{
			\includegraphics[width=0.48\linewidth]{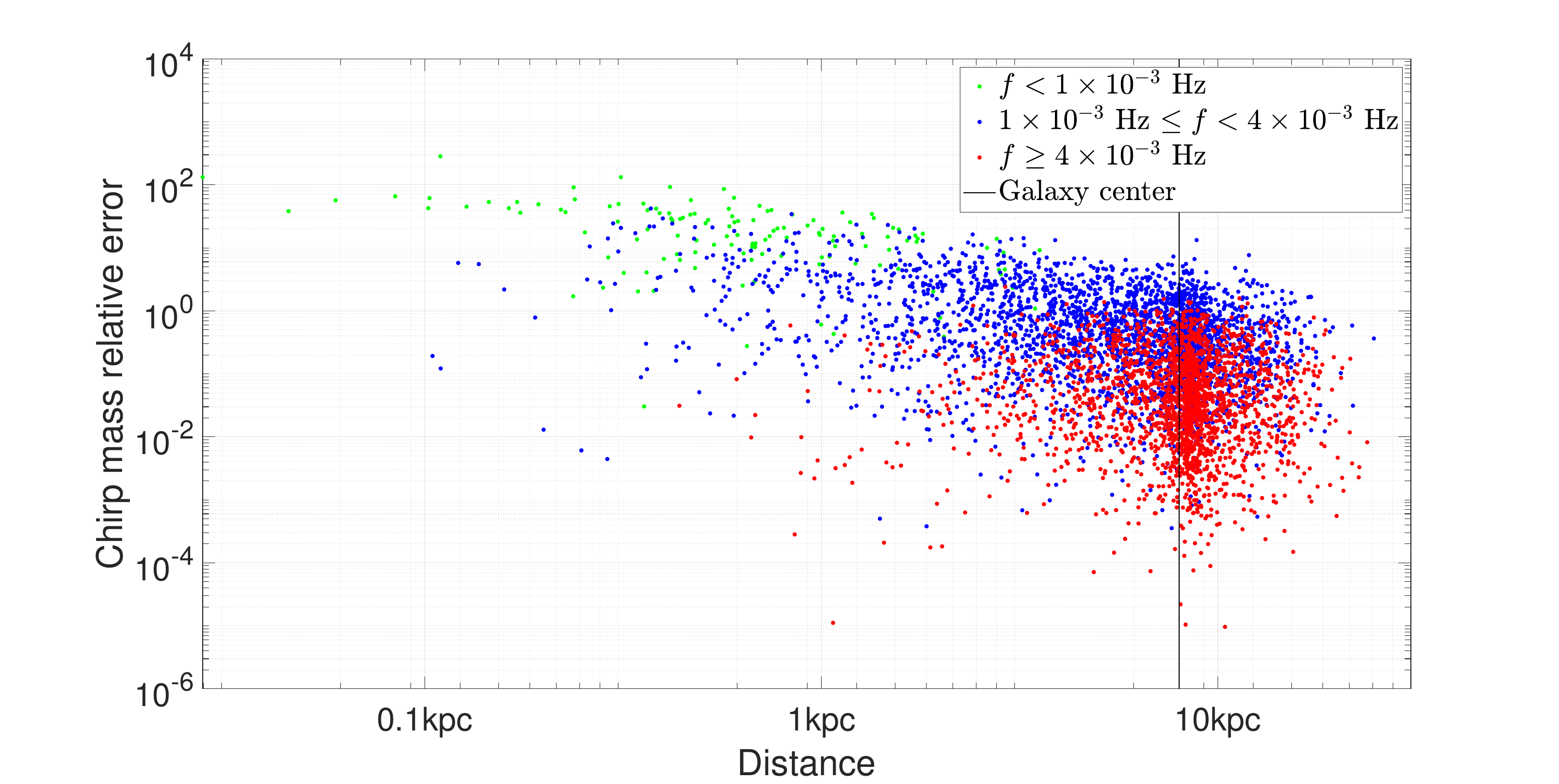}}\\
		\subfigure[\label{fig:p_2y_distance_error}2y]{
			\includegraphics[width=0.48\linewidth]{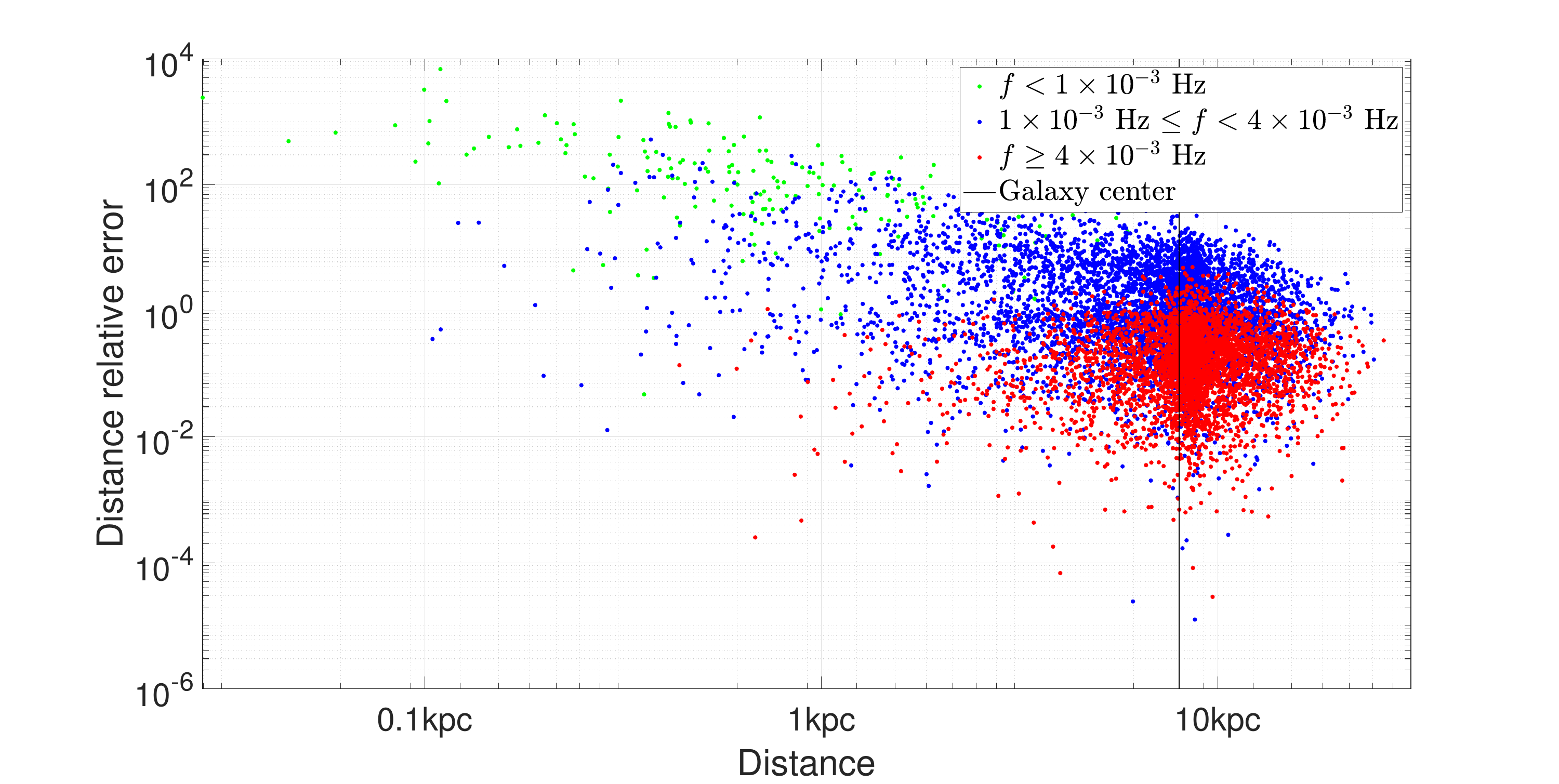}}
		\subfigure[\label{fig:p_3m_2y_distance_error}3m\_2y]{
			\includegraphics[width=0.48\linewidth]{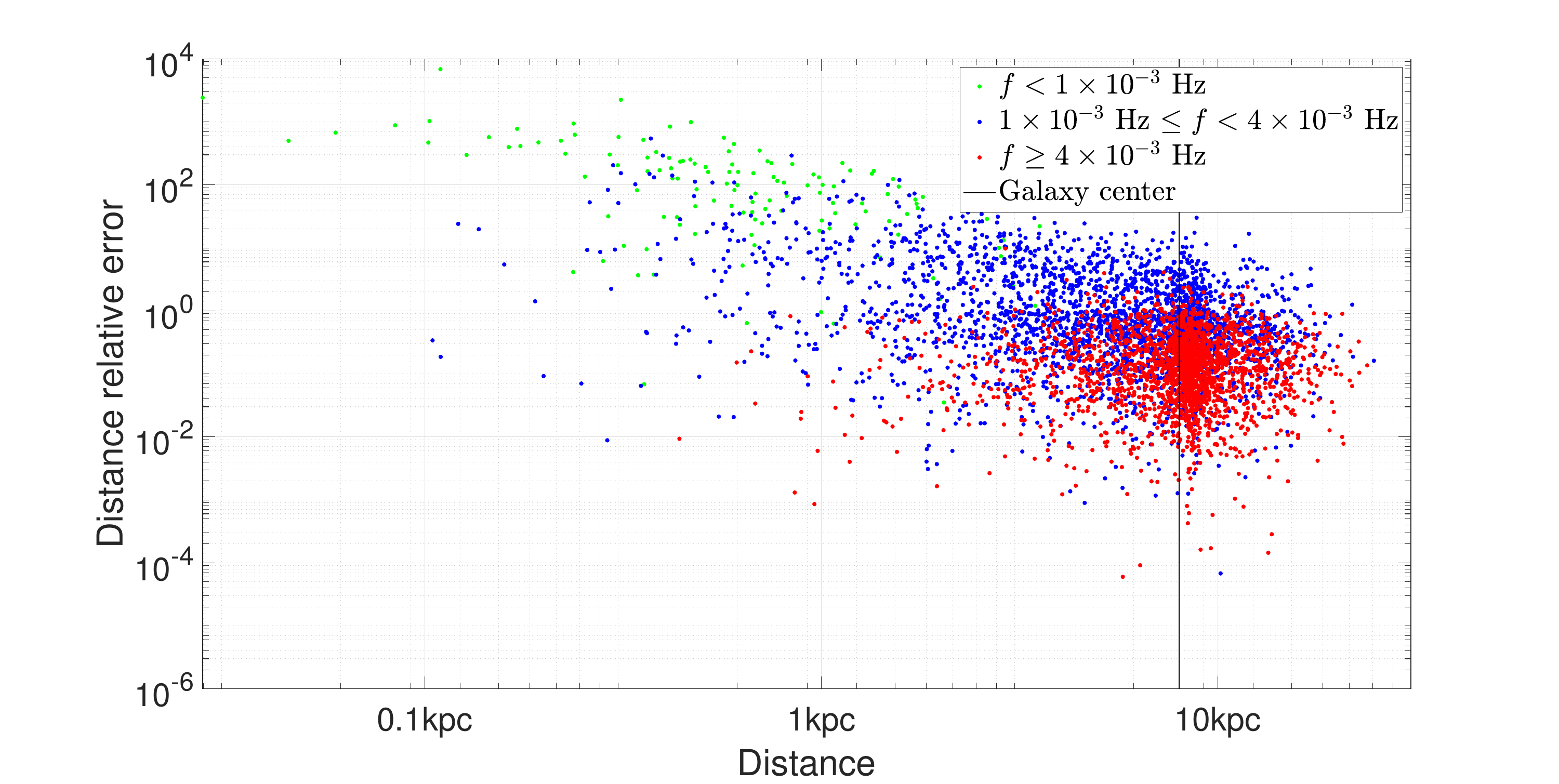}}
		\caption{\label{fig:error}The relative error of chirp mass and distance of the confirmed 
			source obtained through direct-search and accelerate-search to the two-year data.
			The x-axis is the distance between the source and the sun. The black line represents 
			the location of the galactic center.}
	\end{figure*}

	Fig.~\ref{fig:error} shows the relative error of distance and chirp mass for direct-search 
	and accelerate-search to the two-year data. The x-axis is the distance between the source and 
	the sun. 
	The y-axis is $|\frac{\Delta \mathcal{M}}{\mathcal{M}_{inj}}|$ or $|\frac{\Delta D}{D_{inj}}|$, 
	where $\Delta \mathcal{M} = |\mathcal{M}_{injection}-\mathcal{M}_{confirmed}|$ and $\Delta D = 
	|D_{injection}-D_{confirmed}|$. 
	For the sources with high frequency ($f\ge4\times10^{-3}$ Hz), the relative error
	of the average chirp mass and distance in the direct-search are 14.5\% and 28.2\%, 
	respectively, and the accelerate-search 12.0\% and 24.0\%.
	Sources with medium frequency ($1\times10^{-3}$ Hz $\le f<4\times10^{-3}$ Hz) have a relatively 
	poor performance in parameters estimation. The relative error of the average chirp mass and 
	distance in the direct-search are respectively 145.7\% and 392.0\%, and the accelerate-search 
	124.4\% and 359.3\%.
	Sources with the low-frequency range ($f<1\times10^{-3}$ Hz) have the worst accuracy in 
	parameters estimation. 
	In general, for the two-year data, the direct-search is not as good as the accelerate-search in 
	the estimation of distance and chirp mass.

	\section{The detection SNR threshold}
	\label{sec:results}
	In realistic space-based gravitational wave detection, we cannot calculate R for 
	reported sources because of the unknown injection source parameter sets. The main evidence to 
	determine whether a DWD reported source is a real signal is its SNR. 
	Being targeted at finding the sources confirmed with R $\ge0.9$ in Sec.~\ref{sec:result} as 
	much as possible, we try to seek out the best SNR threshold for searching DWD binary with LISA.
	
	We take confirmed sources as the standard answer, 
	using the Receiver Operating Characteristic Curve (ROC) to discuss the relation
	between SNR and R through four search results. In Tab.~\ref{table:3}, R 
	takes 0.9 as the threshold, and the SNR threshold $\eta$ is a positive integer ranging from 7 
	to 21. We grouped all reporting sources into four categories. True Positive (TP) represents the 
	reported source that both SNR and R are greater than the threshold. False Negative (FN) 
	represents the reported source that meets 
	SNR $\ge\eta$ and R $<0.9$. False Positive (FP) represents the reported source that meets 
	SNR $<\eta$ and R $\ge0.9$. True Negative (TN) represents the reported source that both SNR and 
	R are smaller than their threshold. 
	Eqs.~\eqref{format3} and \eqref{format4} mean the detection ability (DB) and the false 
	alarm rate (FAR) when we determine the threshold $\eta$ of SNR. That is the proportion of 
	reported sources of SNR $\ge\eta$ to the confirmed sources, and the proportion of reported 
	sources of SNR $\ge\eta$ to the unconfirmed sources. The larger the DB is, as long as the 
	smaller the FAR is, the closer the results of the SNR screening and R screening is. 
	That is to say, a larger DB and a smaller FAR corresponds to a better SNR threshold.
	\begin{table}
		\begin{center}   
			\begin{tabular}{|c|c|c|c|}
				\hline
				\multicolumn{2}{|c}{\multirow{2}{*}{}} & \multicolumn{2}{|c|}{R} \\
				\cline{3-4}
				\multicolumn{2}{|c|}{} & $\ge0.9$ & $<0.9$ \\
				\hline
				\multirow{2}{*}{SNR} & $\ge\eta$ & TP & FN \\
				\cline{2-4}
				& $<\eta$ & FP & TN \\
				\hline
			\end{tabular}
			\caption{All reported sources can be divided into four parts using R and SNR. 
				There are true positive, false positive, false negative, and true negative.} 
			\label{table:3} 
		\end{center}   
	\end{table}
	\begin{equation}
		\label{format3}
		DB = \frac{TP}{TP+FP},
	\end{equation}
	\begin{equation}
		\label{format4}
		FAR = \frac{FN}{FN+TN}.
	\end{equation}
	
\begin{figure}
		\centering  
		\subfigbottomskip=2pt 
		\subfigcapskip=-5pt 
		\subfigure[\label{fig:p_roc_1_1491}All frequency]{
			\includegraphics[width=1\linewidth]{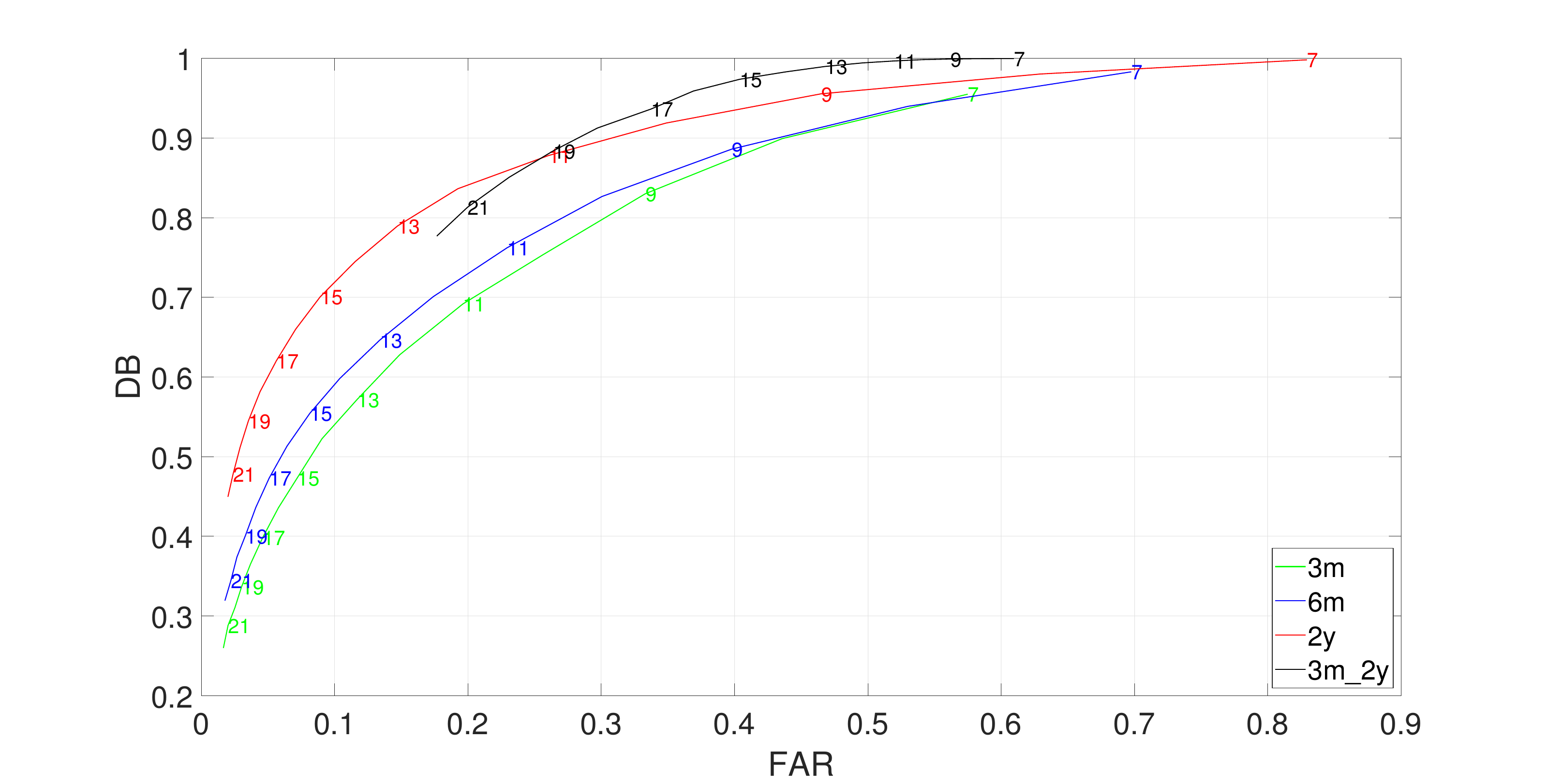}}\\
		\subfigure[\label{fig:p_roc_1_391}Low frequency + Medium frequency]{
			\includegraphics[width=1\linewidth]{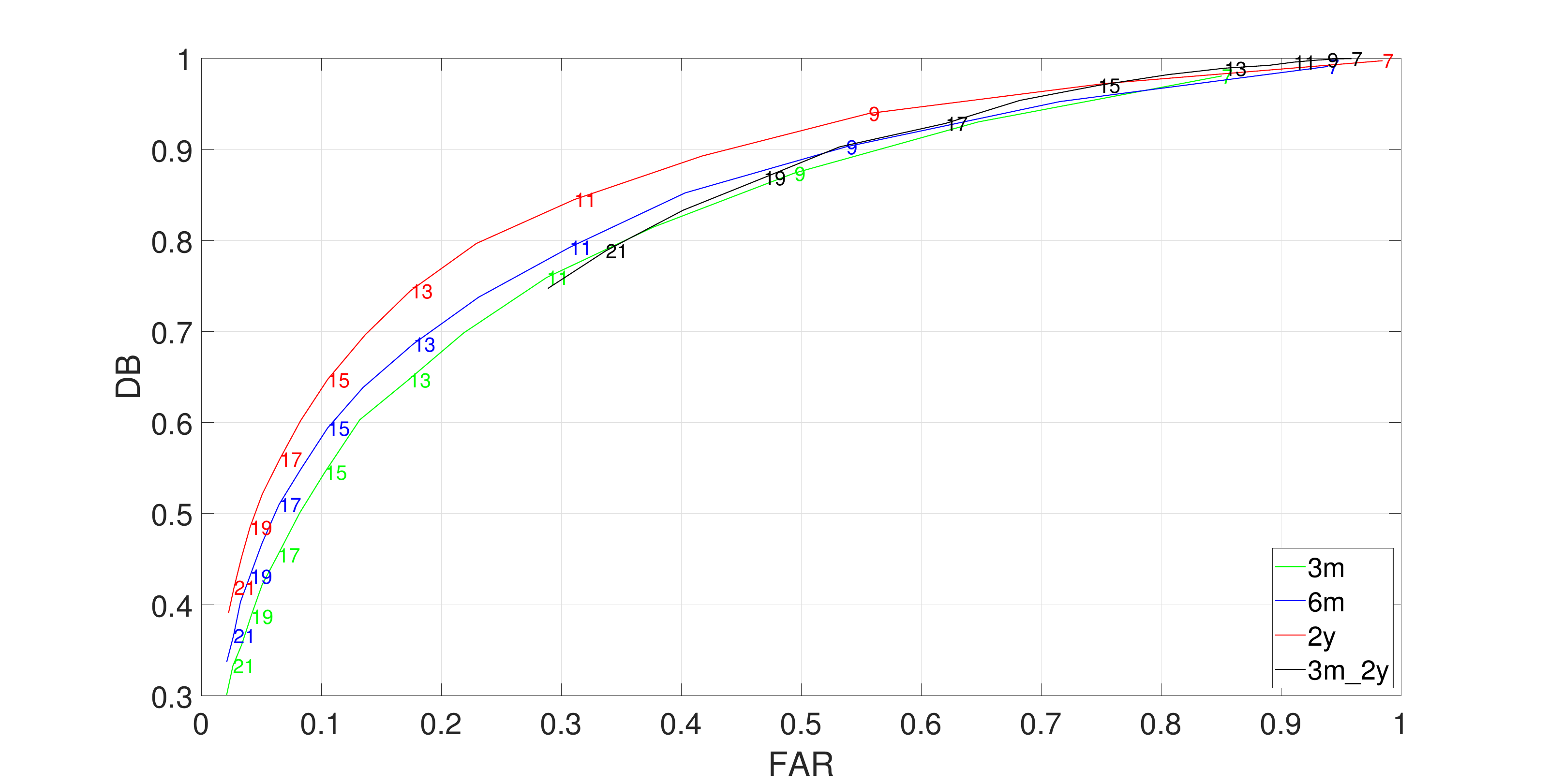}}
		\subfigure[\label{fig:p_roc_392_1491}High frequency]{
			\includegraphics[width=1\linewidth]{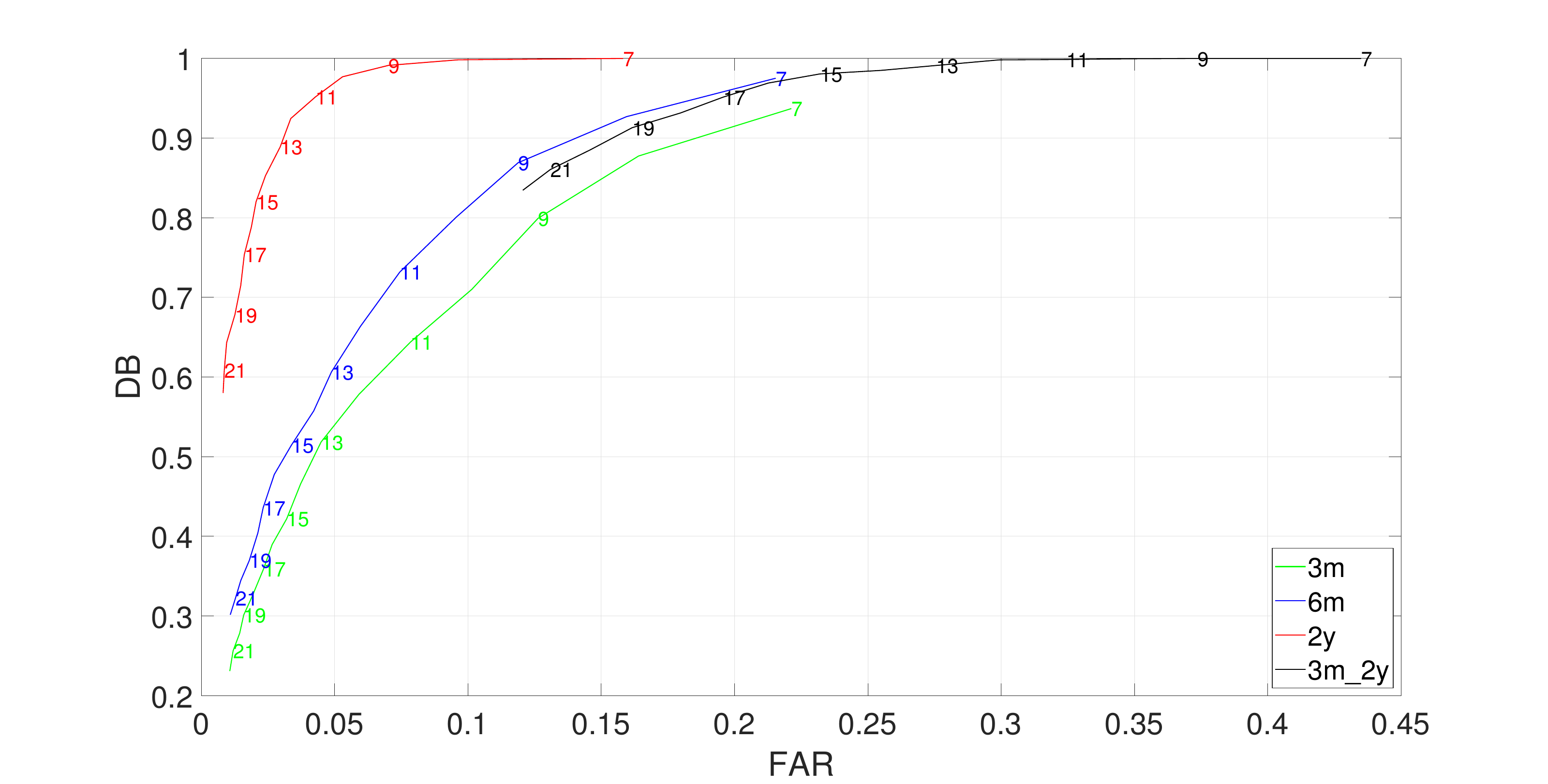}}
		\caption{\label{ROC}The top, middle, and bottom figures are the ROC diagrams for all 
			frequencies ($1\times10^{-4}$ Hz$-1.5\times10^{-2}$ Hz, Fig.~\ref{fig:p_roc_1_1491}), 
			low and medium frequencies ($1\times10^{-4}$ Hz$-4\times10^{-3}$ Hz, 
			Fig.~\ref{fig:p_roc_1_391}), high frequencies ($4\times10^{-3}$ Hz$-1.5\times10^{-2}$ 
			Hz, Fig.~\ref{fig:p_roc_392_1491}), respectively.
			Each diagram contains four search situations. The color numbers in the diagram indicate 
			the SNR threshold $\eta$ of the corresponding points. The closer the point is to the 
			upper left corner of the diagram, the closer the reported sources filtered from the 
			corresponding SNR threshold $\eta$ are to the confirmed sources.}
	\end{figure}

	Fig.~\ref{ROC} presents the ROC diagram in different frequency ranges.
	The lower SNR threshold will lead to a higher DB but bring a bigger FAR problem.  
	Moreover, the higher SNR threshold leads to a lower FAR and a lower DB. Therefore, finding a 
	proper SNR threshold to balance the DB and FAR seems complicated.
	For direct-search situations, the longer the detection time, the closer the search results 
	screened out by the SNR to the confirmed sources screened out by the R (closer to the 
	upper left corner in the figure). 
	In the case of accelerate-search to the two-year data, the overall DB is higher than the other 
	three cases because it shields a lot of unconfirmed sources (R $<0.9$) in the low and 
	medium-frequency domain ($f<4\times10^{-3}$ Hz). However, its 
	FAR is also high, mainly because the number of unconfirmed sources is smaller than that of 
	other cases (as can be seen from the fourth line in Tab.~\ref{table:1}), and the denominator in 
	Eq.~\eqref{format4} is small, which makes a big FAR.
	
\begin{figure}
		\centering
		\includegraphics[width=1\linewidth]{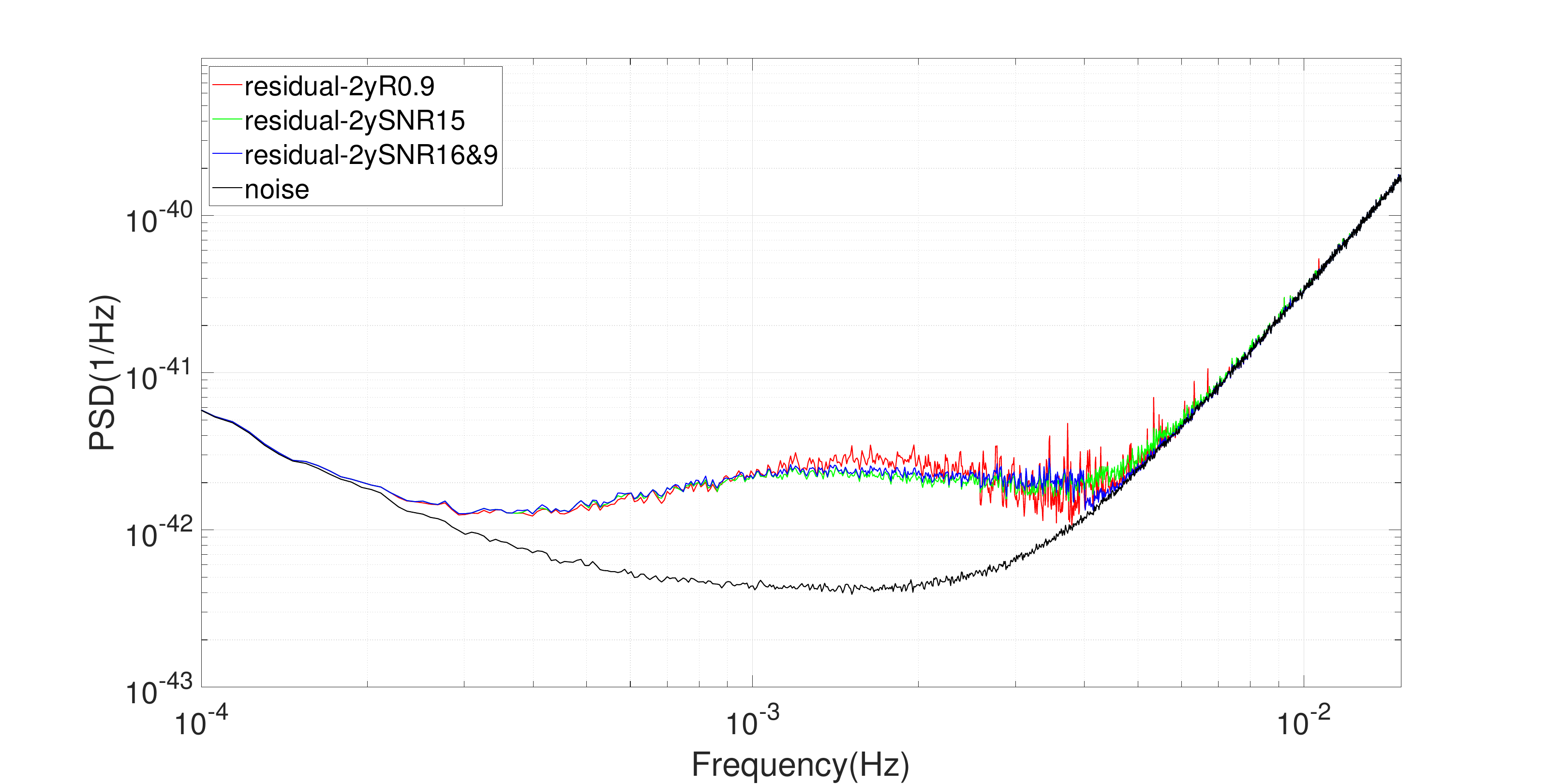}
		\caption{The red line represents residual from confirmed sources, the same as the 
			red line in Fig.~\ref{fig:p_residual_2years1}. The green line is the residual
			obtained from sources with SNR $\ge15$ in all frequency domains ($1\times10^{-4}$ 
			Hz$-1.5\times10^{-2}$ Hz). And the blue line is 
			the residual obtained from sources with SNR $\ge16$ in low and medium-frequency domains 
			($1\times10^{-4}$ Hz$-4\times10^{-3}$ Hz)
			together with SNR $\ge9$ in high-frequency domains ($4\times10^{-3}$ 
			Hz$-1.5\times10^{-2}$ Hz). The black line denotes LISA 
			instrumental noise PSD.}
		\label{fig:presidual2yearscompare}
	\end{figure}

	Fig.~\ref{fig:p_roc_1_1491} contains the data in all frequency domains ($1\times10^{-4}$ 
	Hz$-1.5\times10^{-2}$ Hz). 
	Taking the result of direct-search to the two-year data for an example (the red 
	line), when the SNR threshold $\eta=15$, the FPR is lower than 0.1, and the TPR is about 
	0.7. The ROC diagram in Fig.~\ref{fig:p_roc_1_391} corresponds the data in low and 
	medium-frequency domains ($1\times10^{-4}$ Hz$-4\times10^{-3}$ Hz). For the result of 
	direct-search to two-year data, when the SNR 
	threshold $\eta=16$, the FPR is lower than 0.1, and the TPR is about 0.63. 	
	Fig.~\ref{fig:p_roc_392_1491} is the ROC diagram corresponding to the data in high-frequency 
	domains ($4\times10^{-3}$ Hz$-1.5\times10^{-2}$ Hz). For the result of direct-search to 
	two-year data, when the SNR 
	threshold $\eta=9$, the FPR is lower than 0.1, and the TPR is about 0.98.
	As shown in Fig.~\ref{fig:presidual2yearscompare}, based on the result of direct-search to 
	the two-year data, the red line denotes the residual of the confirmed source, and the green 
	line is the residual of the source screened by SNR $\ge15$ throughout the frequency domain. 
	The	blue line represents the residual of the source screened by SNR $\ge16$ in the low and 
	medium-frequency domains and SNR $\ge 9$ in the high-frequency domains.
	There is a slight difference between the second type source (green line) and the third type 
	source (blue line) on the high-frequency domains. The blue line almost coincides with the 
	instrument noise curve, while the green line has many small bumps.	

	\section{Conclusions}
	\label{sec:conclusions}
	An accelerated search method is proposed depending on the early search results from the 
	previous LISA data.
	We demonstrate this approach with the GBSIEVER method.
	For the LDC1-4 two year data, this accelerate-search strategy obtains about half of the 
	confirmed sources within one-fortieth time of direct-search (Fig.~\ref{fig:p_time_number}), and 
	almost identical parameter estimation on intrinsic parameters ($f$, $\dot{f}$, $\beta$ and 
	$\lambda$)  (Fig.~\ref{fig:paerror}).
	
	To verify the DWD binaries' foreground noise effect when searching for other types of sources, 
	we calculate the SNR of the MBHB signal from LDC1-1 data ($m_1=2.8\times10^{5} M_{\odot}$, 
	$m_2=2.8\times10^{6} M_{\odot}$, $z=6.1$ and 1.33 years from the merger, 
	Fig.~\ref{fig:p_sensitivity}) in LISA detector TDI A 
	under different noise assumptions (Fig.~\ref{fig:p_residual1}). When only LISA instrument noise 
	is considered, the SNR is 135.7. For original LDC1-4 data without subtracting the signals, 
	the SNR is 99.8. In the case of direct-search and accelerate-search to the two-year data, the 
	SNR of MBHB is 121.3 and 112.0, respectively (Tab.~\ref{table:2}). 
	
	In the context of the iterative extraction algorithm, imprecise subtraction can result in 
	residual data contamination that could subsequently affect the accuracy of future searches. We 
	have observed that the degree of contamination tends to rise in proportion to the number of 
	subtraction sources and the reduction in data length (Fig.~\ref{fig:contimation}), and the SNR 
	of DWD (Tab.~\ref{table:4}) is more sensitive to different noise assumptions than that of MBHB. 
	Additionally, several questions remain unanswered, such as the potential impact of 
	contamination on the four intrinsic parameters. Our team intends to delve deeper into this 
	topic in future research endeavors.
	
	We discuss LISA's ability to explore the Milky Way structure by the detected DWD binaries 
	(Fig.~\ref{fig:p_results} and Fig.~\ref{fig:p_gala}). The distribution of selected
	injection sources (SNR $\ge3$ over two-year detection) reveals the most promising detected 
	area of the Milky Way by LISA (Fig.~\ref{fig:p_gala_all_sel0}). Fig.~\ref{fig:p_gala_2yR0.9} 
	and Fig.~\ref{fig:p_gala_3m_2yR0.9} present the confirmed sources distribution from 
	direct-search and accelerate-search to the two-year data, respectively.
	For confirmed sources (Fig.~\ref{fig:error}) in the high-frequency range 
	($f\ge4\times10^{-3}$ Hz), the latitude and longitude errors are $1.3^{\circ}$ and 
	$0.3^{\circ}$, respectively. Moreover, the relative error of chirp mass and distance can 
	reach about 20\%. It implies that detecting the DWD binary through LISA could be a 
	powerful tool for studying the Milky Way and could be complementary to 
	electromagnetic wave detection methods. 
	
	To prepare for the future real observational stage, we also propose an SNR detection threshold 
	for the blind search in the case of unknown injection parameter sets. 
	As we can see from Fig.~\ref{fig:p_roc_1_391} and Fig.~\ref{fig:p_roc_392_1491}, the SNR 
	threshold $\eta$ should be set to 16 for the sources with $f<4\times10^{-3}$ Hz and 9 for 
	those with $f\ge4\times10^{-3}$ Hz. Thus a low FAR and a high DB can be ensured simultaneously.

	\section*{Acknowledgments}
	We acknowledge valuable input from our anonymous referee.
	This work is supported by National Key R\&D Program of China (2020YFC2201400), and by the NSFC 
	(No.~11920101003, 	No.~11922303, No.~12021003 and No.~12005016) . 
	X. Fan. is supported by Hubei province Natural Science Fund for the Distinguished Young 
	Scholars 
	(2019CFA052). Z. Cao was supported by ``the Interdiscipline Research Funds of Beijing Normal 
	University" and CAS Project for Young Scientists in Basic Research YSBR-006.
	
	\bibliography{paper.bbl}

\begin{thebibliography}{49}%
\makeatletter
\providecommand \@ifxundefined [1]{%
 \@ifx{#1\undefined}
}%
\providecommand \@ifnum [1]{%
 \ifnum #1\expandafter \@firstoftwo
 \else \expandafter \@secondoftwo
 \fi
}%
\providecommand \@ifx [1]{%
 \ifx #1\expandafter \@firstoftwo
 \else \expandafter \@secondoftwo
 \fi
}%
\providecommand \natexlab [1]{#1}%
\providecommand \enquote  [1]{``#1''}%
\providecommand \bibnamefont  [1]{#1}%
\providecommand \bibfnamefont [1]{#1}%
\providecommand \citenamefont [1]{#1}%
\providecommand \href@noop [0]{\@secondoftwo}%
\providecommand \href [0]{\begingroup \@sanitize@url \@href}%
\providecommand \@href[1]{\@@startlink{#1}\@@href}%
\providecommand \@@href[1]{\endgroup#1\@@endlink}%
\providecommand \@sanitize@url [0]{\catcode `\\12\catcode `\$12\catcode
  `\&12\catcode `\#12\catcode `\^12\catcode `\_12\catcode `\%12\relax}%
\providecommand \@@startlink[1]{}%
\providecommand \@@endlink[0]{}%
\providecommand \url  [0]{\begingroup\@sanitize@url \@url }%
\providecommand \@url [1]{\endgroup\@href {#1}{\urlprefix }}%
\providecommand \urlprefix  [0]{URL }%
\providecommand \Eprint [0]{\href }%
\providecommand \doibase [0]{https://doi.org/}%
\providecommand \selectlanguage [0]{\@gobble}%
\providecommand \bibinfo  [0]{\@secondoftwo}%
\providecommand \bibfield  [0]{\@secondoftwo}%
\providecommand \translation [1]{[#1]}%
\providecommand \BibitemOpen [0]{}%
\providecommand \bibitemStop [0]{}%
\providecommand \bibitemNoStop [0]{.\EOS\space}%
\providecommand \EOS [0]{\spacefactor3000\relax}%
\providecommand \BibitemShut  [1]{\csname bibitem#1\endcsname}%
\let\auto@bib@innerbib\@empty
\bibitem [{\citenamefont {Abbott}\ \emph {et~al.}(2016)\citenamefont {Abbott},
  \citenamefont {Abbott}, \citenamefont {Abbott}, \citenamefont {Abernathy},
  \citenamefont {Acernese}, \citenamefont {Ackley}, \citenamefont {Adams},
  \citenamefont {Adams}, \citenamefont {Addesso}, \citenamefont {Adhikari},\
  and\ \citenamefont {Adya}}]{PhysRevLett.116.061102}%
  \BibitemOpen
  \bibfield  {author} {\bibinfo {author} {\bibfnamefont {B.~P.}\ \bibnamefont
  {Abbott}}, \bibinfo {author} {\bibfnamefont {R.}~\bibnamefont {Abbott}},
  \bibinfo {author} {\bibfnamefont {T.~D.}\ \bibnamefont {Abbott}}, \bibinfo
  {author} {\bibfnamefont {M.~R.}\ \bibnamefont {Abernathy}}, \bibinfo {author}
  {\bibfnamefont {F.}~\bibnamefont {Acernese}}, \bibinfo {author}
  {\bibfnamefont {K.}~\bibnamefont {Ackley}}, \bibinfo {author} {\bibfnamefont
  {C.}~\bibnamefont {Adams}}, \bibinfo {author} {\bibfnamefont
  {T.}~\bibnamefont {Adams}}, \bibinfo {author} {\bibfnamefont
  {P.}~\bibnamefont {Addesso}}, \bibinfo {author} {\bibfnamefont {R.~X.}\
  \bibnamefont {Adhikari}},\ and\ \bibinfo {author} {\bibfnamefont {V.~B.}\
  \bibnamefont {Adya}} (\bibinfo {collaboration} {LIGO Scientific Collaboration
  and Virgo Collaboration}),\ }\bibfield  {title} {\bibinfo {title}
  {Observation of gravitational waves from a binary black hole merger},\ }\href
  {https://doi.org/10.1103/PhysRevLett.116.061102} {\bibfield  {journal}
  {\bibinfo  {journal} {Phys. Rev. Lett.}\ }\textbf {\bibinfo {volume} {116}},\
  \bibinfo {pages} {061102} (\bibinfo {year} {2016})}\BibitemShut {NoStop}%
\bibitem [{noa(2017)}]{noauthor_lisa_nodate}%
  \BibitemOpen
  \href@noop {} {\bibinfo {title} {{LISA} {Laser} {Interferometer} {Space}
  {Antenna}.pdf}},\ \bibinfo {howpublished}
  {\url{https://arxiv.org/ftp/arxiv/papers/1702/1702.00786.pdf}} (\bibinfo
  {year} {2017})\BibitemShut {NoStop}%
\bibitem [{\citenamefont {Baker}\ \emph {et~al.}(2019)\citenamefont {Baker},
  \citenamefont {Bellovary}, \citenamefont {Bender} \emph
  {et~al.}}]{baker_laser_2019}%
  \BibitemOpen
  \bibfield  {author} {\bibinfo {author} {\bibfnamefont {J.}~\bibnamefont
  {Baker}}, \bibinfo {author} {\bibfnamefont {J.}~\bibnamefont {Bellovary}},
  \bibinfo {author} {\bibfnamefont {P.~L.}\ \bibnamefont {Bender}}, \emph
  {et~al.},\ }\href {http://arxiv.org/abs/1907.06482} {\bibinfo {title} {The
  {Laser} {Interferometer} {Space} {Antenna}: {Unveiling} the {Millihertz}
  {Gravitational} {Wave} {Sky}}} (\bibinfo {year} {2019}),\ \bibinfo {note}
  {arXiv:1907.06482 [astro-ph, physics:gr-qc]}\BibitemShut {NoStop}%
\bibitem [{\citenamefont {Gong}\ \emph {et~al.}(2011)\citenamefont {Gong},
  \citenamefont {Xu}, \citenamefont {Bai}, \citenamefont {Cao} \emph
  {et~al.}}]{gong_scientific_2011}%
  \BibitemOpen
  \bibfield  {author} {\bibinfo {author} {\bibfnamefont {X.}~\bibnamefont
  {Gong}}, \bibinfo {author} {\bibfnamefont {S.}~\bibnamefont {Xu}}, \bibinfo
  {author} {\bibfnamefont {S.}~\bibnamefont {Bai}}, \bibinfo {author}
  {\bibfnamefont {Z.}~\bibnamefont {Cao}}, \emph {et~al.},\ }\bibfield  {title}
  {\bibinfo {title} {A scientific case study of an advanced {LISA} mission},\
  }\href {https://doi.org/10.1088/0264-9381/28/9/094012} {\bibfield  {journal}
  {\bibinfo  {journal} {Classical and Quantum Gravity}\ }\textbf {\bibinfo
  {volume} {28}},\ \bibinfo {pages} {094012} (\bibinfo {year}
  {2011})}\BibitemShut {NoStop}%
\bibitem [{\citenamefont {{Ruan}}\ \emph {et~al.}(2020)\citenamefont {{Ruan}},
  \citenamefont {{Guo}}, \citenamefont {{Cai}},\ and\ \citenamefont
  {{Zhang}}}]{2020IJMPA..3550075R}%
  \BibitemOpen
  \bibfield  {author} {\bibinfo {author} {\bibfnamefont {W.-H.}\ \bibnamefont
  {{Ruan}}}, \bibinfo {author} {\bibfnamefont {Z.-K.}\ \bibnamefont {{Guo}}},
  \bibinfo {author} {\bibfnamefont {R.-G.}\ \bibnamefont {{Cai}}},\ and\
  \bibinfo {author} {\bibfnamefont {Y.-Z.}\ \bibnamefont {{Zhang}}},\
  }\bibfield  {title} {\bibinfo {title} {{Taiji program: Gravitational-wave
  sources}},\ }\href {https://doi.org/10.1142/S0217751X2050075X} {\bibfield
  {journal} {\bibinfo  {journal} {International Journal of Modern Physics A}\
  }\textbf {\bibinfo {volume} {35}},\ \bibinfo {eid} {2050075} (\bibinfo {year}
  {2020})}\BibitemShut {NoStop}%
\bibitem [{\citenamefont {Hu}\ and\ \citenamefont {Wu}(2017)}]{hu_taiji_2017}%
  \BibitemOpen
  \bibfield  {author} {\bibinfo {author} {\bibfnamefont {W.-R.}\ \bibnamefont
  {Hu}}\ and\ \bibinfo {author} {\bibfnamefont {Y.-L.}\ \bibnamefont {Wu}},\
  }\bibfield  {title} {\bibinfo {title} {The {Taiji} {Program} in {Space} for
  gravitational wave physics and the nature of gravity},\ }\href
  {https://doi.org/10.1093/nsr/nwx116} {\bibfield  {journal} {\bibinfo
  {journal} {National Science Review}\ }\textbf {\bibinfo {volume} {4}},\
  \bibinfo {pages} {685} (\bibinfo {year} {2017})}\BibitemShut {NoStop}%
\bibitem [{\citenamefont {{The Taiji Scientific Collaboration}}\ \emph
  {et~al.}(2021)\citenamefont {{The Taiji Scientific Collaboration}},
  \citenamefont {Wu}, \citenamefont {Luo}, \citenamefont {Wang} \emph
  {et~al.}}]{the_taiji_scientific_collaboration_chinas_2021}%
  \BibitemOpen
  \bibfield  {author} {\bibinfo {author} {\bibnamefont {{The Taiji Scientific
  Collaboration}}}, \bibinfo {author} {\bibfnamefont {Y.-L.}\ \bibnamefont
  {Wu}}, \bibinfo {author} {\bibfnamefont {Z.-R.}\ \bibnamefont {Luo}},
  \bibinfo {author} {\bibfnamefont {J.-Y.}\ \bibnamefont {Wang}}, \emph
  {et~al.},\ }\bibfield  {title} {\bibinfo {title} {China’s first step
  towards probing the expanding universe and the nature of gravity using a
  space borne gravitational wave antenna},\ }\href
  {https://doi.org/10.1038/s42005-021-00529-z} {\bibfield  {journal} {\bibinfo
  {journal} {Communications Physics}\ }\textbf {\bibinfo {volume} {4}},\
  \bibinfo {pages} {34} (\bibinfo {year} {2021})}\BibitemShut {NoStop}%
\bibitem [{\citenamefont {Luo}\ \emph {et~al.}(2016)\citenamefont {Luo},
  \citenamefont {Chen}, \citenamefont {Duan} \emph {et~al.}}]{Luo_2016}%
  \BibitemOpen
  \bibfield  {author} {\bibinfo {author} {\bibfnamefont {J.}~\bibnamefont
  {Luo}}, \bibinfo {author} {\bibfnamefont {L.-S.}\ \bibnamefont {Chen}},
  \bibinfo {author} {\bibfnamefont {H.-Z.}\ \bibnamefont {Duan}}, \emph
  {et~al.},\ }\bibfield  {title} {\bibinfo {title} {{TianQin}: a space-borne
  gravitational wave detector},\ }\href
  {https://doi.org/10.1088/0264-9381/33/3/035010} {\bibfield  {journal}
  {\bibinfo  {journal} {Classical and Quantum Gravity}\ }\textbf {\bibinfo
  {volume} {33}},\ \bibinfo {pages} {035010} (\bibinfo {year}
  {2016})}\BibitemShut {NoStop}%
\bibitem [{\citenamefont {Mei}\ \emph {et~al.}(2020)\citenamefont {Mei},
  \citenamefont {Bai}, \citenamefont {Bao} \emph {et~al.}}]{mei_tianqin_2020}%
  \BibitemOpen
  \bibfield  {author} {\bibinfo {author} {\bibfnamefont {J.}~\bibnamefont
  {Mei}}, \bibinfo {author} {\bibfnamefont {Y.-Z.}\ \bibnamefont {Bai}},
  \bibinfo {author} {\bibfnamefont {J.}~\bibnamefont {Bao}}, \emph {et~al.},\
  }\bibfield  {title} {\bibinfo {title} {The {TianQin} project: current
  progress on science and technology},\ }\bibfield  {journal} {\bibinfo
  {journal} {Progress of Theoretical and Experimental Physics}\ }\href
  {https://doi.org/10.1093/ptep/ptaa114} {10.1093/ptep/ptaa114} (\bibinfo
  {year} {2020}),\ \bibinfo {note} {arXiv: 2008.10332}\BibitemShut {NoStop}%
\bibitem [{\citenamefont {Luo}\ \emph {et~al.}(2020)\citenamefont {Luo},
  \citenamefont {Bai}, \citenamefont {Cai} \emph {et~al.}}]{luo_first_2020}%
  \BibitemOpen
  \bibfield  {author} {\bibinfo {author} {\bibfnamefont {J.}~\bibnamefont
  {Luo}}, \bibinfo {author} {\bibfnamefont {Y.-Z.}\ \bibnamefont {Bai}},
  \bibinfo {author} {\bibfnamefont {L.}~\bibnamefont {Cai}}, \emph {et~al.},\
  }\bibfield  {title} {\bibinfo {title} {The first round result from the
  {TianQin}-1 satellite},\ }\href {https://doi.org/10.1088/1361-6382/aba66a}
  {\bibfield  {journal} {\bibinfo  {journal} {Classical and Quantum Gravity}\
  }\textbf {\bibinfo {volume} {37}},\ \bibinfo {pages} {185013} (\bibinfo
  {year} {2020})}\BibitemShut {NoStop}%
\bibitem [{\citenamefont {Robson}\ \emph
  {et~al.}(2019{\natexlab{a}})\citenamefont {Robson}, \citenamefont {Cornish},\
  and\ \citenamefont {Liu}}]{Robson_2019}%
  \BibitemOpen
  \bibfield  {author} {\bibinfo {author} {\bibfnamefont {T.}~\bibnamefont
  {Robson}}, \bibinfo {author} {\bibfnamefont {N.~J.}\ \bibnamefont
  {Cornish}},\ and\ \bibinfo {author} {\bibfnamefont {C.}~\bibnamefont {Liu}},\
  }\bibfield  {title} {\bibinfo {title} {The construction and use of {LISA}
  sensitivity curves},\ }\href {https://doi.org/10.1088/1361-6382/ab1101}
  {\bibfield  {journal} {\bibinfo  {journal} {Classical and Quantum Gravity}\
  }\textbf {\bibinfo {volume} {36}},\ \bibinfo {pages} {105011} (\bibinfo
  {year} {2019}{\natexlab{a}})}\BibitemShut {NoStop}%
\bibitem [{\citenamefont {Cornish}\ and\ \citenamefont
  {Robson}(2017)}]{Cornish_2017}%
  \BibitemOpen
  \bibfield  {author} {\bibinfo {author} {\bibfnamefont {N.}~\bibnamefont
  {Cornish}}\ and\ \bibinfo {author} {\bibfnamefont {T.}~\bibnamefont
  {Robson}},\ }\bibfield  {title} {\bibinfo {title} {Galactic binary science
  with the new {LISA} design},\ }\href
  {https://doi.org/10.1088/1742-6596/840/1/012024} {\bibfield  {journal}
  {\bibinfo  {journal} {Journal of Physics: Conference Series}\ }\textbf
  {\bibinfo {volume} {840}},\ \bibinfo {pages} {012024} (\bibinfo {year}
  {2017})}\BibitemShut {NoStop}%
\bibitem [{\citenamefont {Arnaud}\ \emph {et~al.}(2006)\citenamefont {Arnaud},
  \citenamefont {Babak}, \citenamefont {Baker}, \citenamefont {Benacquista},
  \citenamefont {Cornish}, \citenamefont {Cutler}, \citenamefont {Larson},
  \citenamefont {Sathyaprakash}, \citenamefont {Vallisneri}, \citenamefont
  {Vecchio},\ and\ \citenamefont {Vinet}}]{arnaud_mock_2006}%
  \BibitemOpen
  \bibfield  {author} {\bibinfo {author} {\bibfnamefont {K.~A.}\ \bibnamefont
  {Arnaud}}, \bibinfo {author} {\bibfnamefont {S.}~\bibnamefont {Babak}},
  \bibinfo {author} {\bibfnamefont {J.~G.}\ \bibnamefont {Baker}}, \bibinfo
  {author} {\bibfnamefont {M.~J.}\ \bibnamefont {Benacquista}}, \bibinfo
  {author} {\bibfnamefont {N.~J.}\ \bibnamefont {Cornish}}, \bibinfo {author}
  {\bibfnamefont {C.}~\bibnamefont {Cutler}}, \bibinfo {author} {\bibfnamefont
  {S.~L.}\ \bibnamefont {Larson}}, \bibinfo {author} {\bibfnamefont {B.~S.}\
  \bibnamefont {Sathyaprakash}}, \bibinfo {author} {\bibfnamefont
  {M.}~\bibnamefont {Vallisneri}}, \bibinfo {author} {\bibfnamefont
  {A.}~\bibnamefont {Vecchio}},\ and\ \bibinfo {author} {\bibfnamefont {J.-Y.}\
  \bibnamefont {Vinet}},\ }\bibfield  {title} {\bibinfo {title} {The {Mock}
  {LISA} {Data} {Challenges}: {An} overview},\ }\href
  {https://doi.org/10.1063/1.2405108} {\bibfield  {journal} {\bibinfo
  {journal} {AIP Conference Proceedings}\ }\textbf {\bibinfo {volume} {873}},\
  \bibinfo {pages} {619} (\bibinfo {year} {2006})},\ \bibinfo {note} {arXiv:
  gr-qc/0609105}\BibitemShut {NoStop}%
\bibitem [{\citenamefont {Arnaud}\ \emph {et~al.}(2007)\citenamefont {Arnaud},
  \citenamefont {Babak}, \citenamefont {Baker}, \citenamefont {Benacquista},
  \citenamefont {Cornish}, \citenamefont {Cutler}, \citenamefont {Finn},
  \citenamefont {Larson}, \citenamefont {Littenberg}, \citenamefont {Porter},
  \citenamefont {Vallisneri}, \citenamefont {Vecchio},\ and\ \citenamefont
  {Vinet}}]{arnaud_overview_2007}%
  \BibitemOpen
  \bibfield  {author} {\bibinfo {author} {\bibfnamefont {K.~A.}\ \bibnamefont
  {Arnaud}}, \bibinfo {author} {\bibfnamefont {S.}~\bibnamefont {Babak}},
  \bibinfo {author} {\bibfnamefont {J.~G.}\ \bibnamefont {Baker}}, \bibinfo
  {author} {\bibfnamefont {M.~J.}\ \bibnamefont {Benacquista}}, \bibinfo
  {author} {\bibfnamefont {N.~J.}\ \bibnamefont {Cornish}}, \bibinfo {author}
  {\bibfnamefont {C.}~\bibnamefont {Cutler}}, \bibinfo {author} {\bibfnamefont
  {L.~S.}\ \bibnamefont {Finn}}, \bibinfo {author} {\bibfnamefont {S.~L.}\
  \bibnamefont {Larson}}, \bibinfo {author} {\bibfnamefont {T.}~\bibnamefont
  {Littenberg}}, \bibinfo {author} {\bibfnamefont {E.~K.}\ \bibnamefont
  {Porter}}, \bibinfo {author} {\bibfnamefont {M.}~\bibnamefont {Vallisneri}},
  \bibinfo {author} {\bibfnamefont {A.}~\bibnamefont {Vecchio}},\ and\ \bibinfo
  {author} {\bibfnamefont {J.-Y.}\ \bibnamefont {Vinet}},\ }\bibfield  {title}
  {\bibinfo {title} {An overview of the second round of the {Mock} {LISA}
  {Data} {Challenges}},\ }\href {https://doi.org/10.1088/0264-9381/24/19/S18}
  {\bibfield  {journal} {\bibinfo  {journal} {Classical and Quantum Gravity}\
  }\textbf {\bibinfo {volume} {24}},\ \bibinfo {pages} {S551} (\bibinfo {year}
  {2007})},\ \bibinfo {note} {arXiv: gr-qc/0701170}\BibitemShut {NoStop}%
\bibitem [{\citenamefont {Babak}\ \emph {et~al.}(2008)\citenamefont {Babak},
  \citenamefont {Baker}, \citenamefont {Benacquista}, \citenamefont {Cornish},
  \citenamefont {Crowder} \emph {et~al.}}]{Babak_2008}%
  \BibitemOpen
  \bibfield  {author} {\bibinfo {author} {\bibfnamefont {S.}~\bibnamefont
  {Babak}}, \bibinfo {author} {\bibfnamefont {J.~G.}\ \bibnamefont {Baker}},
  \bibinfo {author} {\bibfnamefont {M.~J.}\ \bibnamefont {Benacquista}},
  \bibinfo {author} {\bibfnamefont {N.~J.}\ \bibnamefont {Cornish}}, \bibinfo
  {author} {\bibfnamefont {J.}~\bibnamefont {Crowder}}, \emph {et~al.},\
  }\bibfield  {title} {\bibinfo {title} {Report on the second mock {LISA} data
  challenge},\ }\href {https://doi.org/10.1088/0264-9381/25/11/114037}
  {\bibfield  {journal} {\bibinfo  {journal} {Classical and Quantum Gravity}\
  }\textbf {\bibinfo {volume} {25}},\ \bibinfo {pages} {114037} (\bibinfo
  {year} {2008})}\BibitemShut {NoStop}%
\bibitem [{\citenamefont {Babak}\ \emph {et~al.}(2010)\citenamefont {Babak},
  \citenamefont {Baker}, \citenamefont {Benacquista}, \citenamefont {Cornish},
  \citenamefont {Larson} \emph {et~al.}}]{babak_mock_2010}%
  \BibitemOpen
  \bibfield  {author} {\bibinfo {author} {\bibfnamefont {S.}~\bibnamefont
  {Babak}}, \bibinfo {author} {\bibfnamefont {J.~G.}\ \bibnamefont {Baker}},
  \bibinfo {author} {\bibfnamefont {M.~J.}\ \bibnamefont {Benacquista}},
  \bibinfo {author} {\bibfnamefont {N.~J.}\ \bibnamefont {Cornish}}, \bibinfo
  {author} {\bibfnamefont {S.~L.}\ \bibnamefont {Larson}}, \emph {et~al.},\
  }\bibfield  {title} {\bibinfo {title} {The {Mock} {LISA} {Data} {Challenges}:
  from challenge 3 to challenge 4},\ }\href
  {https://doi.org/10.1088/0264-9381/27/8/084009} {\bibfield  {journal}
  {\bibinfo  {journal} {Classical and Quantum Gravity}\ }\textbf {\bibinfo
  {volume} {27}},\ \bibinfo {pages} {084009} (\bibinfo {year}
  {2010})}\BibitemShut {NoStop}%
\bibitem [{noa(2020)}]{noauthor_ldc-manual-sangriapdf_nodate}%
  \BibitemOpen
  \href@noop {} {\bibinfo {title} {{LDC}-manual-{Sangria}.pdf}},\ \bibinfo
  {howpublished} {\url{https://lisa-ldc.lal.in2p3.fr/}} (\bibinfo {year}
  {2020})\BibitemShut {NoStop}%
\bibitem [{\citenamefont {Mohanty}\ and\ \citenamefont
  {Nayak}(2006)}]{mohanty_tomographic_2006}%
  \BibitemOpen
  \bibfield  {author} {\bibinfo {author} {\bibfnamefont {S.~D.}\ \bibnamefont
  {Mohanty}}\ and\ \bibinfo {author} {\bibfnamefont {R.~K.}\ \bibnamefont
  {Nayak}},\ }\bibfield  {title} {\bibinfo {title} {Tomographic approach to
  resolving the distribution of {LISA} {Galactic} binaries},\ }\href
  {https://doi.org/10.1103/PhysRevD.73.083006} {\bibfield  {journal} {\bibinfo
  {journal} {Physical Review D}\ }\textbf {\bibinfo {volume} {73}},\ \bibinfo
  {pages} {083006} (\bibinfo {year} {2006})}\BibitemShut {NoStop}%
\bibitem [{\citenamefont {Stroeer}\ \emph {et~al.}(2007)\citenamefont
  {Stroeer}, \citenamefont {Veitch}, \citenamefont {Roever} \emph
  {et~al.}}]{stroeer_inference_2007}%
  \BibitemOpen
  \bibfield  {author} {\bibinfo {author} {\bibfnamefont {A.}~\bibnamefont
  {Stroeer}}, \bibinfo {author} {\bibfnamefont {J.}~\bibnamefont {Veitch}},
  \bibinfo {author} {\bibnamefont {Roever}}, \emph {et~al.},\ }\bibfield
  {title} {\bibinfo {title} {Inference on white dwarf binary systems using the
  first round {Mock} {LISA} {Data} {Challenges} data sets},\ }\href
  {https://doi.org/10.1088/0264-9381/24/19/S17} {\bibfield  {journal} {\bibinfo
   {journal} {Classical and Quantum Gravity}\ }\textbf {\bibinfo {volume}
  {24}},\ \bibinfo {pages} {S541} (\bibinfo {year} {2007})},\ \bibinfo {note}
  {arXiv: 0704.0048}\BibitemShut {NoStop}%
\bibitem [{\citenamefont {Crowder}\ and\ \citenamefont
  {Cornish}(2007)}]{crowder_extracting_2007-1}%
  \BibitemOpen
  \bibfield  {author} {\bibinfo {author} {\bibfnamefont {J.}~\bibnamefont
  {Crowder}}\ and\ \bibinfo {author} {\bibfnamefont {N.~J.}\ \bibnamefont
  {Cornish}},\ }\bibfield  {title} {\bibinfo {title} {Extracting galactic
  binary signals from the first round of {Mock} {LISA} {Data} {Challenges}},\
  }\href {https://doi.org/10.1088/0264-9381/24/19/S20} {\bibfield  {journal}
  {\bibinfo  {journal} {Classical and Quantum Gravity}\ }\textbf {\bibinfo
  {volume} {24}},\ \bibinfo {pages} {S575} (\bibinfo {year}
  {2007})}\BibitemShut {NoStop}%
\bibitem [{\citenamefont {Prix}\ and\ \citenamefont
  {Whelan}(2007)}]{prix_f-statistic_2007}%
  \BibitemOpen
  \bibfield  {author} {\bibinfo {author} {\bibfnamefont {R.}~\bibnamefont
  {Prix}}\ and\ \bibinfo {author} {\bibfnamefont {J.~T.}\ \bibnamefont
  {Whelan}},\ }\bibfield  {title} {\bibinfo {title} {F-statistic search for
  white-dwarf binaries in the first {Mock} {LISA} {Data} {Challenge}},\ }\href
  {https://doi.org/10.1088/0264-9381/24/19/S19} {\bibfield  {journal} {\bibinfo
   {journal} {Classical and Quantum Gravity}\ }\textbf {\bibinfo {volume}
  {24}},\ \bibinfo {pages} {S565} (\bibinfo {year} {2007})},\ \bibinfo {note}
  {arXiv: 0707.0128}\BibitemShut {NoStop}%
\bibitem [{\citenamefont {Zhang}\ \emph {et~al.}(2021)\citenamefont {Zhang},
  \citenamefont {Mohanty}, \citenamefont {Zou},\ and\ \citenamefont
  {Liu}}]{zhang_resolving_2021-1}%
  \BibitemOpen
  \bibfield  {author} {\bibinfo {author} {\bibfnamefont {X.-H.}\ \bibnamefont
  {Zhang}}, \bibinfo {author} {\bibfnamefont {S.~D.}\ \bibnamefont {Mohanty}},
  \bibinfo {author} {\bibfnamefont {X.-B.}\ \bibnamefont {Zou}},\ and\ \bibinfo
  {author} {\bibfnamefont {Y.-X.}\ \bibnamefont {Liu}},\ }\bibfield  {title}
  {\bibinfo {title} {Resolving {Galactic} binaries in {LISA} data using
  particle swarm optimization and cross-validation},\ }\href
  {https://doi.org/10.1103/PhysRevD.104.024023} {\bibfield  {journal} {\bibinfo
   {journal} {Physical Review D}\ }\textbf {\bibinfo {volume} {104}},\ \bibinfo
  {pages} {024023} (\bibinfo {year} {2021})}\BibitemShut {NoStop}%
\bibitem [{\citenamefont {Littenberg}(2011)}]{littenberg_detection_2011}%
  \BibitemOpen
  \bibfield  {author} {\bibinfo {author} {\bibfnamefont {T.~B.}\ \bibnamefont
  {Littenberg}},\ }\bibfield  {title} {\bibinfo {title} {Detection pipeline for
  {Galactic} binaries in {LISA} data},\ }\href
  {https://doi.org/10.1103/PhysRevD.84.063009} {\bibfield  {journal} {\bibinfo
  {journal} {Physical Review D}\ }\textbf {\bibinfo {volume} {84}},\ \bibinfo
  {pages} {063009} (\bibinfo {year} {2011})}\BibitemShut {NoStop}%
\bibitem [{\citenamefont {Błaut}\ \emph {et~al.}(2010)\citenamefont {Błaut},
  \citenamefont {Babak},\ and\ \citenamefont {Królak}}]{blaut_mock_2010}%
  \BibitemOpen
  \bibfield  {author} {\bibinfo {author} {\bibfnamefont {A.}~\bibnamefont
  {Błaut}}, \bibinfo {author} {\bibfnamefont {S.}~\bibnamefont {Babak}},\ and\
  \bibinfo {author} {\bibfnamefont {A.}~\bibnamefont {Królak}},\ }\bibfield
  {title} {\bibinfo {title} {Mock {LISA} data challenge for the {Galactic}
  white dwarf binaries},\ }\href {https://doi.org/10.1103/PhysRevD.81.063008}
  {\bibfield  {journal} {\bibinfo  {journal} {Physical Review D}\ }\textbf
  {\bibinfo {volume} {81}},\ \bibinfo {pages} {063008} (\bibinfo {year}
  {2010})}\BibitemShut {NoStop}%
\bibitem [{\citenamefont {Littenberg}\ \emph {et~al.}(2020)\citenamefont
  {Littenberg}, \citenamefont {Cornish}, \citenamefont {Lackeos},\ and\
  \citenamefont {Robson}}]{littenberg_global_2020}%
  \BibitemOpen
  \bibfield  {author} {\bibinfo {author} {\bibfnamefont {T.~B.}\ \bibnamefont
  {Littenberg}}, \bibinfo {author} {\bibfnamefont {N.~J.}\ \bibnamefont
  {Cornish}}, \bibinfo {author} {\bibfnamefont {K.}~\bibnamefont {Lackeos}},\
  and\ \bibinfo {author} {\bibfnamefont {T.}~\bibnamefont {Robson}},\
  }\bibfield  {title} {\bibinfo {title} {Global analysis of the gravitational
  wave signal from {Galactic} binaries},\ }\href
  {https://doi.org/10.1103/PhysRevD.101.123021} {\bibfield  {journal} {\bibinfo
   {journal} {Physical Review D}\ }\textbf {\bibinfo {volume} {101}},\ \bibinfo
  {pages} {123021} (\bibinfo {year} {2020})}\BibitemShut {NoStop}%
\bibitem [{\citenamefont {Strub}\ \emph {et~al.}(2022)\citenamefont {Strub},
  \citenamefont {Ferraioli}, \citenamefont {Schmelzbach}, \citenamefont
  {Stähler},\ and\ \citenamefont {Giardini}}]{strub_bayesian_2022}%
  \BibitemOpen
  \bibfield  {author} {\bibinfo {author} {\bibfnamefont {S.~H.}\ \bibnamefont
  {Strub}}, \bibinfo {author} {\bibfnamefont {L.}~\bibnamefont {Ferraioli}},
  \bibinfo {author} {\bibfnamefont {C.}~\bibnamefont {Schmelzbach}}, \bibinfo
  {author} {\bibfnamefont {S.~C.}\ \bibnamefont {Stähler}},\ and\ \bibinfo
  {author} {\bibfnamefont {D.}~\bibnamefont {Giardini}},\ }\bibfield  {title}
  {\bibinfo {title} {Bayesian parameter-estimation of {Galactic} binaries in
  {LISA} data with {Gaussian} {Process} {Regression}},\ }\href
  {http://arxiv.org/abs/2204.04467} {\bibfield  {journal} {\bibinfo  {journal}
  {arXiv:2204.04467 [astro-ph, physics:gr-qc, physics:physics]}\ } (\bibinfo
  {year} {2022})},\ \bibinfo {note} {arXiv: 2204.04467}\BibitemShut {NoStop}%
\bibitem [{\citenamefont {Lu}\ \emph {et~al.}(2023)\citenamefont {Lu},
  \citenamefont {Li}, \citenamefont {Hu}, \citenamefont {dong Zhang},\ and\
  \citenamefont {Mei}}]{Lu_2023}%
  \BibitemOpen
  \bibfield  {author} {\bibinfo {author} {\bibfnamefont {Y.}~\bibnamefont
  {Lu}}, \bibinfo {author} {\bibfnamefont {E.-K.}\ \bibnamefont {Li}}, \bibinfo
  {author} {\bibfnamefont {Y.-M.}\ \bibnamefont {Hu}}, \bibinfo {author}
  {\bibfnamefont {J.}~\bibnamefont {dong Zhang}},\ and\ \bibinfo {author}
  {\bibfnamefont {J.}~\bibnamefont {Mei}},\ }\bibfield  {title} {\bibinfo
  {title} {An implementation of galactic white dwarf binary data analysis for
  mldc-3.1},\ }\href {https://doi.org/10.1088/1674-4527/aca8ed} {\bibfield
  {journal} {\bibinfo  {journal} {Research in Astronomy and Astrophysics}\
  }\textbf {\bibinfo {volume} {23}},\ \bibinfo {pages} {015022} (\bibinfo
  {year} {2023})}\BibitemShut {NoStop}%
\bibitem [{\citenamefont {Zhang}\ \emph {et~al.}(2022)\citenamefont {Zhang},
  \citenamefont {Zhao}, \citenamefont {Mohanty},\ and\ \citenamefont
  {Liu}}]{zhang_resolving_2022}%
  \BibitemOpen
  \bibfield  {author} {\bibinfo {author} {\bibfnamefont {X.-H.}\ \bibnamefont
  {Zhang}}, \bibinfo {author} {\bibfnamefont {S.-D.}\ \bibnamefont {Zhao}},
  \bibinfo {author} {\bibfnamefont {S.~D.}\ \bibnamefont {Mohanty}},\ and\
  \bibinfo {author} {\bibfnamefont {Y.-X.}\ \bibnamefont {Liu}},\ }\href
  {http://arxiv.org/abs/2206.12083} {\bibinfo {title} {Resolving {Galactic}
  binaries using a network of space-borne gravitational wave detectors}}
  (\bibinfo {year} {2022}),\ \bibinfo {note} {arXiv:2206.12083
  [gr-qc]}\BibitemShut {NoStop}%
\bibitem [{\citenamefont {Han}\ \emph {et~al.}(2020)\citenamefont {Han},
  \citenamefont {Ge}, \citenamefont {Chen},\ and\ \citenamefont
  {Chen}}]{han_binary_2020}%
  \BibitemOpen
  \bibfield  {author} {\bibinfo {author} {\bibfnamefont {Z.-W.}\ \bibnamefont
  {Han}}, \bibinfo {author} {\bibfnamefont {H.-W.}\ \bibnamefont {Ge}},
  \bibinfo {author} {\bibfnamefont {X.-F.}\ \bibnamefont {Chen}},\ and\
  \bibinfo {author} {\bibfnamefont {H.-L.}\ \bibnamefont {Chen}},\ }\bibfield
  {title} {\bibinfo {title} {Binary {Population} {Synthesis}},\ }\href
  {https://doi.org/10.1088/1674-4527/20/10/161} {\bibfield  {journal} {\bibinfo
   {journal} {Research in Astronomy and Astrophysics}\ }\textbf {\bibinfo
  {volume} {20}},\ \bibinfo {pages} {161} (\bibinfo {year} {2020})}\BibitemShut
  {NoStop}%
\bibitem [{\citenamefont {Postnov}\ and\ \citenamefont
  {Yungelson}(2014)}]{postnov_evolution_2014}%
  \BibitemOpen
  \bibfield  {author} {\bibinfo {author} {\bibfnamefont {K.~A.}\ \bibnamefont
  {Postnov}}\ and\ \bibinfo {author} {\bibfnamefont {L.~R.}\ \bibnamefont
  {Yungelson}},\ }\bibfield  {title} {\bibinfo {title} {The {Evolution} of
  {Compact} {Binary} {Star} {Systems}},\ }\href
  {https://doi.org/10.12942/lrr-2014-3} {\bibfield  {journal} {\bibinfo
  {journal} {Living Reviews in Relativity}\ }\textbf {\bibinfo {volume} {17}},\
  \bibinfo {pages} {3} (\bibinfo {year} {2014})}\BibitemShut {NoStop}%
\bibitem [{\citenamefont {Marsh}\ \emph {et~al.}(2004)\citenamefont {Marsh},
  \citenamefont {Nelemans},\ and\ \citenamefont {Steeghs}}]{marsh_mass_2004}%
  \BibitemOpen
  \bibfield  {author} {\bibinfo {author} {\bibfnamefont {T.~R.}\ \bibnamefont
  {Marsh}}, \bibinfo {author} {\bibfnamefont {G.}~\bibnamefont {Nelemans}},\
  and\ \bibinfo {author} {\bibfnamefont {D.}~\bibnamefont {Steeghs}},\
  }\bibfield  {title} {\bibinfo {title} {Mass {Transfer} between {Double}
  {White} {Dwarfs}},\ }\href {https://doi.org/10.1111/j.1365-2966.2004.07564.x}
  {\bibfield  {journal} {\bibinfo  {journal} {Monthly Notices of the Royal
  Astronomical Society}\ }\textbf {\bibinfo {volume} {350}},\ \bibinfo {pages}
  {113} (\bibinfo {year} {2004})},\ \bibinfo {note}
  {arXiv:astro-ph/0312577}\BibitemShut {NoStop}%
\bibitem [{\citenamefont {Sberna}\ \emph {et~al.}(2021)\citenamefont {Sberna},
  \citenamefont {Toubiana},\ and\ \citenamefont {Miller}}]{Sberna_2021}%
  \BibitemOpen
  \bibfield  {author} {\bibinfo {author} {\bibfnamefont {L.}~\bibnamefont
  {Sberna}}, \bibinfo {author} {\bibfnamefont {A.}~\bibnamefont {Toubiana}},\
  and\ \bibinfo {author} {\bibfnamefont {M.~C.}\ \bibnamefont {Miller}},\
  }\bibfield  {title} {\bibinfo {title} {Golden galactic binaries for {LISA}:
  Mass-transferring white dwarf black hole binaries},\ }\href
  {https://doi.org/10.3847/1538-4357/abccc7} {\bibfield  {journal} {\bibinfo
  {journal} {The Astrophysical Journal}\ }\textbf {\bibinfo {volume} {908}},\
  \bibinfo {pages} {1} (\bibinfo {year} {2021})}\BibitemShut {NoStop}%
\bibitem [{\citenamefont {Stroeer}\ and\ \citenamefont
  {Vecchio}(2006)}]{noauthor_lisa_nodate1}%
  \BibitemOpen
  \bibfield  {author} {\bibinfo {author} {\bibfnamefont {A.}~\bibnamefont
  {Stroeer}}\ and\ \bibinfo {author} {\bibfnamefont {A.}~\bibnamefont
  {Vecchio}},\ }\bibfield  {title} {\bibinfo {title} {The {LISA} verification
  binaries},\ }\href {https://doi.org/10.1088/0264-9381/23/19/s19} {\bibfield
  {journal} {\bibinfo  {journal} {Classical and Quantum Gravity}\ }\textbf
  {\bibinfo {volume} {23}},\ \bibinfo {pages} {S809} (\bibinfo {year}
  {2006})}\BibitemShut {NoStop}%
\bibitem [{\citenamefont {Martens}\ and\ \citenamefont
  {Joffre}(2021)}]{martens_trajectory_2021}%
  \BibitemOpen
  \bibfield  {author} {\bibinfo {author} {\bibfnamefont {W.}~\bibnamefont
  {Martens}}\ and\ \bibinfo {author} {\bibfnamefont {E.}~\bibnamefont
  {Joffre}},\ }\bibfield  {title} {\bibinfo {title} {Trajectory {Design} for
  the {ESA} {LISA} {Mission}},\ }\href {http://arxiv.org/abs/2101.03040}
  {\bibfield  {journal} {\bibinfo  {journal} {arXiv:2101.03040 [gr-qc]}\ }
  (\bibinfo {year} {2021})},\ \bibinfo {note} {arXiv: 2101.03040}\BibitemShut
  {NoStop}%
\bibitem [{\citenamefont {Armstrong}\ \emph {et~al.}(1999)\citenamefont
  {Armstrong}, \citenamefont {Estabrook},\ and\ \citenamefont
  {Tinto}}]{armstrong_timedelay_1999}%
  \BibitemOpen
  \bibfield  {author} {\bibinfo {author} {\bibfnamefont {J.~W.}\ \bibnamefont
  {Armstrong}}, \bibinfo {author} {\bibfnamefont {F.~B.}\ \bibnamefont
  {Estabrook}},\ and\ \bibinfo {author} {\bibfnamefont {M.}~\bibnamefont
  {Tinto}},\ }\bibfield  {title} {\bibinfo {title} {Time‐{Delay}
  {Interferometry} for {Space}‐based {Gravitational} {Wave} {Searches}},\
  }\href {https://doi.org/10.1086/308110} {\bibfield  {journal} {\bibinfo
  {journal} {The Astrophysical Journal}\ }\textbf {\bibinfo {volume} {527}},\
  \bibinfo {pages} {814} (\bibinfo {year} {1999})}\BibitemShut {NoStop}%
\bibitem [{\citenamefont {Tinto}\ and\ \citenamefont
  {Dhurandhar}(2014)}]{tinto_time-delay_2014}%
  \BibitemOpen
  \bibfield  {author} {\bibinfo {author} {\bibfnamefont {M.}~\bibnamefont
  {Tinto}}\ and\ \bibinfo {author} {\bibfnamefont {S.~V.}\ \bibnamefont
  {Dhurandhar}},\ }\bibfield  {title} {\bibinfo {title} {Time-{Delay}
  {Interferometry}},\ }\href {https://doi.org/10.12942/lrr-2014-6} {\bibfield
  {journal} {\bibinfo  {journal} {Living Reviews in Relativity}\ }\textbf
  {\bibinfo {volume} {17}},\ \bibinfo {pages} {6} (\bibinfo {year}
  {2014})}\BibitemShut {NoStop}%
\bibitem [{\citenamefont {Tinto}\ \emph {et~al.}(2004)\citenamefont {Tinto},
  \citenamefont {Estabrook},\ and\ \citenamefont
  {Armstrong}}]{tinto_time_2004}%
  \BibitemOpen
  \bibfield  {author} {\bibinfo {author} {\bibfnamefont {M.}~\bibnamefont
  {Tinto}}, \bibinfo {author} {\bibfnamefont {F.~B.}\ \bibnamefont
  {Estabrook}},\ and\ \bibinfo {author} {\bibfnamefont {J.~W.}\ \bibnamefont
  {Armstrong}},\ }\bibfield  {title} {\bibinfo {title} {Time delay
  interferometry with moving spacecraft arrays},\ }\href
  {https://doi.org/10.1103/PhysRevD.69.082001} {\bibfield  {journal} {\bibinfo
  {journal} {Physical Review D}\ }\textbf {\bibinfo {volume} {69}},\ \bibinfo
  {pages} {082001} (\bibinfo {year} {2004})}\BibitemShut {NoStop}%
\bibitem [{\citenamefont {Jaranowski}\ \emph {et~al.}(1998)\citenamefont
  {Jaranowski}, \citenamefont {Królak},\ and\ \citenamefont
  {Schutz}}]{jaranowski_data_1998}%
  \BibitemOpen
  \bibfield  {author} {\bibinfo {author} {\bibfnamefont {P.}~\bibnamefont
  {Jaranowski}}, \bibinfo {author} {\bibfnamefont {A.}~\bibnamefont
  {Królak}},\ and\ \bibinfo {author} {\bibfnamefont {B.~F.}\ \bibnamefont
  {Schutz}},\ }\bibfield  {title} {\bibinfo {title} {Data analysis of
  gravitational-wave signals from spinning neutron stars: {The} signal and its
  detection},\ }\href {https://doi.org/10.1103/PhysRevD.58.063001} {\bibfield
  {journal} {\bibinfo  {journal} {Physical Review D}\ }\textbf {\bibinfo
  {volume} {58}},\ \bibinfo {pages} {063001} (\bibinfo {year}
  {1998})}\BibitemShut {NoStop}%
\bibitem [{\citenamefont {Petiteau}\ \emph {et~al.}(2008)\citenamefont
  {Petiteau}, \citenamefont {Auger}, \citenamefont {Halloin}, \citenamefont
  {Jeannin}, \citenamefont {Plagnol}, \citenamefont {Pireaux}, \citenamefont
  {Regimbau},\ and\ \citenamefont {Vinet}}]{petiteau_lisacode_2008}%
  \BibitemOpen
  \bibfield  {author} {\bibinfo {author} {\bibfnamefont {A.}~\bibnamefont
  {Petiteau}}, \bibinfo {author} {\bibfnamefont {G.}~\bibnamefont {Auger}},
  \bibinfo {author} {\bibfnamefont {H.}~\bibnamefont {Halloin}}, \bibinfo
  {author} {\bibfnamefont {O.}~\bibnamefont {Jeannin}}, \bibinfo {author}
  {\bibfnamefont {E.}~\bibnamefont {Plagnol}}, \bibinfo {author} {\bibfnamefont
  {S.}~\bibnamefont {Pireaux}}, \bibinfo {author} {\bibfnamefont
  {T.}~\bibnamefont {Regimbau}},\ and\ \bibinfo {author} {\bibfnamefont
  {J.-Y.}\ \bibnamefont {Vinet}},\ }\bibfield  {title} {\bibinfo {title}
  {{LISACode}: {A} scientific simulator of {LISA}},\ }\href
  {https://doi.org/10.1103/PhysRevD.77.023002} {\bibfield  {journal} {\bibinfo
  {journal} {Physical Review D}\ }\textbf {\bibinfo {volume} {77}},\ \bibinfo
  {pages} {023002} (\bibinfo {year} {2008})}\BibitemShut {NoStop}%
\bibitem [{\citenamefont {Kennedy'}\ and\ \citenamefont
  {Eberhart}(1995)}]{kennedy_particle_nodate}%
  \BibitemOpen
  \bibfield  {author} {\bibinfo {author} {\bibfnamefont {J.}~\bibnamefont
  {Kennedy'}}\ and\ \bibinfo {author} {\bibfnamefont {R.}~\bibnamefont
  {Eberhart}},\ }\bibfield  {title} {\bibinfo {title} {Particle {Swarm}
  {Optimization}},\ }\href@noop {} {\  (\bibinfo {year} {1995})}\BibitemShut
  {NoStop}%
\bibitem [{Note1()}]{Note1}%
  \BibitemOpen
  \bibinfo {note} {Note that, since there are always DWD GW singals in the
  data, this detection definition is not the same as the transient signal
  search.}\BibitemShut {Stop}%
\bibitem [{\citenamefont {Robson}\ \emph
  {et~al.}(2019{\natexlab{b}})\citenamefont {Robson}, \citenamefont {Cornish},\
  and\ \citenamefont {Liu}}]{robson_construction_2019}%
  \BibitemOpen
  \bibfield  {author} {\bibinfo {author} {\bibfnamefont {T.}~\bibnamefont
  {Robson}}, \bibinfo {author} {\bibfnamefont {N.~J.}\ \bibnamefont
  {Cornish}},\ and\ \bibinfo {author} {\bibfnamefont {C.}~\bibnamefont {Liu}},\
  }\bibfield  {title} {\bibinfo {title} {The construction and use of {LISA}
  sensitivity curves},\ }\href {https://doi.org/10.1088/1361-6382/ab1101}
  {\bibfield  {journal} {\bibinfo  {journal} {Classical and Quantum Gravity}\
  }\textbf {\bibinfo {volume} {36}},\ \bibinfo {pages} {105011} (\bibinfo
  {year} {2019}{\natexlab{b}})}\BibitemShut {NoStop}%
\bibitem [{\citenamefont {Moore}\ \emph {et~al.}(2015)\citenamefont {Moore},
  \citenamefont {Cole},\ and\ \citenamefont
  {Berry}}]{moore_gravitational-wave_2015}%
  \BibitemOpen
  \bibfield  {author} {\bibinfo {author} {\bibfnamefont {C.~J.}\ \bibnamefont
  {Moore}}, \bibinfo {author} {\bibfnamefont {R.~H.}\ \bibnamefont {Cole}},\
  and\ \bibinfo {author} {\bibfnamefont {C.~P.~L.}\ \bibnamefont {Berry}},\
  }\bibfield  {title} {\bibinfo {title} {Gravitational-wave sensitivity
  curves},\ }\href {https://doi.org/10.1088/0264-9381/32/1/015014} {\bibfield
  {journal} {\bibinfo  {journal} {Classical and Quantum Gravity}\ }\textbf
  {\bibinfo {volume} {32}},\ \bibinfo {pages} {015014} (\bibinfo {year}
  {2015})}\BibitemShut {NoStop}%
\bibitem [{\citenamefont {Korol}\ \emph {et~al.}(2017)\citenamefont {Korol},
  \citenamefont {Rossi}, \citenamefont {Groot}, \citenamefont {Nelemans},
  \citenamefont {Toonen},\ and\ \citenamefont {Brown}}]{korol_prospects_2017}%
  \BibitemOpen
  \bibfield  {author} {\bibinfo {author} {\bibfnamefont {V.}~\bibnamefont
  {Korol}}, \bibinfo {author} {\bibfnamefont {E.~M.}\ \bibnamefont {Rossi}},
  \bibinfo {author} {\bibfnamefont {P.~J.}\ \bibnamefont {Groot}}, \bibinfo
  {author} {\bibfnamefont {G.}~\bibnamefont {Nelemans}}, \bibinfo {author}
  {\bibfnamefont {S.}~\bibnamefont {Toonen}},\ and\ \bibinfo {author}
  {\bibfnamefont {A.~G.~A.}\ \bibnamefont {Brown}},\ }\bibfield  {title}
  {\bibinfo {title} {Prospects for detection of detached double white dwarf
  binaries with {Gaia}, {LSST} and {LISA}},\ }\href
  {https://doi.org/10.1093/mnras/stx1285} {\bibfield  {journal} {\bibinfo
  {journal} {Monthly Notices of the Royal Astronomical Society}\ }\textbf
  {\bibinfo {volume} {470}},\ \bibinfo {pages} {1894} (\bibinfo {year}
  {2017})}\BibitemShut {NoStop}%
\bibitem [{\citenamefont {Korol}\ \emph {et~al.}(2018)\citenamefont {Korol},
  \citenamefont {Koop},\ and\ \citenamefont
  {Rossi}}]{korol_detectability_2018}%
  \BibitemOpen
  \bibfield  {author} {\bibinfo {author} {\bibfnamefont {V.}~\bibnamefont
  {Korol}}, \bibinfo {author} {\bibfnamefont {O.}~\bibnamefont {Koop}},\ and\
  \bibinfo {author} {\bibfnamefont {E.~M.}\ \bibnamefont {Rossi}},\ }\bibfield
  {title} {\bibinfo {title} {Detectability of {Double} {White} {Dwarfs} in the
  {Local} {Group} with {LISA}},\ }\href
  {https://doi.org/10.3847/2041-8213/aae587} {\bibfield  {journal} {\bibinfo
  {journal} {The Astrophysical Journal}\ }\textbf {\bibinfo {volume} {866}},\
  \bibinfo {pages} {L20} (\bibinfo {year} {2018})}\BibitemShut {NoStop}%
\bibitem [{\citenamefont {Lamberts}\ \emph {et~al.}(2019)\citenamefont
  {Lamberts}, \citenamefont {Blunt}, \citenamefont {Littenberg}, \citenamefont
  {Garrison-Kimmel}, \citenamefont {Kupfer},\ and\ \citenamefont
  {Sanderson}}]{lamberts_predicting_2019}%
  \BibitemOpen
  \bibfield  {author} {\bibinfo {author} {\bibfnamefont {A.}~\bibnamefont
  {Lamberts}}, \bibinfo {author} {\bibfnamefont {S.}~\bibnamefont {Blunt}},
  \bibinfo {author} {\bibfnamefont {T.}~\bibnamefont {Littenberg}}, \bibinfo
  {author} {\bibfnamefont {S.}~\bibnamefont {Garrison-Kimmel}}, \bibinfo
  {author} {\bibfnamefont {T.}~\bibnamefont {Kupfer}},\ and\ \bibinfo {author}
  {\bibfnamefont {R.}~\bibnamefont {Sanderson}},\ }\bibfield  {title} {\bibinfo
  {title} {Predicting the {LISA} white dwarf binary population in the {Milky}
  {Way} with cosmological simulations},\ }\href
  {https://doi.org/10.1093/mnras/stz2834} {\bibfield  {journal} {\bibinfo
  {journal} {Monthly Notices of the Royal Astronomical Society}\ }\textbf
  {\bibinfo {volume} {490}},\ \bibinfo {pages} {5888} (\bibinfo {year}
  {2019})},\ \bibinfo {note} {arXiv: 1907.00014}\BibitemShut {NoStop}%
\bibitem [{\citenamefont {Korol}\ \emph {et~al.}(2019)\citenamefont {Korol},
  \citenamefont {Rossi},\ and\ \citenamefont
  {Barausse}}]{korol_multimessenger_2019}%
  \BibitemOpen
  \bibfield  {author} {\bibinfo {author} {\bibfnamefont {V.}~\bibnamefont
  {Korol}}, \bibinfo {author} {\bibfnamefont {E.~M.}\ \bibnamefont {Rossi}},\
  and\ \bibinfo {author} {\bibfnamefont {E.}~\bibnamefont {Barausse}},\
  }\bibfield  {title} {\bibinfo {title} {A multimessenger study of the {Milky}
  {Way}’s stellar disc and bulge with {LISA}, \textit{{Gaia}} , and {LSST}},\
  }\href {https://doi.org/10.1093/mnras/sty3440} {\bibfield  {journal}
  {\bibinfo  {journal} {Monthly Notices of the Royal Astronomical Society}\
  }\textbf {\bibinfo {volume} {483}},\ \bibinfo {pages} {5518} (\bibinfo {year}
  {2019})}\BibitemShut {NoStop}%
\bibitem [{\citenamefont {Korol}\ \emph {et~al.}(2020)\citenamefont {Korol},
  \citenamefont {Toonen}, \citenamefont {Klein}, \citenamefont {Belokurov},
  \citenamefont {Vincenzo}, \citenamefont {Buscicchio}, \citenamefont {Gerosa},
  \citenamefont {Moore}, \citenamefont {Roebber}, \citenamefont {Rossi},\ and\
  \citenamefont {Vecchio}}]{korol_populations_2020}%
  \BibitemOpen
  \bibfield  {author} {\bibinfo {author} {\bibfnamefont {V.}~\bibnamefont
  {Korol}}, \bibinfo {author} {\bibfnamefont {S.}~\bibnamefont {Toonen}},
  \bibinfo {author} {\bibfnamefont {A.}~\bibnamefont {Klein}}, \bibinfo
  {author} {\bibfnamefont {V.}~\bibnamefont {Belokurov}}, \bibinfo {author}
  {\bibfnamefont {F.}~\bibnamefont {Vincenzo}}, \bibinfo {author}
  {\bibfnamefont {R.}~\bibnamefont {Buscicchio}}, \bibinfo {author}
  {\bibfnamefont {D.}~\bibnamefont {Gerosa}}, \bibinfo {author} {\bibfnamefont
  {C.~J.}\ \bibnamefont {Moore}}, \bibinfo {author} {\bibfnamefont
  {E.}~\bibnamefont {Roebber}}, \bibinfo {author} {\bibfnamefont {E.~M.}\
  \bibnamefont {Rossi}},\ and\ \bibinfo {author} {\bibfnamefont
  {A.}~\bibnamefont {Vecchio}},\ }\bibfield  {title} {\bibinfo {title}
  {Populations of double white dwarfs in {Milky} {Way} satellites and their
  detectability with {LISA}},\ }\href
  {https://doi.org/10.1051/0004-6361/202037764} {\bibfield  {journal} {\bibinfo
   {journal} {Astronomy \& Astrophysics}\ }\textbf {\bibinfo {volume} {638}},\
  \bibinfo {pages} {A153} (\bibinfo {year} {2020})}\BibitemShut {NoStop}%
\bibitem [{\citenamefont {Georgousi}\ \emph {et~al.}(2022)\citenamefont
  {Georgousi}, \citenamefont {Karnesis}, \citenamefont {Korol}, \citenamefont
  {Pieroni},\ and\ \citenamefont {Stergioulas}}]{georgousi_gravitational_2022}%
  \BibitemOpen
  \bibfield  {author} {\bibinfo {author} {\bibfnamefont {M.}~\bibnamefont
  {Georgousi}}, \bibinfo {author} {\bibfnamefont {N.}~\bibnamefont {Karnesis}},
  \bibinfo {author} {\bibfnamefont {V.}~\bibnamefont {Korol}}, \bibinfo
  {author} {\bibfnamefont {M.}~\bibnamefont {Pieroni}},\ and\ \bibinfo {author}
  {\bibfnamefont {N.}~\bibnamefont {Stergioulas}},\ }\bibfield  {title}
  {\bibinfo {title} {Gravitational {Waves} from {Double} {White} {Dwarfs} as
  probes of the {Milky} {Way}},\ }\href {http://arxiv.org/abs/2204.07349}
  {\bibfield  {journal} {\bibinfo  {journal} {arXiv:2204.07349 [astro-ph,
  physics:gr-qc, physics:physics]}\ } (\bibinfo {year} {2022})},\ \bibinfo
  {note} {arXiv: 2204.07349}\BibitemShut {NoStop}%
\end{thebibliography}%
	
\end{document}